\documentclass[12pt]{article}
\pdfoutput=1

\usepackage{amsmath,amsfonts}
\usepackage{graphicx}
\usepackage{feynmp}
\usepackage{multirow}
\usepackage{setspace}
\usepackage{bigstrut}        
\usepackage[numbers,sort&compress]{natbib}

\usepackage{nicematrix}       
\usepackage{subcaption}   

\usepackage{calligra}
\usepackage{bbm}
\usepackage{physics}
 \usepackage{float}
 \usepackage{tikz}

\usepackage{xcolor}

\usepackage{hyperref}
\hypersetup{
pdfstartview={FitH},
pdftitle={Monodromy defects in Chern-Simons theory and Holography},
pdfauthor={F. Ambrosino, J. Gomis, S. R. Kannagi},
colorlinks=true,
citecolor=blue,
linkcolor=blue,
urlcolor=blue,
}

\usepackage{amsthm}
\usepackage{amssymb}
\usepackage{epic}
\usepackage[matrix,arrow]{xy}
\usepackage{array}
\usepackage{graphicx}
\usepackage{tikz}
\usepackage{cancel}
\usepackage{float}
\usepackage{mathrsfs}
\usetikzlibrary{decorations.markings,decorations.pathmorphing,arrows.meta,calc,decorations.pathreplacing,positioning}
\usepackage[export]{adjustbox}

\usepackage{psfrag}
\usepackage{float}
\usepackage{textcomp}

\makeatletter\@addtoreset{equation}{section}\makeatother

\newcommand{\beq}{\begin{equation}}
\newcommand{\eeq}{\end{equation}}
\newcommand{\bea}{\begin{eqnarray}}
\newcommand{\eea}{\end{eqnarray}}

\newcommand{\that}{\widehat t\, }
\newcommand{\gs}{g_s}
\newcommand{\D}{\mathcal A}
\newcommand{\AntiP}{\mathcal B}
\newcommand{\SymP}{\overline{\mathcal B}}

\renewcommand{\title}[1]{\vbox{\center\LARGE{#1}}\vspace{5mm}}
\renewcommand{\author}[1]{\vbox{\center\large#1}\vspace{5mm}}
\newcommand{\address}[1]{\vbox{\center\em#1}}

\usepackage{booktabs}
\newcommand{\Li}{\operatorname{Li}}
\newcommand{\cA}{{\mathcal A}}
\newcommand{\cB}{{\mathcal B}}
\newcommand{\cF}{{\mathcal F}}
\newcommand{\sumeven}{{\sum_{\substack{m\ge1\\m\ {\rm even}}}
}}
\newcommand{\sumodd}{{\sum_{\substack{m\ge1\\m\ {\rm odd}}}
}}
\newcommand{\sumall}{{\sum_{{m\ge1}}
}}

\newsavebox{\MrowYoungbox}
\newsavebox{\MrowCYbox}
\newsavebox{\MrowTextbox}
\newsavebox{\ProwYoungbox}
\newsavebox{\ProwCYbox}
\newsavebox{\ProwTextbox}
\newlength{\Mrowheight}
\newlength{\Prowheight}
\newlength{\Rowheighttmp}
\newcommand{\updrowheight}[2]{%
  \setlength{\Rowheighttmp}{\dimexpr\ht#2+\dp#2\relax}%
  \ifdim\Rowheighttmp>#1
    \setlength{#1}{\Rowheighttmp}%
  \fi
}
\newcommand{\Rowcell}[3]{%
  \begin{minipage}[c][#1][c]{#2}%
    \parbox{#2}{\centering #3}%
  \end{minipage}%
}

\newsavebox{\MrowFigbox}
\newsavebox{\ProwFigbox}

\newcommand{\dynsp}{.62}
\tikzset{dynnode/.style={circle,draw,thick,fill=white,inner sep=0pt,minimum size=5.5pt}}

\newcommand{\dynmk}[2]{\node[scale=.72] at ($(#1)+(0,.30)$) {$#2$};}      
\newcommand{\dynmkb}[2]{\node[scale=.72] at ($(#1)+(0,-.30)$) {$#2$};}    
\newcommand{\dynsimple}[2]{\draw[thick] (#1)--(#2);}
\newcommand{\dyndoubleR}[2]{%
  \draw[thick] ($(#1)+(0,1.4pt)$)--($(#2)+(0,1.4pt)$);
  \draw[thick] ($(#1)-(0,1.4pt)$)--($(#2)-(0,1.4pt)$);
  \coordinate (mm) at ($(#1)!.5!(#2)$);
  \draw[thick] ($(mm)+(-2.6pt,3pt)$)--($(mm)+(2.2pt,0)$)--($(mm)+(-2.6pt,-3pt)$);}
\newcommand{\dyndoubleL}[2]{%
  \draw[thick] ($(#1)+(0,1.4pt)$)--($(#2)+(0,1.4pt)$);
  \draw[thick] ($(#1)-(0,1.4pt)$)--($(#2)-(0,1.4pt)$);
  \coordinate (mm) at ($(#1)!.5!(#2)$);
  \draw[thick] ($(mm)+(2.6pt,3pt)$)--($(mm)+(-2.2pt,0)$)--($(mm)+(2.6pt,-3pt)$);}
\newcommand{\dynname}[1]{\rule{0pt}{2.6ex}\large $#1$}
\colorlet{affcol}{blue}
\tikzset{affnode/.style={dynnode,draw=affcol}}

\newcommand{\diagAtwoN}{\begin{tikzpicture}[baseline=(o2.base)]
  \foreach \i in {0,1,2} {\coordinate (a\i) at (\i*\dynsp,0);}
  \coordinate (cd) at (3*\dynsp,0); \coordinate (b1) at (4*\dynsp,0); \coordinate (b0) at (5*\dynsp,0);
  \dynsimple{a0}{a1}\dynsimple{a1}{a2}\draw[thick](a2)--($(cd)+(-.16,0)$);
  \draw[thick]($(cd)+(.16,0)$)--(b1);\dynsimple{b1}{b0}
  \node[scale=.85] at (cd) {$\cdots$};
  \node[dynnode](o0)at(a0){};\node[dynnode](o1)at(a1){};\node[dynnode](o2)at(a2){};
  \node[dynnode](p1)at(b1){};\node[dynnode](p0)at(b0){};
  \dynmk{a0}{1}\dynmk{a1}{1}\dynmk{a2}{1}\dynmk{b1}{1}\dynmk{b0}{1}
  \dynmkb{a0}{1}\dynmkb{a1}{1}\dynmkb{a2}{1}\dynmkb{b1}{1}\dynmkb{b0}{1}
\end{tikzpicture}}

\newcommand{\diagAtwoNm}{\begin{tikzpicture}[baseline=(o2.base)]
  \foreach \i in {0,1,2} {\coordinate (a\i) at (\i*\dynsp,0);}
  \coordinate (cd) at (3*\dynsp,0); \coordinate (b1) at (4*\dynsp,0); \coordinate (b0) at (5*\dynsp,0);
  \dynsimple{a0}{a1}\dynsimple{a1}{a2}\draw[thick](a2)--($(cd)+(-.16,0)$);
  \draw[thick]($(cd)+(.16,0)$)--(b1);\dynsimple{b1}{b0}
  \node[scale=.85] at (cd) {$\cdots$};
  \node[dynnode](o0)at(a0){};\node[dynnode](o1)at(a1){};\node[dynnode](o2)at(a2){};
  \node[dynnode](p1)at(b1){};\node[dynnode](p0)at(b0){};
  \dynmk{a0}{1}\dynmk{a1}{1}\dynmk{a2}{1}\dynmk{b1}{1}\dynmk{b0}{1}
  \dynmkb{a0}{1}\dynmkb{a1}{1}\dynmkb{a2}{1}\dynmkb{b1}{1}\dynmkb{b0}{1}
\end{tikzpicture}}

\newcommand{\diagAtwoNaff}{\begin{tikzpicture}[baseline=(o0.base)]
  \coordinate (n0) at (0,0); \coordinate (n1) at (\dynsp,0); \coordinate (n2) at (2*\dynsp,0);
  \coordinate (cd) at (3*\dynsp,0); \coordinate (nm) at (4*\dynsp,0); \coordinate (nN) at (5*\dynsp,0);
  \dyndoubleL{n0}{n1}\dynsimple{n1}{n2}\draw[thick](n2)--($(cd)+(-.16,0)$);
  \draw[thick]($(cd)+(.16,0)$)--(nm);\dyndoubleL{nm}{nN}
  \node[scale=.85] at (cd) {$\cdots$};
  \node[affnode](o0)at(n0){};\node[dynnode](o1)at(n1){};\node[dynnode](o2)at(n2){};
  \node[dynnode](om)at(nm){};\node[dynnode](oN)at(nN){};
  \dynmk{n0}{{\color{affcol}1}}\dynmk{n1}{2}\dynmk{n2}{2}\dynmk{nm}{2}\dynmk{nN}{2}
  \dynmkb{n0}{\color{blue} 2 }\dynmkb{n1}{2}\dynmkb{n2}{2}\dynmkb{nm}{2}\dynmkb{nN}{1}
\end{tikzpicture}}

\newcommand{\diagAtwoNmaff}{\begin{tikzpicture}[baseline=(o2.base)]
  \coordinate (n0) at (.14,.28); \coordinate (n1) at (.14,-.28); \coordinate (n2) at (.72,0);
  \coordinate (n3) at (.72+\dynsp,0); \coordinate (cd) at (.72+2*\dynsp,0);
  \coordinate (nm) at (.72+3*\dynsp,0); \coordinate (nN) at (.72+4*\dynsp,0);
  \dynsimple{n0}{n2}\dynsimple{n1}{n2}\dynsimple{n2}{n3}\draw[thick](n3)--($(cd)+(-.16,0)$);
  \draw[thick]($(cd)+(.16,0)$)--(nm);\dyndoubleL{nm}{nN}
  \node[scale=.85] at (cd) {$\cdots$};
  \node[affnode](o0)at(n0){};\node[dynnode](o1)at(n1){};\node[dynnode](o2)at(n2){};
  \node[dynnode](o3)at(n3){};\node[dynnode](om)at(nm){};\node[dynnode](oN)at(nN){};
  \node[scale=.72] at ($(n0)+(-.18,.16)$) {\textcolor{affcol}{$1$}};
  \node[scale=.72] at ($(n0)+(-.18,-.16)$) {\textcolor{affcol}{$1$}};
  \node[scale=.72] at ($(n1)+(-.18,.16)$) {$1$};
  \node[scale=.72] at ($(n1)+(-.18,-.16)$) {$1$};
  \dynmk{n2}{2}\dynmk{n3}{2}\dynmk{nm}{2}\dynmk{nN}{2}
  \dynmkb{n2}{2}\dynmkb{n3}{2}\dynmkb{nm}{2}\dynmkb{nN}{1}
\end{tikzpicture}}

\newcommand{\diagDaff}{\begin{tikzpicture}[baseline=(o0.base)]
  \coordinate (n0) at (0,0); \coordinate (n1) at (\dynsp,0); \coordinate (n2) at (2*\dynsp,0);
  \coordinate (cd) at (3*\dynsp,0); \coordinate (nm) at (4*\dynsp,0); \coordinate (nN) at (5*\dynsp,0);
  \dyndoubleL{n0}{n1}
  \dynsimple{n1}{n2}\draw[thick](n2)--($(cd)+(-.16,0)$);
  \draw[thick]($(cd)+(.16,0)$)--(nm);\dyndoubleR{nm}{nN}
  \node[scale=.85] at (cd) {$\cdots$};
  \node[affnode](o0)at(n0){};\node[dynnode](o1)at(n1){};\node[dynnode](o2)at(n2){};
  \node[dynnode](om)at(nm){};\node[dynnode](oN)at(nN){};
  \dynmk{n0}{{\color{affcol}1}}\dynmk{n1}{2}\dynmk{n2}{2}\dynmk{nm}{2}\dynmk{nN}{1}
  \dynmkb{n0}{\color{ affcol} 1}\dynmkb{n1}{1}\dynmkb{n2}{1}\dynmkb{nm}{1}\dynmkb{nN}{1}
\end{tikzpicture}}

\newcommand{\diagBN}{\begin{tikzpicture}[baseline=(o0.base)]
  \coordinate (a0) at (0,0); \coordinate (a1) at (\dynsp,0);
  \coordinate (cd) at (2*\dynsp,0); \coordinate (b1) at (3*\dynsp,0); \coordinate (b0) at (4*\dynsp,0);
  \dynsimple{a0}{a1}\draw[thick](a1)--($(cd)+(-.16,0)$);
  \draw[thick]($(cd)+(.16,0)$)--(b1);\dyndoubleR{b1}{b0}
  \node[scale=.85] at (cd) {$\cdots$};
  \node[dynnode](o0)at(a0){};\node[dynnode](o1)at(a1){};\node[dynnode](p1)at(b1){};\node[dynnode](p0)at(b0){};
  \dynmk{a0}{1}\dynmk{a1}{2}\dynmk{p1}{2}\dynmk{b0}{1}
  \dynmkb{a0}{1}\dynmkb{a1}{2}\dynmkb{p1}{2}\dynmkb{b0}{2}
\end{tikzpicture}}

\newcommand{\diagCN}{\begin{tikzpicture}[baseline=(o0.base)]
  \coordinate (a0) at (0,0); \coordinate (a1) at (\dynsp,0);
  \coordinate (cd) at (2*\dynsp,0); \coordinate (b1) at (3*\dynsp,0); \coordinate (b0) at (4*\dynsp,0);
  \dynsimple{a0}{a1}\draw[thick](a1)--($(cd)+(-.16,0)$);
  \draw[thick]($(cd)+(.16,0)$)--(b1);\dyndoubleL{b1}{b0}
  \node[scale=.85] at (cd) {$\cdots$};
  \node[dynnode](o0)at(a0){};\node[dynnode](o1)at(a1){};\node[dynnode](p1)at(b1){};\node[dynnode](p0)at(b0){};
  \dynmk{a0}{1}\dynmk{a1}{1}\dynmk{p1}{1}\dynmk{b0}{1}
  \dynmkb{a0}{2}\dynmkb{a1}{2}\dynmkb{p1}{2}\dynmkb{b0}{1}
\end{tikzpicture}}

\newcommand{\diagAoneExt}{%
\begin{tikzpicture}[baseline=(a0.base)]
  \coordinate (a0) at (0,0);
  \foreach \i in {1,2} {\coordinate (a\i) at (\i*\dynsp,0);}
  \coordinate (cd) at (3*\dynsp,0);
  \coordinate (b1) at (4*\dynsp,0);
  \coordinate (b0) at (5*\dynsp,0);
  \coordinate (n0) at (2.5*\dynsp,-.30);
  \dynsimple{a0}{a1}\dynsimple{a1}{a2}
  \draw[thick](a2)--($(cd)+(-.16,0)$);
  \draw[thick]($(cd)+(.16,0)$)--(b1);\dynsimple{b1}{b0}
  \dynsimple{n0}{a0}\dynsimple{n0}{b0}
  \node[scale=.85] at (cd) {$\cdots$};
  \node[affnode](o0)at(n0){};\node[dynnode]at(a0){};\node[dynnode]at(a1){};\node[dynnode]at(a2){};
  \node[dynnode]at(b1){};\node[dynnode]at(b0){};
\end{tikzpicture}}

\newcommand{\diagAtwoNaffExt}{%
\begin{tikzpicture}[baseline=(o0.base)]
  \coordinate (n0) at (0,0); \coordinate (n1) at (\dynsp,0); \coordinate (n2) at (2*\dynsp,0);
  \coordinate (cd) at (3*\dynsp,0); \coordinate (nm) at (4*\dynsp,0); \coordinate (nN) at (5*\dynsp,0);
  \dyndoubleL{n0}{n1}\dynsimple{n1}{n2}\draw[thick](n2)--($(cd)+(-.16,0)$);
  \draw[thick]($(cd)+(.16,0)$)--(nm);\dyndoubleL{nm}{nN}
  \node[scale=.85] at (cd) {$\cdots$};
  \node[affnode](o0)at(n0){};\node[dynnode]at(n1){};\node[dynnode]at(n2){};
  \node[dynnode]at(nm){};\node[dynnode]at(nN){};
\end{tikzpicture}}

\newcommand{\diagAtwoNmaffExt}{%
\begin{tikzpicture}[baseline=(o2.base)]
  \coordinate (n0) at (.14,.28); \coordinate (n1) at (.14,-.28); \coordinate (n2) at (.72,0);
  \coordinate (n3) at (.72+\dynsp,0); \coordinate (cd) at (.72+2*\dynsp,0);
  \coordinate (nm) at (.72+3*\dynsp,0); \coordinate (nN) at (.72+4*\dynsp,0);
  \dynsimple{n0}{n2}\dynsimple{n1}{n2}\dynsimple{n2}{n3}\draw[thick](n3)--($(cd)+(-.16,0)$);
  \draw[thick]($(cd)+(.16,0)$)--(nm);\dyndoubleL{nm}{nN}
  \node[scale=.85] at (cd) {$\cdots$};
  \node[affnode](o0)at(n0){};\node[dynnode](o1)at(n1){};\node[dynnode](o2)at(n2){};
  \node[dynnode]at(n3){};\node[dynnode]at(nm){};\node[dynnode]at(nN){};
\end{tikzpicture}}

\newcommand{\diagDaffExt}{%
\begin{tikzpicture}[baseline=(o0.base)]
  \coordinate (n0) at (0,0); \coordinate (n1) at (\dynsp,0); \coordinate (n2) at (2*\dynsp,0);
  \coordinate (cd) at (3*\dynsp,0); \coordinate (nm) at (4*\dynsp,0); \coordinate (nN) at (5*\dynsp,0);
  \dyndoubleL{n0}{n1}\dynsimple{n1}{n2}\draw[thick](n2)--($(cd)+(-.16,0)$);
  \draw[thick]($(cd)+(.16,0)$)--(nm);\dyndoubleR{nm}{nN}
  \node[scale=.85] at (cd) {$\cdots$};
  \node[affnode](o0)at(n0){};\node[dynnode]at(n1){};\node[dynnode]at(n2){};
  \node[dynnode]at(nm){};\node[dynnode]at(nN){};
\end{tikzpicture}}

\newcommand{\diagDoneExt}{%
\begin{tikzpicture}[baseline=(n0.base)]
  \coordinate (n0) at (0,0); \coordinate (n1) at (\dynsp,0);
  \coordinate (cd) at (2*\dynsp,0); \coordinate (nm) at (3*\dynsp,0);
  \coordinate (nN) at (3*\dynsp+.58,.28); \coordinate (nB) at (3*\dynsp+.58,-.28);
  \dynsimple{n0}{n1}\draw[thick](n1)--($(cd)+(-.16,0)$);
  \draw[thick]($(cd)+(.16,0)$)--(nm);
  \dynsimple{nm}{nN}\dynsimple{nm}{nB}
  \node[scale=.85] at (cd) {$\cdots$};
  \node[affnode](o0)at(n0){};\node[dynnode]at(n1){};\node[dynnode]at(nm){};
  \node[dynnode]at(nN){};\node[dynnode]at(nB){};
\end{tikzpicture}}

\newcommand{\diagEoneExt}{%
\begin{tikzpicture}[baseline=(n4.base)]
  \coordinate (n4) at (2*\dynsp,0);
  \coordinate (n3) at (3*\dynsp,0);
  \coordinate (n1) at (4*\dynsp,0);
  \coordinate (n5) at (2*\dynsp-.48,.18);
  \coordinate (n6) at (2*\dynsp-.96,.36);
  \coordinate (n2) at (2*\dynsp-.48,-.18);
  \coordinate (n0) at (2*\dynsp-.96,-.36);
  \dynsimple{n1}{n3}\dynsimple{n3}{n4}
  \dynsimple{n4}{n5}\dynsimple{n5}{n6}
  \dynsimple{n4}{n2}\dynsimple{n2}{n0}
  \node[affnode](o0)at(n0){};\node[dynnode]at(n1){};\node[dynnode]at(n2){};\node[dynnode]at(n3){};
  \node[dynnode](o4)at(n4){};\node[dynnode]at(n5){};\node[dynnode]at(n6){};
\end{tikzpicture}}

\newcommand{\diagEtwoExt}{%
\begin{tikzpicture}[baseline=(o0.base)]
  \coordinate (n0) at (0,0); \coordinate (n1) at (\dynsp,0);
  \coordinate (n2) at (2*\dynsp,0); \coordinate (n3) at (3*\dynsp,0); \coordinate (n4) at (4*\dynsp,0);
  \dynsimple{n0}{n1}\dynsimple{n1}{n2}\dyndoubleL{n2}{n3}\dynsimple{n3}{n4}
  \node[affnode](o0)at(n0){};\node[dynnode]at(n1){};\node[dynnode]at(n2){};\node[dynnode]at(n3){};\node[dynnode]at(n4){};
\end{tikzpicture}}

\usepackage{comment}

\tikzset{
  line/.style={line width=1.0pt},
  brane/.style={line width=1.25pt,blue!70!black},
  antibrane/.style={line width=1.25pt,blue!70!black},
  oplane/.style={line width=1.0pt,red!80!black,dashed},
  branepoint/.style={circle,fill=blue!70!black,inner sep=2.25pt},
  stuckbrane/.style={line width=1.6pt,blue!70!black},
  stuckbranepoint/.style={circle,fill=blue!70!black,inner sep=3.0pt},
  redpoint/.style={circle,fill=red!80!black,inner sep=1.4pt},
}

\newcommand{\toricConifoldBaseOrientifold}{%
  \coordinate (L) at (0,0);
  \coordinate (R) at (2.60,0);
  \coordinate (C) at ($(L)!0.5!(R)$);
  \draw[line] (L) -- (R);
  \draw[line] (L) -- ++(0,1.00);
  \draw[line] (L) -- ++(-0.85,-0.72);
  \draw[line] (R) -- ++(0.85,1.00);
  \draw[line] (R) -- ++(0,-1.00);
}

\newcommand{\toricCanvasOrientifold}{%
  \path[use as bounding box] (-1.05,-1.05) rectangle (3.62,1.65);
}

\newcommand{\toricStuckBrane}{%
  \draw[stuckbrane] (C) -- ++(0,1.00);
  \node[stuckbranepoint] at (C) {};
  \draw[oplane] ($(C)+(0,-1.00)$) -- ($(C)+(0,1.00)$);
  \node[redpoint] at (C) {};
}

\usepackage{cancel}

\begin{document}
\bibliographystyle{utphys}
\begin{fmffile}{graphs}

\begin{titlepage}
\begin{center}

\hfill {\tt }\\

\title{\huge{Monodromy defects in Chern--Simons theory and Holography}}

\vspace{6mm}

Federico  Ambrosino,$^{\Omega\sigma_+}\!\!$~\footnote{\href{mailto:federicoambrosino25@gmail.com}{\tt federicoambrosino25@gmail.com}}
Jaume Gomis,$^{\Omega\sigma_+}\!\!$~\footnote{\href{mailto:jgomis@perimeterinstitute.ca}{\tt jgomis@perimeterinstitute.ca}}
Suriyah Rajalingam Kannagi$^{\Omega\sigma_+,\Omega\sigma_-}\!\!$~\footnote{\href{mailto:srajalingamkannagi@perimeterinstitute.ca}{\tt srajalingamkannagi@perimeterinstitute.ca}}
 \vskip 3mm
\address{
$^{\Omega\sigma_+}$Perimeter Institute for Theoretical Physics, \\
Waterloo, Ontario, N2L 2Y5, Canada\\[2ex]
$^{\Omega\sigma_-}$Department of Physics and Astronomy, University of Waterloo,\\ Waterloo, Ontario,N2L 3G1, Canada}

\end{center}
\abstract{ 
\noindent
\normalsize{Wilson loop operators in Chern--Simons theory have revealed profound links
between quantum field theory, the fractional quantum Hall effect, topology, conformal field theory, and string theory.  In Chern--Simons theories with charge conjugation symmetry, we construct a   new class of
observables: codimension-two monodromy defects around which fields return to themselves up to charge
conjugation. Whereas Wilson loops
are labeled by integrable representations of an untwisted affine Lie
algebra, monodromy defects are labeled by those of the corresponding
twisted affine algebra.
The modular and fusion data of these two algebras  
determine the exact correlation functions of Wilson lines and monodromy defects,
which together furnish a $\mathbb{Z}_2$-crossed braided tensor category.
 The   spectrum of line defects in Chern--Simons theory thus gives a physical
realization of every algebra  in Kac's classification of affine Lie algebras: untwisted for Wilson loops, twisted for monodromy defects.  We also determine the exact 't~Hooft expansion of monodromy defects in $SU(N)_k$ Chern--Simons theory and identify their holographic duals in
topological string theory.
The insertion of the lightest monodromy defect has a striking effect: it
replaces the resolved conifold background of the Gopakumar--Vafa duality
by a specific orientifold of the resolved conifold, transmuting the dual
theory of oriented strings into one of unoriented strings. Each excited monodromy defect is then realized as a collection of branes
in the orientifold background, with the brane content determined by the
representation of the twisted affine algebra that labels the defect.   }
\noindent
}
\vfill
\end{titlepage}\setcounter{tocdepth}{2}
\setcounter{footnote}{0}

\tableofcontents
 \vfill\eject

\section{Introduction}
\label{sec:intro}

Chern--Simons theory with gauge group $G$ at integer level $k$ ($G_k$ Chern--Simons theory)   is defined by the action
\begin{equation}
S_{\mathrm{CS}}[A]
=
\frac{k}{4\pi}\int_M \operatorname{Tr}\!\left(
A\wedge dA+\frac{2}{3}A\wedge A\wedge A
\right),
\label{action}
\end{equation}
where $M$ is an oriented three-manifold and $A$ is a connection on a principal $G$-bundle over $M$.
Since the classical equations of motion impose  $F=0$, there are no nontrivial local gauge invariant operators in the theory. The basic observables of Chern--Simons theory are instead nonlocal Wilson loop operators,
\begin{equation}
W_R(K)=\operatorname{Tr}_R\,\mathcal{P}\exp\left(i\oint_K A\right),
\end{equation}
where $K\subset M$ is an oriented knot and $R$ is a representation of $G$.\footnote{$G$ is taken to be compact, simple, connected, and simply connected, with Lie algebra denoted by $\mathfrak g$.}

Chern--Simons theory holds a privileged status in quantum field theory. It is an interacting theory in which every known observable --  the partition function on any closed three-manifold, and the expectation value of Wilson loops along arbitrary knots and links --  is exactly computable~\cite{Witten:1988hf,Reshetikhin:1991tc}: here the analytic difficulties of quantum field theory surrender to the rigid structures of topology and two-dimensional conformal field theory. Nor is this solubility a purely theoretical delight. As the effective field theory of the fractional quantum Hall effect~\cite{Zhang:1989,Wen:1995qn}, Chern--Simons theory makes hallmark predictions --  fractional charge, anyonic statistics, quantized Hall conductance --   which have been confirmed experimentally~\cite{dePicciotto:1997,Bartolomei:2020}.

\medskip
Chern--Simons theory has forged profound connections with   mathematics. Witten demonstrated~\cite{Witten:1988hf} that expectation values of Wilson loops yield topological invariants of knots and links -- for $G=SU(2)$, the Jones polynomial~\cite{Jones:1985} -- while the partition function on a closed oriented three-manifold defines the Reshetikhin--Turaev invariant~\cite{Reshetikhin:1991tc,Turaev:1992hq}. Canonical quantization on a Riemann surface $\Sigma$ identifies the Hilbert space of the theory with the space of conformal blocks of the $\mathfrak{g}_k$  affine Lie algebra~\cite{Witten:1988hf,Elitzur:1989nr}, weaving three-dimensional topology, two-dimensional conformal field theory, affine Lie algebras, and modular tensor categories~\cite{Moore:1988qv,Turaev:1992hq} into a single tapestry. 
 
 Chern--Simons theory elegantly realizes 
't~Hooft's vision of a dual string theory description of  large $N$ gauge theories~\cite{tHooft:1973alw}. 
The large \(N\) expansion beautifully reorganizes the partition function and Wilson loop amplitudes of Chern--Simons as  
topological string amplitudes on Calabi--Yau geometries. This duality has
provided powerful insights into enumerative geometry, including
Gromov--Witten invariants~\cite{Gopakumar:1998ki,Ooguri:1999bv}.

The far-reaching consequences of Chern--Simons theory -- for knot theory,
modular tensor categories, large \(N\)   duality, and enumerative
geometry  -- have been developed   from a single class of observables:
Wilson loop operators. 

\vspace{1.5pt}
This naturally raises two elementary questions:
\begin{enumerate}
    \item Are Wilson loop operators the only     observables in
    Chern--Simons theory?
    \item If additional   observables exist, what is their holographic dual 
      string theory description?
\end{enumerate}
\vspace{2pt}

\begin{figure}[H]
  \centering
  \begin{tikzpicture}
    \draw[dashed, line width=1pt] (3,3) circle (1.5cm);
    \path[
      postaction={decorate},
      decoration={
        markings,
        mark=at position 0.18 with {\arrow{Stealth[length=3mm]}}
      }
    ]
    (4.5,3) arc[start angle=0, end angle=360, radius=1.5cm];
    \node[inner sep=1pt, line width=1pt] at (3,2.4) {$\mathsf M_a$};
    \node[draw, circle, inner sep=1pt, line width=1pt] at (3,3) {$\times$};
    \draw[
      thick,
      decorate,
      decoration={
        snake,
        amplitude=0.1cm,
        segment length=4mm
      }
    ]
    (3,3) -- (5.5,5.5);
    \node at (3,5.5) {$\varphi\bigl(z\,e^{2\pi i}\bigr)=g\:\cdot\:\varphi(z)$};
  \end{tikzpicture}
  \caption{A $g$-monodromy defect  $\mathsf M_a$ imposing $g$-twisted boundary conditions. $z$ is a complex coordinate in the plane transverse to the codimension-two defect. The wavy line is a branch cut implementing the action of $g$.}
  \label{fig:branchcut}
\end{figure}
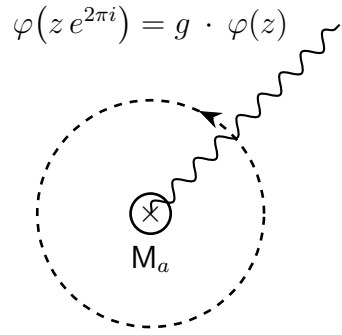

In this paper we show that Wilson loops are not the end of the story.  Chern--Simons theories with charge conjugation symmetry,\footnote{That is,  for   $\mathfrak{su}(N)$, $\mathfrak{so}(2N)$ and $\mathfrak e_6$. $\mathsf C$ denotes the charge conjugation symmetry generator. $\mathfrak{so}(8)$  has an additional order $3$ outer  automorphism symmetry generator, denoted by $\sigma$, see footnote \ref{sign}.} also known as outer automorphism symmetry,   
admit a rich collection of codimension-two defects:\footnote{We use the terms   defects  and operators for nonlocal observables interchangeably: a “defect” is   extended in time, and modifies the Hilbert space, while “operator” sits   at fixed time and acts on the Hilbert space. This is because we view these observables as insertions in the Euclidean path integral.}   $\mathsf C$-monodromy defects $\mathsf M_a$.\footnote{Monodromy defects are often referred to as disorder operators, especially when described as partial symmetry transformations on a spatial region  (see Section \ref{sec:dos}).}
 These are defined by  twisting: upon encircling the defect, fields are transformed by the 
corresponding   action of    charge conjugation (Figure \ref{fig:branchcut}). We develop the geometric and algebraic characterization of these operators within Chern--Simons theory.
 Charge conjugation monodromy (surface) defects in four-dimensional gauge  theories were studied in \cite{Gomis:2025adsCPT}.

We show  that $G_k$ Chern--Simons theory with charge conjugation symmetry   admits two distinct classes of line defects, Wilson lines and monodromy defects, labeled by:
\begin{itemize}
    \item \(W_R\)\,:     $R$ is an integrable representation of the untwisted affine Lie algebra \(\mathfrak g_k\)\,.
    \item 
     \(\mathsf M_a\)\,: $a$ is an integrable representation of the ${\mathsf C}$-twisted affine Lie algebra \(\mathfrak g^{(2)}_k\).
\end{itemize}
Mathematically, these defects organize into a tensor category:   they furnish the simple objects of a $\mathbb Z_2^{\mathsf C}$-crossed braided tensor category~\cite{Turaev2000,Kirillov2004,Mueger2005,Barkeshli2019},
\begin{equation}
    \mathcal C^\times
    =
    \mathcal C_{\mathbf 1}
    \oplus
    \mathcal C_{\mathsf C}\, .
\end{equation}
Wilson lines are the simple objects of $\mathcal C_{\mathbf 1}$, while the $\mathsf C$-monodromy defects are the simple objects of $\mathcal C_{\mathsf C}$.

\medskip 
The algebraic structure underlying the
\(\mathbb Z_2^{\mathsf C}\)-crossed braided tensor category of
Chern--Simons theory with charge conjugation symmetry is governed by the
interplay between the theory of untwisted and twisted affine Lie algebras  developed by Kac~\cite{kac_1990}.
The untwisted
affine algebra \(\mathfrak g_k\) controls the ordinary Wilson lines and their
fusion. The \(\mathsf C\)-twisted sector controls monodromy defects,  labeled by
integrable representations of the corresponding  \(\mathsf C\)-twisted affine algebra \(\mathfrak g^{(2)}_k\), which is constructed by twisting the currents generating \(\mathfrak g_k\) by the action of the outer automorphism  $\mathsf C$. In this way, the categorical data of the
charge conjugation extension --  including the action of $\mathsf C$ on Wilson
lines, the spectrum of monodromy defects,  the modular transformations
between twisted sectors, and the fusion of defects -- admit a concrete realization in the representation
theory of affine Lie algebras, knitting together untwisted and twisted affine algebras. This data can be used to compute --  exactly --  any correlation function of Wilson lines and monodromy defects 

In particular, the  expectation values of a Wilson line $W_R$ and of a monodromy defect $\mathsf M_a$ on an unknot in $S^3$ are governed by two distinct  modular $S$-matrices: $S^{(1,1)}_{RS}$ and $S^{(\mathsf C,1)}_{a\mu}$. The exact normalized expectation values are given by the   quantum dimensions of representations of 
\(\mathfrak g_k\) and  
\(\mathfrak g^{(2)}_k\) respectively:
\begin{equation}
    \langle W_R\rangle_{\mathrm{norm}}
    =\frac{S^{(1,1)}_{R0}}{S^{(1,1)}_{00}}\,,
    \qquad\qquad
    \langle \mathsf M_a\rangle_{\mathrm{norm}}
    =\frac{S^{(\mathsf C,1)}_{a0}}{S^{(1,1)}_{00}}\,.
\end{equation}
This implementation of charge conjugation symmetry in Chern--Simons theory
provides a particularly explicit and computable realization of
symmetry-enriched topological order~\cite{Barkeshli2019}.

Having identified these additional defects in Chern--Simons theory, we turn to question $2$ above.  The very existence of   nontrivial monodromy defects provides a sharp stress-test
of the holographic duality of Gopakumar--Vafa~\cite{Gopakumar:1998ki}: the duality should not only realize the
  string theory  dual of ordinary Chern--Simons observables, but also account
for the monodromy defects constructed in this paper. In the rest of the paper we accomplish this and   determine the string theory dual of monodromy defects in $SU(N)_k$ Chern--Simons theory. 

 The  original   discovery of Gopakumar--Vafa~\cite{Gopakumar:1998ki} was to realize that the partition function of $SU(N)_k$ Chern--Simons theory on $S^3$ coincides with the    A-model topological closed string partition function on
the resolved conifold geometry:
\[
    X=\mathcal O(-1)\oplus \mathcal O(-1)\longrightarrow \mathbb P^1\,.
\]
Inserting a Wilson loop operator $W_R$ in Chern--Simons theory  admits an elegant interpretation in the dual string theory~\cite{Ooguri:1999bv}: it corresponds to adding a very specific collection of branes or antibranes in the resolved conifold geometry, the collection being completely determined by the choice of representation $R$~\cite{Gomis:2006sb}.\footnote{In this paper, we use the brane realization of analytically continued
Chern--Simons theory in topological string theory~\cite{Witten:1992fb} to develop a small \(q\)
expansion for Chern--Simons observables, distinct from the conventional
\(q\to 1\) expansion of unitary Chern--Simons theory. This expansion not only
substantially simplifies computations, but also provides a new perspective into the
resurgence properties of the perturbative series and the nonperturbative
completion of topological string theory~\cite{Pasquetti:2010bps,Hatsuda:2013oxa, Alexandrov:2023wdj,Alim:2021mhp}. The reader, may refer to Section~\ref{constantmaps} for a discussion on the latter, and to Section~\ref{partitionfunctiondual} for the small $q$ expansion, and to  Appendix~\ref{smallgs} for its comparison with the small $\gs$.  }

We show that   the insertion of even the lightest vacuum monodromy defect $\mathsf M_{a_*}$ in  $SU(N)_k$ Chern--Simons theory on $S^3$ has a dramatic effect: it modifies the dual closed string background to be a very specific orientifold background $\Omega \sigma_+^{(\cdot)}$ of topological string theory on the resolved conifold, where $\Omega$ denotes worldsheet parity and $\sigma_+$ is an antiholomorphic involution of the resolved conifold that has fixed locus $\mathbb R^2\times S^1$. 
The precise    orientifold background dual to vacuum monodromy defects is summarized in Table \ref{tab:ts-backgroundsa}.\footnote{The superscript in $\Omega \sigma_+^{(\cdot)}$  encodes the two-fold  discrete choice of crosscap sign,
or equivalently, the choice of orientifold projection on open strings.} 
\begin{table}[H]
\centering
{\renewcommand{\arraystretch}{1.2}
\begin{tabular}{|c|c|c|c|}
\hline
{Theory} & {Parity of $k$} & {Background} & Planes \\
\hline
\hline
$SU(2N)_k$, & $k$ odd
&
$\Omega\sigma^{ Sp}_+$& $O^+$\\
\hline
$SU(2N)_k$, & $k$ even, $2N\leq k$
&
$\Omega\sigma^{ Sp}_+$&
$ O^+$
\\
$SU(2N)_k$, & $k$ even, $2N>k+2$
&
$\Omega\sigma^{ SO}_+$&
$ O^-$\\
\hline
$SU(2N+1)_k$, & $k$ even
&
$\Omega\sigma^{ SO}_+$&
$ O^-$
\\
\hline
$SU(2N+1)_k$, & $k$ odd
&
$\Omega\sigma^{ SO}_+ \oplus \mathcal B$&
$O^- \oplus \mathcal B $
\\
\hline
\end{tabular}}
\caption{String backgrounds dual to vacuum monodromy defects.}
\label{tab:ts-backgroundsa}
\end{table}
How about for non-vacuum monodromy defects $\mathsf M_a$ with $a\neq a_*$?  We show that the non-vacuum, excited defects, labeled by integrable representations of the twisted affine algebra
$\mathfrak{su}(N)^{(2)}_k$,  correspond in the dual string theory to  a specific collection  of  
branes or antibranes to the   \(\Omega\sigma^{{(\cdot)}}_+\)
orientifold background of the resolved conifold that describes the vacuum monodromy defect. We find that the    collection of branes or antibranes is completely determined by the choice of representation $a$, as summarized  in Table \ref{tab:summarynitro}:
\begin{table}[H]
  \centering
  \begin{tabular}{|>{\centering\arraybackslash}m{3.5cm}
                  | >{\centering\arraybackslash}m{7cm}
                  | >{\centering\arraybackslash}m{3.5cm}|}
                  \hline
                  \vspace{0.2cm}
    \(\langle \mathsf{M}_a\rangle = S^{(\mathsf{C},1)}_{a0}\) \vspace{0.2cm}
    &\vspace{0.2cm} dual string theory background
     \vspace{0.2cm}
    &
    \vspace{0.2cm} toric diagram decription
    \vspace{0.2cm}\\
    \hline\hline
    \vspace{0.5cm}
    vacuum  & \vspace{0.2cm}\parbox{6.2cm }{\centering $\Omega\sigma^{{\color{red}SO}/{\color{blue}Sp}}_+$ orientifold of   conifold:\\[+2pt]
    $t = N\gs {{\mathbin{\substack{{\color{red}\scalebox{1.05}{$-$}}\\[-1ex]{\color{blue}\scalebox{1.05}{$+$}}}}} } \gs$
       \vspace{0.2cm}}& {\begin{tikzpicture}[scale=0.8,baseline={(current bounding box.center)},
line/.style={line width=1.0pt},
brane/.style={line width=1.25pt,blue!70!black},
branepoint/.style={circle,fill=blue!70!black,inner sep=2.25pt},
redpoint/.style={circle,fill=red!80!black,inner sep=1.4pt},
axis/.style={line width=1.0pt,red!80!black,dashed},
every node/.style={font=\small}
]
\coordinate (L) at (0,0);
\coordinate (R) at (2.60,0);
\coordinate (C) at ($(L)!0.5!(R)$);
\draw[line] (L) -- (R);
\draw[line] (L) -- ++(0,1.00);
\draw[line] (L) -- ++(-0.85,-0.72);
\draw[line] (R) -- ++(0.85,1.00);
\draw[line] (R) -- ++(0,-1.00);
\draw[axis] ($(C)+(0,-1.05)$) -- ($(C)+(0,1.05)$);
\node[redpoint] at (C) {};
\end{tikzpicture}}\\
    \hline
        \vspace{0.5cm}
    antisymmetric \(\Lambda^\ell\)
    & \parbox{6.2cm}{\centering
\vspace{0.2cm}
      brane on internal leg at \(x=\gs(\ell+\tfrac12)\), image at
      \(\widehat t-x\);\quad \(\widehat t=t+\gs\)
      \vspace{0.2cm}
    }
    & {\begin{tikzpicture}[scale=0.8,baseline={(current bounding box.center)},
line/.style={line width=1.0pt},
brane/.style={line width=1.25pt,blue!70!black},
branepoint/.style={circle,fill=blue!70!black,inner sep=2.25pt},
redpoint/.style={circle,fill=red!80!black,inner sep=1.4pt},
axis/.style={line width=1.0pt,red!80!black,dashed},
every node/.style={font=\small}
]
\coordinate (L) at (0,0);
\coordinate (R) at (2.60,0);
\coordinate (C) at ($(L)!0.5!(R)$);
\draw[line] (L) -- (R);
\draw[line] (L) -- ++(0,1.00);
\draw[line] (L) -- ++(-0.85,-0.72);
\draw[line] (R) -- ++(0.85,1.00);
\draw[line] (R) -- ++(0,-1.00);
\draw[axis] ($(C)+(0,-1.05)$) -- ($(C)+(0,1.05)$);
\node[redpoint] at (C) {};
\coordinate (Bpos) at ($(L)!0.30!(R)$);
\coordinate (IBpos) at ($(L)!0.70!(R)$);
\draw[brane] (Bpos) -- ++(0,0.84);
\node[branepoint] at (Bpos) {};
\draw[brane] (IBpos) -- ++(0,0.84);
\node[branepoint] at (IBpos) {};
\end{tikzpicture}}\\
    \hline 
    \vspace{0.6cm}
    symmetric \(\mathrm{Sym}^w\)
    & \parbox{6.2cm}{\centering
    \vspace{0.2cm}
       snti-brane on external leg at \(y=\gs(w+\tfrac12)\), image at
      \(\widehat t+y\);\newline \vspace{0.2cm} \(\widehat t=t-\gs\)
    }
    & {\begin{tikzpicture}[scale=0.8,baseline={(current bounding box.center)},
line/.style={line width=1.0pt},
brane/.style={line width=1.25pt,blue!70!black},
branepoint/.style={circle,fill=blue!70!black,inner sep=2.25pt},
redpoint/.style={circle,fill=red!80!black,inner sep=1.4pt},
axis/.style={line width=1.0pt,red!80!black,dashed},
every node/.style={font=\small}
]
\coordinate (L) at (0,0);
\coordinate (R) at (2.60,0);
\coordinate (Lext) at ($(L)+(-0.85,-0.72)$);
\coordinate (Rext) at ($(R)+(0.85,1.00)$);
\coordinate (C) at ($(L)!0.5!(R)$);
\draw[line] (L) -- (R);
\draw[line] (L) -- ++(0,1.00);
\draw[line] (L) -- ++(-0.85,-0.72);
\draw[line] (R) -- ++(0.85,1.00);
\draw[line] (R) -- ++(0,-1.00);
\draw[axis] ($(C)+(0,-1.05)$) -- ($(C)+(0,1.05)$);
\node[redpoint] at (C) {};
\coordinate (BposL) at ($(L)!0.50!(Lext)$);
\coordinate (BposR) at ($(R)!0.50!(Rext)$);
\draw[brane] (BposL) -- ++(0,0.84);
\node[branepoint] at (BposL) {};
\draw[brane] (BposR) -- ++(0,0.84);
\node[branepoint] at (BposR) {};
\end{tikzpicture}}\\
\hline
  \end{tabular}

\caption{Summary of dual string description   of  simplest monodromy defects $\mathsf{M}_{a}$. For the most general monodromy defect, see  Table~\ref{tab:summarySUmono}.}
\label{tab:summarynitro}
\end{table}

\begin{figure}[H]
    \centering
\includegraphics[width=\linewidth]{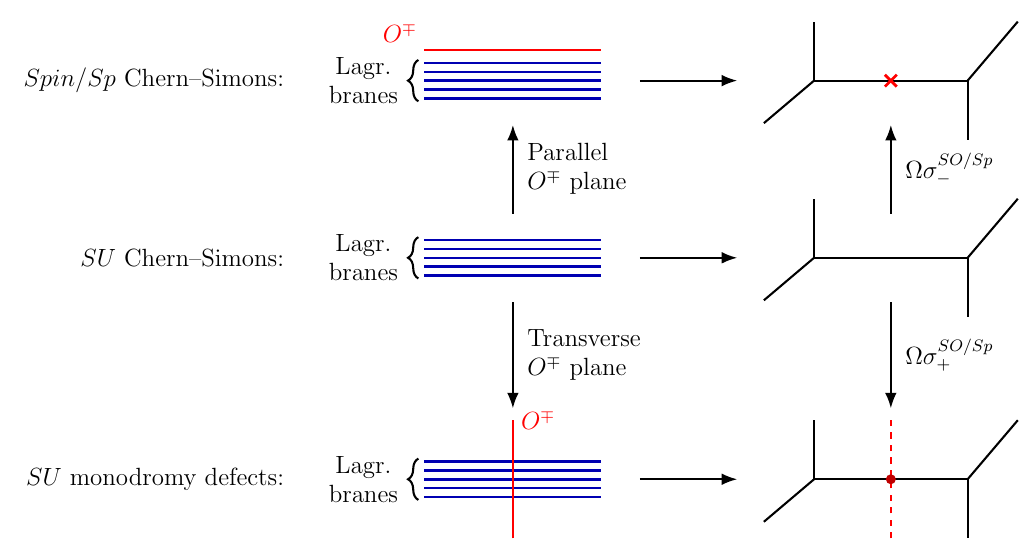}
    \caption{
    Summary of the two orientifold projections on the deformed and
resolved conifold.  On the deformed conifold \(T^*S^3\), the orientifold
\(\Omega\sigma_-\) introduces an \(O^{\mp}\)-plane parallel to the stack of
Lagrangian branes wrapping \(S^3\), thereby engineering \(Spin/Sp\)
Chern--Simons theory~\cite{Sinha:2000ap}.  By contrast, \(\Omega\sigma_+\)
introduces an \(O^{\mp}\)-plane transverse to the branes; this configuration
realizes \(\mathsf C\)-monodromy defects in \(SU(N)\) Chern--Simons theory.
The diagrams on the right show the corresponding orientifold backgrounds after
the geometric transition to the resolved conifold.}
    \label{fig:diagrambranes}
\end{figure}

We would like to emphasize a pleasing consequence of our analysis. It is known that the resolved conifold geometry admits two natural antiholomorphic involutions: $\sigma_+$ and $\sigma_-$ \cite{Acharya:2002kv,Hori:2005bk,Krefl:2009md}. While it was known since the work of Sinha and Vafa \cite{Sinha:2000ap} that the closed string partition function of the $\Omega \sigma_-^{(\cdot)}$
orientifold of the resolved conifold is dual to $Spin(N)$ or $Sp(N/2)$ Chern--Simons theory, it was not   known whether there is a dual  gauge theory interpretation of the closed string
partition function on the $\Omega \sigma_+^{(\cdot)}$
orientifold of the resolved conifold. In this paper, we have   answered   this question in the affirmative:  the  A-model closed string 
partition functions on the $\Omega \sigma_+^{(\cdot)}$
orientifold of the resolved conifold are dual to the vacuum monodromy defects of $SU(N)_k$ Chern--Simons theory. The holographic realization of these two orientifold backgrounds is summarized in Figure \ref{fig:diagrambranes}. 

As a bonus, we also provide  an explicit dual description of arbitrary Wilson lines $W_R$ in  Chern--Simons theory with $Spin/Sp$ gauge group -- including Wilson lines in spinor representations\footnote{This gives the Chern--Simons/topological string analogue of the
\(AdS_5\times S^5\) realization of spinor Wilson lines in
\({\cal N}=4\) super-Yang--Mills theory \cite{Witten:1998xy}.} -- in terms of branes and antibranes in the \(\Omega\sigma_-^{(\cdot)}\) orientifold of the resolved conifold. 

\paragraph
{Outlook.}
We conclude by outlining several directions for future work.   The realization
of charge conjugation symmetry developed here provides a particularly explicit
and computable example of symmetry-enriched topological order
\cite{Barkeshli2019} in terms of twisted affine Lie algebras.  In light of the long-standing expectation, going back
to Moore and Seiberg~\cite{Moore:1989vd}, that many, and perhaps all, \(2+1\)-dimensional TQFTs
admit a realization as Chern--Simons theories for a suitable choice of gauge
group, it is natural to ask whether symmetry enrichment
more generally admits an equally algebraic description.   

It would be interesting to study general knots and links decorated by both
Wilson lines and monodromy defects.  The natural organizing principle for such
observables is the \(\mathbb Z_2^{\mathsf C}\)-crossed braided tensor category, whose crossed braiding and defect fusion  
control the corresponding link amplitudes.  This raises the intriguing question
of whether these correlators define new invariants of knots and links, refined
by charge conjugation monodromy and by the data of twisted affine Lie
algebras.

A natural direction for future work is to extend our construction to monodromy defects associated with quantum symmetries of $SU(N)_k$ Chern--Simons theory (Section \ref{sec:dos}), and to identify their dual description in topological string theory.  Another natural direction is to place monodromy defects on more general
three-manifolds and identify their dual string backgrounds.  A particularly
timely example is Chern--Simons theory on \(\mathbb R\times S^2\), whose large
\(N\) dual has recently been studied in~\cite{Gaiotto:2025nrd}.

Another open problem is to formulate a purely closed string, geometric, ``bubbling Calabi--Yau''
description of the full family of monodromy defects \(\mathsf M_a\) in
\(SU(N)_k\) Chern--Simons theory, incorporating the backreaction of the dual
branes identified here. This would provide the analogue, for monodromy
defects, of the bubbling Calabi--Yau description of Wilson lines \(W_R\)
developed in~\cite{Gomis:2006sb,Gomis:2007kz}. Another challenge is
to identify the   topological string theory duals of
\(\mathsf C\)-monodromy defects in \(Spin(2N)_k\) Chern--Simons theory.

\paragraph{ Outline of the paper.} The plan of the rest of the paper is as follows.  
  
  In \textbf{Section~\ref{sec:dos}}, 
  We give a complete characterization of monodromy defects in Chern--Simons
theory with charge conjugation symmetry.  Quantizing the theory on a torus in
the presence of charge conjugation topological defects, we identify the monodromy defects
with integrable representations of the twisted affine algebra
\(\mathfrak g^{(2)}_k\).  We present the relevant modular
\(S\)-matrices and fusion coefficients, thereby making the defect sector
explicitly computable.

In \textbf{Section~\ref{sec:suNdual}} we review the correspondence between the partition function and Wilson loop
observables of \(SU(N)_k\) Chern--Simons theory and their holographic
description in topological string theory.  Instead of using the conventional
small \(g_s\) expansion of the original Gopakumar--Vafa duality
\cite{Gopakumar:1998ki}, we present an alternative and considerably simpler
approach based on analytically continued Chern--Simons theory.  We expand the
Chern--Simons partition function in the \(q\to 0\) regime, which corresponds to
a strong string-coupling expansion, \(g_s\gg 1\).  The resulting expression
reproduces the closed topological string partition function on the resolved
conifold, as obtained from worldsheet instanton counting.

In \textbf{Section~\ref{sec:four}} we study monodromy defects in \(SU(N)_k\)
Chern--Simons theory in detail.  We use the twisted modular \(S\)-matrices to
compute exact unknot expectation values for the lightest monodromy defects,
and from these amplitudes identify the dual topological string background as
a \(\Omega\sigma^{(\cdot)}_+\) orientifold of the resolved conifold.  We then extend
the analysis to excited monodromy defects, showing that they are realized by
brane configurations in this orientifold background.  Finally, we develop the
corresponding orientifold version of the topological vertex needed for the
computations in this paper.

The \textbf{Appendix} complements the main text and is composed of 4 sections. In Appendix~\ref{App:su-so-sp-branes-geometry}, we review the $SU(N)$, $Spin(N)$ and $Sp(N)$ Chern--Simons theory duality. In particular, we include a detailed discussion of the A-model dual for the expectation value  of  Wilson lines in generic representation  of the gauge group. This also includes the cases of spinors, that were not discussed in the literature.
Appendix~\ref{app:monodef}, presents  further details on how to identify the lightest  monodromy defect. In Appendix~\ref{smallgs}, we compare the match of the small $\gs$ expansion with the $q\ll 1$ expansion. In Appendix~\ref{app:constant-maps}, we conclude by discussing the constant maps contributions to the topological string partition function in all the cases.

\section{Monodromy Defects in Chern--Simons Theory}
\label{sec:dos}

A fundamental lesson of the modern understanding of   quantum field theory is that a (zero-form) global symmetry~\cite{Gaiotto:2014kfa}
gives rise to a collection of codimension-two  
defects. These are defined by specifying a symmetry twist around 
the defect: upon encircling it, fields are transformed by the 
corresponding symmetry action. This notion is well-defined 
because loops linking a codimension-two defect can carry 
nontrivial monodromy.

A codimension-two   defect implementing a twist by a group element $g$ -- a 
$g$-monodromy defect -- naturally lives at the boundary of a 
codimension-one topological defect, which implements the symmetry 
action as the $g$-monodromy defect is encircled. Given a symmetry 
element $g$, a rich spectrum of   $g$-monodromy defects $\mathsf{M}_a$ 
  can be defined, where $a$    labels the distinct monodromy defects associated to the symmetry $g$ for fixed codimension-two surface.  
These defects   define a new class of correlation functions in the theory and capture important   physical information (e.g. \cite{KadanoffCeva1971,Wang:2021disorderU1}).\footnote{Monodromy defects  have been studied, for example,  as twist fields in $2d$ theories~\cite{Dixon:1985jw,Dixon:1986qv}, as a line  
defect associated to the $\mathbb Z_2$ symmetry of  the Ising $3d$  model \cite{Billo:2013jda},  and  $O(N)$ flavor symmetry  monodromy defects in  free scalars and fermions (see e.g.~\cite{Lauria:2020emq,Giombi:2021uae,Bianchi:2021snj,Bashmakov:2024suh}) and charge conjugation monodromy defects in gauge theories~\cite{Gomis:2025adsCPT}.} 

A complementary description of a \(g\)-monodromy operator, especially on the
lattice, is as a partial symmetry transformation: one acts with the global
symmetry \(g\) only on the degrees of freedom contained in a region of space,
at fixed time (e.g. \cite{KadanoffCeva1971,Chen:2022upe}).  Since the symmetry is applied only on a subregion, the
operation is indistinguishable from the identity in the interior and exterior, but it fails
to cancel along the boundary of the region.  The boundary therefore supports a
codimension-two defect.  Equivalently, the partial symmetry transformation
creates a topological branch cut across which any operator  crossing the
cut is acted on by \(g\).  The endpoint of this branch cut is precisely the
\(g\)-monodromy defect. Physically, the endpoint can carry additional localized
  degrees of freedom, and the possible choices of such degrees of
freedom give the different \(g\)-monodromy defects.

We begin  our investigation of monodromy defects in Chern--Simons theory by addressing the    prerequisite question: does the Chern--Simons action  admit unitary, internal (zero-form) symmetries? This can be answered by considering the most general linear transformation  
\beq {\mathsf C}: A^\alpha \longrightarrow O^\alpha{}_{\beta} A^\beta\,, \label{ctrans} \eeq where $\alpha,\beta$ are Lie algebra indices. This transformation leaves  the Chern--Simons action \eqref{action} invariant  if and only if \beq \begin{aligned} O^\alpha{}_{\delta} O^\beta{}_{\epsilon} f_{\alpha\beta}{}^\gamma &= O^\gamma{}_{\zeta} f_{\delta\epsilon}{}^\zeta\,, \\[2pt] \kappa_{\alpha\beta} O^\alpha{}_{\gamma} O^\beta{}_{\delta} &= \kappa_{\gamma\delta}\,, \label{automor} \end{aligned} \eeq where $f_{\alpha\beta}{}^\gamma$ are the structure constants of $\mathfrak g$, and $\kappa_{\alpha\beta}$ is the invariant bilinear form  of $\mathfrak g$. Therefore, the action is invariant under \eqref{ctrans}   when $O^\alpha{}_{\beta}$ defines an automorphism of the Lie algebra $\mathfrak g$.\footnote{ Every automorphism of a finite dimensional Lie algebra preserves the Killing form.}
 
Since inner automorphisms act trivially on gauge invariant operators, the transformation ${\mathsf C}$ defines a faithfully acting global symmetry of Chern--Simons theory only when it represents a nontrivial outer automorphism of $\mathfrak{g}$, that is  when ${\mathsf C}\in \operatorname{Out}(\mathfrak{g})$. This global symmetry is commonly referred to as charge conjugation symmetry. Table~\ref{tab:outer-auto} lists the outer automorphism groups of the simple Lie algebras. These are precisely the symmetry groups of the corresponding Dynkin diagrams.
\begin{table}[H]
\centering
\includegraphics[width=0.75\linewidth]{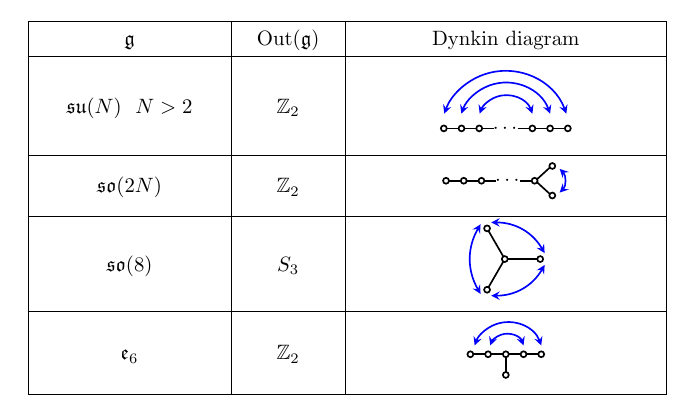}
\caption{Outer automorphism groups of   simple Lie algebras.}
\label{tab:outer-auto}
\end{table}    
Even though Chern--Simons theory has no local operators, the action of $\mathsf{C}$ is realized on Wilson line operators:
\begin{equation}
   \mathsf{C}: W_R \mapsto W_{\mathsf{C}(R)} \, . 
\end{equation}
Here $\mathsf{C}(R)$ denotes the representation obtained from $R$ by the action of the outer automorphism $\mathsf{C}$, which permutes the nodes of the Dynkin diagram (see Table \ref{tab:outer-auto}). For an irreducible representation $R$ with Dynkin labels $R_i$, the Dynkin labels of $\mathsf{C}(R)$ are given by:\footnote{\label{sign}
The transformation in Table \ref{tab:outer-auto-dynkin} obeys $\mathsf{C}^2 = 1$. For $D_4$, there is an additional $\mathbb{Z}_3$ generator of the $S_3$ outer automorphism group, which acts as $\sigma: (R_1, R_2, R_3, R_4) \mapsto (R_4, R_2, R_1, R_3)$.}
\begin{table}[H]
\centering
\[
\begin{array}{c|c|c}
  \mathfrak g & R   & {\mathsf C}(R)   \\[+2pt]
\hline
\noalign{\vskip 2pt}
A_N\  
& (R_1,R_2,\dots,R_{N-1},R_{N})
& (R_{N},R_{N-1},\dots,R_2,R_1)
\\[4pt]
D_N\   
& (R_1,R_2,\dots,R_{N-2},R_{N-1},R_N)~
& ~(R_1,R_2,\dots,R_{N-2},R_N,R_{N-1})
\\[4pt]
E_6
& (R_1,R_2,R_3,R_4,R_5,R_6)
& (R_5,R_4,R_3,R_2,R_1,R_6)
\end{array}
\]
\caption{Dynkin labels of the representation \(\mathsf C(R)\) obtained from \(R=(R_1,\ldots,R_{\textrm{rank}(\mathfrak g)})\) by the nontrivial outer automorphism of \(\mathfrak g\). $N\geq 2$ and $N\geq 4$ for $A_N$ and $D_N$ respectively.  }
\label{tab:outer-auto-dynkin}
\end{table}

Chern--Simons theory can also exhibit quantum symmetries: symmetries of its correlation functions that do not descend from the classical action. Mathematically, symmetries are braided autoequivalences of the associated modular tensor category~\cite{ENO2010,Barkeshli2019}, and classifying them  is a nontrivial problem. For Chern--Simons theories based on simple Lie algebras of types $A$, $B$, $C$, and $G$ at level $k$, a complete classification  was obtained in~\cite{Edie-Michell:2022abq}.\footnote{ 
For classification of unitary and anti-unitary quantum symmetries in abelian Chern--Simons theory  
~\cite{Delmastro:2019vnj}.} The existence of   quantum symmetries depends on the  arithmetic properties of   $k$ and the rank of $\mathfrak{g}$. By contrast, whenever charge conjugation is nontrivial, $\mathsf{C}$ is present at every level $k$.

 \subsection{Monodromy Defect  Geometric Data}
 \label{subsec:monodromy-defect-data}
We now turn to the classification of ${\mathsf C}$-monodromy defects in
Chern--Simons theory, identifying the geometric and representation-theoretic
data by which they are labeled.

\begin{figure}
    \centering
\begin{tikzpicture}[baseline={([yshift=-.5ex]current bounding box.center)},scale =2.5]
\fill[blue!30, fill opacity=0.7] (0,0) ellipse (1 and 0.4);
\draw[very thick, blue, line width = 0.1cm] (-1,0) arc (180:360:1 and 0.4);
\draw[very thick, blue, line width = 0.1cm] (1,0) arc (0:180:1 and 0.4);

\draw[->, very thick, >=Latex] (-1.45,0.75) to[bend left=25] (-0.45,0.0);
\node[scale=1] at (-1.85,0.75) {$\mathsf{C}$- Surface};

\node[scale=1.2, blue] at (1.2,0.) {$\mathsf M_a$};
\draw[->, very thick, >=Latex] (1.45,0.75) to[bend right=25] (0.55,0.35);
\node[scale=1,align =center] at (1.9,0.75) {Monodromy \\ Defect};
\end{tikzpicture}
    \caption{Monodromy defect  supported on a knot that is the boundary of a topological surface.}
    \label{fig:surface-knot}
\end{figure}
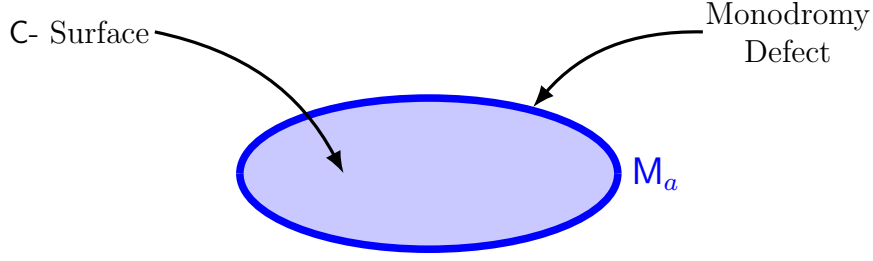

Geometrically, a Chern--Simons monodromy defect for a symmetry $g$ is a
codimension-two defect supported on an oriented knot $K\subset M$. It can
be defined whenever $K$ bounds an oriented Seifert surface $S\subset M$, that is when $[K]=0\in H_1(M,\mathbb Z)$, 
so that the codimension-one topological defect implementing the action of $g$ supported on $S$ can end on
$K$,\footnote{
  There are multiple
surfaces of distinct topology that can bound $K$. The minimum genus among
all such surfaces defines a knot invariant known as the Seifert genus
$g(K)$. For the unknot, the minimal Seifert surface is a disk, with $g(S)=0$, while for the 
 trefoil knot, the minimal Seifert surface is a once-punctured torus, and hence
$
g(S)=1$.} as in Figure \ref{fig:surface-knot}. A \(g\)-monodromy defect is thus defined geometrically by a pair \((K,S)\), where \(K\subset M\) is an oriented knot and \(S\subset M\) is an oriented surface obeying \(\partial S=K\). The dependence on \(S\) is topological: correlation functions are invariant under deformations of \(S\) that do not cross other operator insertions.

\subsection{Monodromy Defect  Representation Data}
\label{mondefectrep}

Our next goal is to determine the representation-theoretic data labeling
$\mathsf{C}$-monodromy defects in Chern--Simons theory.\footnote{A natural direction is to study monodromy defects for quantum symmetries of Chern--Simons theory.} The natural arena for this analysis is the Hilbert space obtained by
quantizing the theory on the torus \(T^2\) (cf.~\cite{Barkeshli2019}).

\begin{figure}[H]
    \centering
    \includegraphics[width=0.6\textwidth]{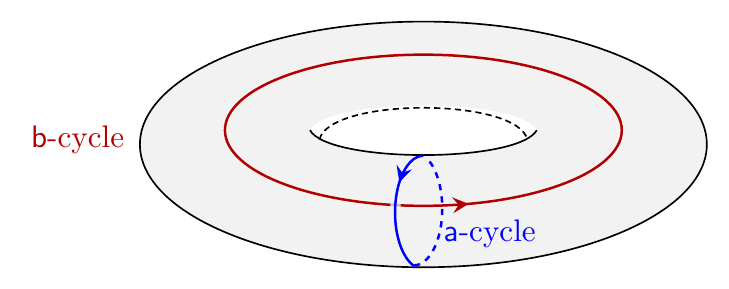}
    \caption{The $\mathsf a$- and $\mathsf b$-cycles of $T^2$.}
    \label{fig:andbcycles}
\end{figure}

When Chern--Simons theory has a (zero-form) symmetry, one can decorate $\mathbb{R} \times T^2$ with codimension-one topological defects implementing the symmetry. We consider topological defects supported  along $\mathbb R$, the ``time coordinate", and wrapping a one-cycle in $T^2$.\footnote{One can also consider the topological defect implementing $\mathsf C$ wrapping $T^2$. This defines a unitary operator in $\mathcal{H}^{(1,1)}$, acting at fixed time. It acts on the basis of  as $|R\rangle\rightarrow |{\mathsf C}(R)\rangle$.}
 Quantizing Chern--Simons theory on such decorated tori yields four Hilbert spaces, 
\begin{equation}
\mathcal{H}^{(1,1)}\,,\qquad
\mathcal{H}^{(1,\mathsf{C})}\,,\qquad
\mathcal{H}^{(\mathsf{C},1)}\,,\qquad
\mathcal{H}^{(\mathsf{C},\mathsf{C})}\,,
\end{equation}
where the two superscript labels indicate the topological defects inserted along the $\mathsf a$- and $\mathsf b$-cycles of $T^2$, respectively (see Figure \ref{fig:andbcycles}).\footnote{We let $z=\sigma_1+\tau \sigma_2$ be a complex coordinate in $T^2$ with modular parameter $\tau$, so that
\begin{equation}
 \mathsf a\textrm{-cycle}: z\simeq z+1\,,\qquad \qquad 
 \mathsf b\textrm{-cycle}: z\simeq z+\tau\,.
 \label{abcycles}
\end{equation}
In $2d$, we choose the quantization channel in which the $\mathsf a$-cycle of the torus is interpreted as the spatial circle and the $\mathsf b$-cycle as Euclidean time.}   Here $1$ denotes the trivial (transparent) defect and $\mathsf{C}$ denotes the charge conjugation topological defect. The well-known Chern--Simons Hilbert space corresponds to $\mathcal{H}^{(1,1)}$, in which no defects are inserted. 

\subsubsection*{$\bullet$ \boldmath Hilbert space $\mathcal{H}^{(1,1)}$.}
Recall   the Hilbert space $\mathcal{H}^{(1,1)}$ of
Chern--Simons theory with gauge group $G$ at level $k$ on $T^2$. Canonical
quantization identifies this Hilbert space with the space of
conformal blocks of the corresponding $\mathfrak{g}_k$ affine Lie algebra\footnote{Throughout the paper, we denote the affine Kac--Moody algebra associated with
\(\mathfrak g\) at level \(k\) by \(\mathfrak g_k\). We also use Kac's notation $\mathfrak g^{(1)}$ for untwisted affine algebras and $\mathfrak g^{(r)}$  with $r=2,3$ for twisted affine algebras.} 
on $T^2$~\cite{Witten:1988hf,Elitzur:1989nr}. A basis of \(\mathcal H^{(1,1)}\) is labeled by the integrable highest weight
representations of the affine algebra \(\mathfrak g_k\).  These
are finite dimensional highest weight representations \(R\) --  with Dynkin labels $R_i$ --  of the horizontal
Lie algebra\footnote{Throughout this paper, we use the term horizontal subalgebra to mean the finite dimensional Lie algebra obtained by deleting the affine node from the affine Dynkin diagram. We denote this Lie algebra by $\mathring{\mathfrak g}$.} \(\mathfrak g\) whose highest weight   lies in the level-\(k\)
alcove:
\beq
 \mathrm{Int}(\mathfrak{g}_k) =\bigl\{\,R_j\geq 0\;\big|\;
\sum_{j=1}^{\mathrm{rank}(\mathfrak{g})} a_j^\vee \, R_j
\leq k\bigl\}\,,
\label{integrable}
\eeq
where $a_j^\vee$ are the comarks of $\mathfrak{g}_k$. 

A basis of states in $\mathcal{H}^{(1,1)}$ is obtained by performing
the path integral over the solid torus, which prepares a state on the
boundary $T^2$. Inserting a Wilson line $W_R$ along the
noncontractible $\mathsf b$-cycle at the center of the solid torus produces a
basis element for each representation $R$   defining an integrable representation 
of $\mathfrak{g}_k$. 
\begin{align}
\mathcal{H}^{(1,1)}\simeq 
\includegraphics[scale =0.7,valign=c]{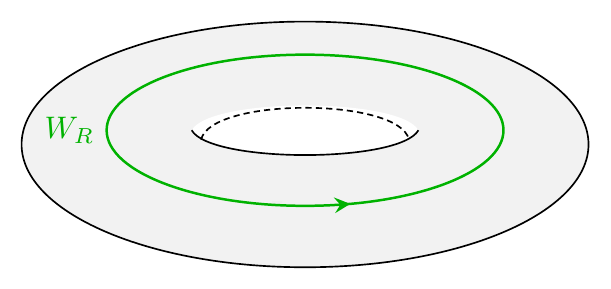}
\end{align} 
The corresponding wavefunction is
the character $\chi_R(\tau)$ of the $\mathfrak{g}_k$-module $L_R$,
defined by
\begin{equation}
  \chi_R(\tau)
  \equiv
  \operatorname{Tr}_{L_R}\!\left(q^{L_0-c/24}\right),
\end{equation}
where $q=e^{2\pi i\tau}$ and $\tau$ is the modular parameter of $T^2$.
Hence,
\begin{equation}
  \mathcal{H}^{(1,1)}
  \cong
  \operatorname{span}\bigl\{
    \chi_R \,\big|\, R \in \mathrm{Int}(\mathfrak{g}_k)
  \bigr\}.
\end{equation}
where $\mathrm{Int}(\mathfrak{g}_k)$ denotes the set of
integrable representations of $\mathfrak{g}_k$.

\subsubsection*{$\bullet$ \boldmath Hilbert space $\mathcal{H}^{(1,\mathsf{C})}$}
We now turn to the Hilbert space $\mathcal{H}^{(1,\mathsf{C})}$,
where a topological defect for the $\mathsf{C}$ symmetry is inserted
along the $\mathsf b$-cycle of the boundary $T^2$. The symmetry $\mathsf{C}$
implements an order-two outer automorphism of the Lie algebra
$\mathfrak{g}$, decomposing it into eigenspaces
\beq
\mathfrak{g} = \mathfrak{g}_+ \oplus \mathfrak{g}_-\,,\qquad
    \mathsf C|_{\mathfrak g_\pm}=\pm 1\,,
    \label{actonLie}
\eeq
where $\mathsf{C}$ acts as $+1$ on $\mathfrak{g}_+$ and as $-1$ on
$\mathfrak{g}_-$. The even part $\mathfrak{g}_+$ is the invariant
subalgebra, and $\mathfrak{g}_-$ transforms as a representation of
$\mathfrak{g}_+$. The $\mathbb{Z}_2$ grading is compatible with the
Lie bracket:
\beq
[\mathfrak{g}_+, \mathfrak{g}_+] \subseteq \mathfrak{g}_+\,,\qquad
[\mathfrak{g}_-, \mathfrak{g}_-] \subseteq \mathfrak{g}_+\,,\qquad
[\mathfrak{g}_+, \mathfrak{g}_-] \subseteq \mathfrak{g}_-\,.
\eeq

The action of \(\mathsf C\) lifts to the affine algebra $\mathfrak{g}_k$  by imposing a
\(\mathsf C\)-twisted boundary condition on the holomorphic currents $J$ -- which generate   $\mathfrak{g}_k$ -- as they
are transported around the \(\mathsf a\)-cycle of the torus:
\begin{equation}
    J(z+1)=\mathsf C\bigl(J(z)\bigr).
\end{equation}
Equivalently, decomposing the finite dimensional Lie algebra as in \eqref{actonLie},
the current decomposes as \(J=J_+ + J_-\), with
\begin{equation}
    J_+(z+1)=J_+(z),
    \qquad
    J_-(z+1)=-J_-(z).
    \label{modetwisted}
\end{equation}
Thus the \(\mathsf C\)-even currents are periodic, while the
\(\mathsf C\)-odd currents are antiperiodic. Their mode expansions therefore
take the form
\begin{equation}
   \alpha\in \mathfrak g_+\implies n \in \mathbb{Z}\,,\qquad  J^\alpha(z)=\sum_{n\in \mathbb Z} J^\alpha_n e^{-2\pi i n z},
\end{equation}
and
\begin{equation}
    ~~~~~~ \alpha\in \mathfrak g_-\implies n \in {\frac 1 2}+ \mathbb{Z}\,,~~~
     J^\alpha(z)=\sum_{n\in \mathbb Z+\frac12} J^\alpha_n e^{-2\pi i n z}.
\end{equation}
In other words, the \(\mathsf C\)-even currents are integrally moded, whereas
the \(\mathsf C\)-odd currents are half-integrally moded.

This  defines a new affine Lie algebra,   the    {\it $\mathsf C$-twisted affine algebra} $\mathfrak{g}^{(2)}_k$. The superscript \((2)\) records the order of the automorphism used in the
twisted affinization, in accordance with the standard notation of
Kac~\cite{kac_1990}.\footnote{Twisting by the order element in footnote \ref{sign} yields the twisted   algebra $D_4^{(3)}$, the unique $\mathfrak g^{(r)}$ with $r=3$.} 
The resulting twisted affine algebra is determined, up to isomorphism, only by the outer automorphism class of the twist, and not by the particular representative chosen within that class~\cite{kac_1990}.

Therefore, the Hilbert space $\mathcal{H}^{(1,\mathsf{C})}$ is no longer described by
the untwisted affine algebra $\mathfrak{g}_k$, which governs $\mathcal{H}^{(1,1)}$,  but by the corresponding
  $\mathsf C$-twisted affine algebra: 
\begin{equation}
\mathfrak{g}^{(2)}_k 
\end{equation}
The twisted Hilbert space $\mathcal{H}^{(1,\mathsf{C})}$ is therefore
the space of $\mathfrak{g}^{(2)}_k$ conformal blocks on $T^2$. A basis of \(\mathcal H^{(1,\mathsf C)}\) is labeled by integrable
representations of the twisted affine algebra \(\mathfrak g^{(2)}_k\).  These
may be viewed as highest weights of the horizontal subalgebra\footnote{For all   twisted affine algebras \(\mathfrak g^{(2)}_k\), the
horizontal finite dimensional Lie algebra \(\mathring{\mathfrak g}\) coincides
with the invariant subalgebra \(\mathfrak g_+\), with one exception. For
\(A_{2N}^{(2)}\), one instead has
$\mathring{\mathfrak g}\simeq {}^{L}\mathfrak g_+$,
where \({}^{L}\mathfrak g_+\) denotes the Langlands dual of
\(\mathfrak g_+\). See~\cite{kac_1990} for details.}
\(\mathfrak g_+\subset \mathfrak g^{(2)}_k\), with Dynkin labels \(a_j\)
lying in the level-\(k\) alcove:
   \beq
 \mathrm{Int}(\mathfrak{g}^{(2)}_k) =\bigl\{\,a_j\geq 0\;\big|\;
\sum_{j=1}^{\mathrm{rank}(\mathfrak{g}_+)} b_j^\vee \, a_j
\leq k\bigl\}\,,
\label{integrable2}
\eeq
where $b_j^\vee$ are the comarks of $\mathfrak{g}^{(2)}_k$. 
A basis of $\mathcal{H}^{(1,\mathsf C)}$ is obtained by inserting, in the
presence of the $\mathsf C$-defect along the $\mathsf b$-cycle of     the boundary torus, a monodromy defect
$\mathsf M_a$ along the noncontractible $\mathsf b$-cycle at the center of the solid torus:
\begin{align}\label{eq:H1C}
    \mathcal{ H}^{(1,\mathsf C)}\simeq \includegraphics[valign =c, scale=0.8]{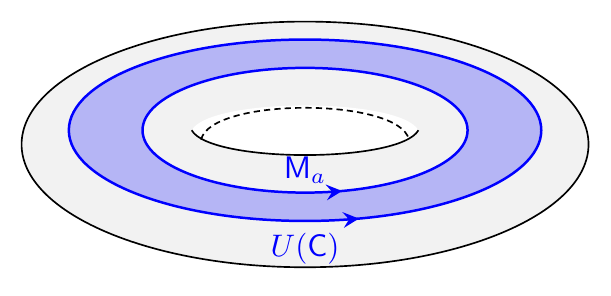}
\end{align}
The resulting wavefunction is
the twisted character $\chi^{(2)}_a(\tau)$ of the corresponding
$\mathfrak{g}^{(2)}_k$-module $L_a$:
\beq
  \chi^{(2)}_a(\tau)
  \equiv \mathrm{Tr}_{L_a}\!\bigl(\,q^{L_0-c/24}\bigr)\,.
\eeq
Therefore, the Hilbert space $\mathcal{H}^{(1,1\mathsf C)}$ is given by 
\begin{equation}
  \mathcal{H}^{(1,\mathsf C )}
  \cong
  \operatorname{span}\bigl\{
    \chi_a \,\big|\, a \in \mathrm{Int}(\mathfrak{g}^{(2)}_k)
  \bigr\}\,,
\end{equation}
where $\mathrm{Int}(\mathfrak{g}^{(2)}_k)$ denotes the set of
integrable representations of $\mathfrak{g}^{(2)}_k$.

We have therefore shown that Chern--Simons theory with charge conjugation
symmetry   admits a new class of observables, namely monodromy
defects $\mathsf M_a$, which are distinct from Wilson lines. Wilson lines $W_R$ are labeled by integrable
representations of the untwisted affine algebra $\mathfrak{g}_k$,
whereas monodromy defects $\mathsf M_a$ are labeled by integrable
representations of the twisted affine algebra $\mathfrak{g}^{(2)}_k$.

\subsubsection*{$\bullet$ \boldmath Hilbert space $\mathcal{H}^{(\mathsf{C},1)}$}
The modular $S$-transformation acts on the cycles of $T^2$ as
$S\colon (\mathsf a,\mathsf b) \to (\mathsf b,-\mathsf a)$, and therefore maps
 $\mathcal{H}^{(1,\mathsf{C})}$ to $\mathcal{H}^{(\mathsf{C},1)}$.
This immediately implies that
\beq
\dim\,\mathcal{H}^{(1,\mathsf{C})}=\dim\,\mathcal{H}^{(\mathsf{C},1)} \,.
\eeq
In the decorated torus defining $\mathcal{H}^{(\mathsf{C},1)}$, the affine algebra $\mathfrak{g}_k$ currents are untwisted along the  
$\mathsf a$-cycle but twisted along the $\mathsf b$-cycle of the boundary torus.A basis of wavefunctions in $\mathcal{H}^{(\mathsf{C},1)}$ is  obtained by inserting 
along the noncontractible $\mathsf b$-cycle  at the center of the solid torus a Wilson line $W_R$, with $R$
   defining an integrable representation of $\mathfrak{g}_k$, but now in the presence of a topological $\mathsf{C}$-defect  $U(\mathsf{C})$ along the $\mathsf a$-cycle of the boundary torus:
\begin{align}
\label{eq:HCI}
\mathcal{H}^{(\mathsf{C},1)}\simeq \includegraphics[valign=c,scale =0.8]{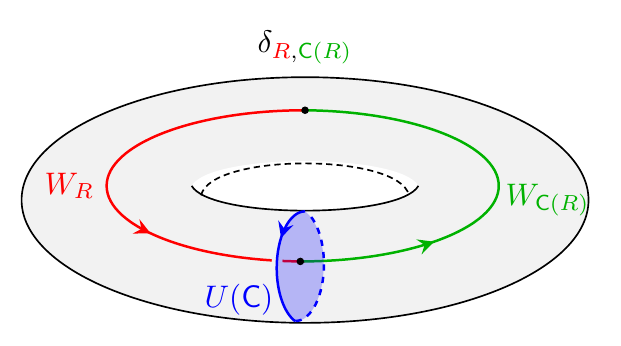}
\end{align}
 \begin{figure}
    \centering
\begin{tikzpicture}[baseline={([yshift=-.5ex]current bounding box.center)},scale =2]
\draw[very thick,red, line width = 0.1cm] (0,-1.3) node[above right, scale =1.7 ] {$W_{R}$}  -- (0,0);

\fill[blue!30, fill opacity=0.7] (0,0) ellipse (1 and 0.4);

\draw[very thick, blue, line width = 0.1cm] (-1,0) arc (180:360:1 and 0.4);
\draw[very thick, blue, line width = 0.1cm] (1,0) arc (0:180:1 and 0.4);

\fill[white] (-0.08,0.372) rectangle (0.08,0.5);

\draw[very thick,  green!70!black,line width = 0.1cm] (0,0) -- (0,1.3) node[below right, scale =1.7  ] {$W_{\mathsf{C}(R)}$} ;

\fill[black] (0,0) circle (0.04);

\draw[->, very thick, >=Latex] (-1.45,0.75) to[bend left=25] (-0.45,0.0);
\node[scale=1] at (-2,0.75) {$\mathsf{C}$- Surface};

\node[scale=1.2, blue] at (1.3,0.) {$ \mathsf M_a$};
\draw[->, very thick, >=Latex] (1.45,0.75) to[bend right=25] (0.55,0.35);
\node[scale=1,align =center] at (2.1,0.75) {Monodromy \\ Defect};
\end{tikzpicture}
\caption{Topological defect implementing the action of charge conjugation on a Wilson line.}
\label{fig:crosstop}
\end{figure}
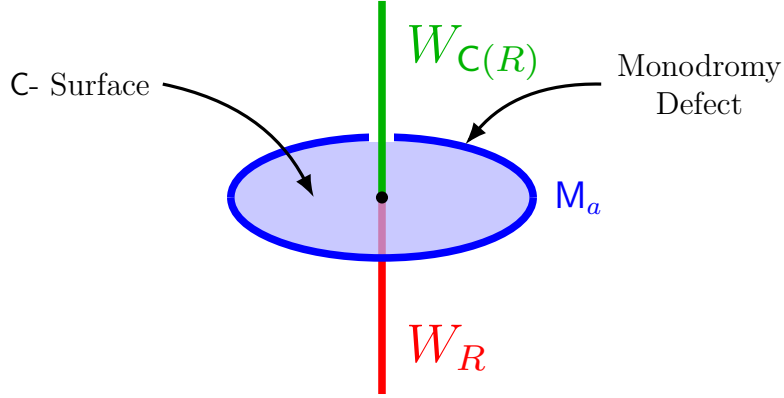
Since crossing  the topological defect   implements the transformation $W_R\rightarrow W_{{\mathsf C}(R)}$, see Figure \ref{fig:crosstop}, the Hilbert space $\mathcal{H}^{(\mathsf{C},1)}$ is the subspace of integrable representations of 
$\mathfrak{g}_k$ that are   $\mathsf C$-invariant.\footnote{
Equivalently, these are precisely the representations
for which there exists a nonzero topological junction between the two Wilson lines, $\delta_{R,\mathsf C(R)}\neq 0$ (see \eqref{eq:HCI}).}
 
The resulting wavefunction is the $\mathsf{C}$-twined character of
$\mathfrak{g}_k$, defined by taking the trace over the $\mathfrak{g}_k$-module $L_R$  
 with an insertion of the action of  
$\mathsf{C}$   on the module
\beq
\mathrm{Tr}_{L_R}\!\left(
U_{\mathsf{C}}\, q^{L_0 - c/24}\right)\,.
\eeq
The twined-trace obviously vanishes unless $R$ is
self-conjugate, that is
\beq
\mathsf{C}(R) = R\,.
\eeq 
Therefore, the number of (simple) monodromy defects in Chern--Simons theory is 
\begin{equation}
\begin{aligned}
   |\mathsf M_a|=  \bigl|\bigl\{R \in \mathrm{Int}(\mathfrak{g}_k)
  \;\big|\; \mathsf{C}(R) = R\bigr\}\bigr|
  \;=\;
  \bigl|\mathrm{Int}(\mathfrak{g}^{(2)}_k)\bigr|\,.
  \label{counting}
\end{aligned}
\end{equation}
The wavefunctions of $\mathcal{H}^{(\mathsf{C},1)}$ can be explicitly  written  down by leveraging a beautiful result in the theory of affine algebras.    The nonvanishing $\mathsf{C}$-twined
characters of $\mathfrak{g}_k$ are themselves characters of
integrable representations of a third affine Lie algebra, the
\emph{orbit Lie algebra} $\breve{\mathfrak{g}}_k$ \cite{fuchs1995affine}.
The 
  Dynkin diagram of $\breve{\mathfrak{g}}_k$
is obtained by folding the Dynkin diagram of
$\mathfrak{g}_k$ under the action of the outer automorphism (see Table \ref{tab:outer-auto}).\footnote{The outer automorphism acts trivially on the extended node.}
Let $\breve{\mathfrak{g}} \subset \breve{\mathfrak{g}}_k$
denote the horizontal subalgebra.  The integrable
representations of $\breve{\mathfrak{g}}_k$ are labeled by
highest weight representations of $\breve{\mathfrak{g}}$
whose Dynkin labels $\mu_j$ satisfy
\beq
\sum_{j=1}^{\mathrm{rank}(\breve{\mathfrak{g}})} c_j^\vee\, \mu_j
\leq k\,,
\label{integrable3}
\eeq
where $c_j^\vee$ are the comarks of $\breve{\mathfrak{g}}_k$. 
A basis of wavefunctions of $\mathcal{H}^{(\mathsf{C},1)}$
are the characters $\breve\chi_\mu(\tau)$
of the   $\breve{\mathfrak{g}}_k$-modules  $L_ \mu$ 
\beq
  \breve\chi_\mu(\tau)
  \equiv \mathrm{Tr}_{L_\mu}\!\bigl(\,q^{L_0-c/24}\bigr)\,.
\eeq
Therefore, the Hilbert space $\mathcal{H}^{(\mathsf C,1)}$ can be written as 
\beq
\mathcal{H}^{(\mathsf{C},1)} \cong
\operatorname{span}\bigl\{\breve{\chi}_\mu \;\big|\; \mu \in
\mathrm{Int}(\breve{\mathfrak{g}}_k)\bigr\}\,,
\eeq
where $\mathrm{Int}(\breve{\mathfrak{g}}_k)$ denotes the set of
integrable representations of $\breve{\mathfrak{g}}_k$.

There is a precise algebraic relation between the twisted affine algebra 
$\mathfrak{g}^{(2)}_k$ and the orbit affine Lie algebra $ \breve{\mathfrak{g}}_k$. Their  horizontal Lie algebras are Langlands dual to each other\footnote{$A_{2N}^{(2)}$  
is self-dual and $ \mathfrak{g}^{(2)}_k \simeq \breve{\mathfrak{g}}_k$.}  
\begin{equation}
    {}^L{\mathfrak g_+} \simeq \breve{\mathfrak{g}}\,.
\end{equation}
From these horizontal subalgebras, we can   define their canonical untwisted affinizations, which we denote by $(\mathfrak{g}_+)_k$ and $\breve{\mathfrak{g}}_k$ respectively. The twisted affine algebra 
$\mathfrak{g}^{(2)}_k$ is then the Langlands dual to 
$\breve{\mathfrak{g}}_k$, while $\breve{\mathfrak{g}}_k$ is Langlands dual\footnote{If $A$ is the generalized Cartan matrix of an affine Lie algebra, then the
generalized Cartan matrix of its Langlands dual is $A^T$.
 } to $(\mathfrak{g}_+)_k$, that is:
\begin{equation}
 \mathfrak{g}^{(2)}_k= {}^L \breve{\mathfrak{g}}_k\,,\qquad  
 \breve{\mathfrak{g}}_k={}^L(\mathfrak{g}_+)_k\,.
\end{equation}

\subsubsection*{$\bullet$ \boldmath Hilbert space $\mathcal {H}^{(\mathsf C, \mathsf C)}$}
Finally, we turn to the Hilbert space $\mathcal{H}^{(\mathsf{C},\mathsf{C})}$. The underlying Hilbert space is the same as  
$\mathcal{H}^{(1,\mathsf{C})}$ -- it is labeled by the integrable
highest weight representations of the twisted affine algebra
$\mathfrak{g}^{(2)}_k$. A wavefunction in $\mathcal{H}^{(\mathsf{C},\mathsf{C})}$ is given by the  Chern--Simons path integral with the insertion of a
monodromy defect $\mathsf{M}_a$ 
along the noncontractible $\mathsf b$-cycle   at the center of the solid torus,   in the presence of a $\mathsf{C}$-defect along the $(\mathsf a+\mathsf b)$-cycle of the boundary torus:
\begin{align}\label{eq:HCC}
    \mathcal{H}^{(\mathsf{C}, \mathsf{C})} \simeq \includegraphics[valign =c, scale=0.8]{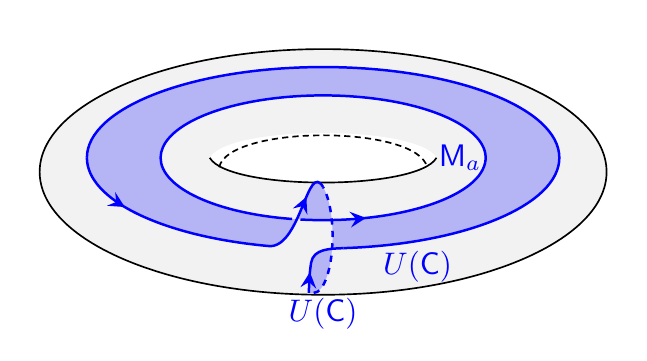}
\end{align} The resulting
wavefunction is the $\mathsf{C}$-twined character of
$\mathfrak{g}^{(2)}_k$, defined by taking the trace over the $\mathfrak{g}^{(2)}_k$-module $L_a$
with an insertion of the action of  
 $\mathsf{C}$   on the module
\[
\chi^{(\mathsf{C})}_a (\tau)\equiv \mathrm{Tr}_{L_a}\!\left(
U_{\mathsf{C}}\, q^{L_0 - c/24}\right)\,.
\]
Choosing a basis of currents adapted to the decomposition
\(\mathfrak g=\mathfrak g_+\oplus\mathfrak g_-\), the action on the currents is (cf. \eqref{modetwisted}) 
\begin{equation}
    U_{\mathsf C} J^\alpha_n U_{\mathsf C}^{-1}
    =
    \begin{cases}
        \phantom{-}J^\alpha_n, & \alpha \in \mathfrak g_+,\\[2pt]
        -J^\alpha_n, & \alpha \in \mathfrak g_- .
    \end{cases}
\end{equation}
The conjugation action of \(\mathsf C\) fixes the operator \(U_{\mathsf C}\)
only up to an overall phase on each irreducible twisted module \(L_a\). This
ambiguity is fixed as follows. Since \(\mathsf C\) preserves the positive mode
subalgebra, \(U_{\mathsf C}|a\rangle\) is again a highest weight state in
\(L_a\). The highest weight space is one-dimensional, so we may choose the
normalization of \(U_{\mathsf C}\) such that
\[
U_{\mathsf C}|a\rangle=|a\rangle .
\]
With this convention, \(U_{\mathsf C}\) is realized on \(L_a\) by
\[
U_{\mathsf C}=e^{2\pi i(L_0-\Delta_a)} .
\]
Indeed, \(L_0-\Delta_a\) measures the excitation level above the
highest weight state: integer-moded oscillators contribute phase \(+1\), while
half-integer-moded oscillators contribute phase \(-1\), precisely reproducing
the action of \(\mathsf C\).

We now use the fact that the modular transformation \(T\) maps the sector
\[
T:\mathcal H^{(1,\mathsf C)}
\longrightarrow
\mathcal H^{(\mathsf C,\mathsf C)} ,
\]
as follows from performing a Dehn twist on \eqref{eq:H1C}; equivalently, this
is the transformation leading to \eqref{eq:HCC}. Since
\[
L_0-\frac{c}{24}
=
\left(\Delta_a-\frac{c}{24}\right)
+
(L_0-\Delta_a),
\]
we obtain
\[
\begin{aligned}
\chi^{(2)}_a(\tau+1)
&=
\mathrm{Tr}_{L_a}\!\left(
e^{2\pi i (L_0-c/24)}\, q^{L_0-c/24}
\right)  \\[4pt]
&=
e^{2\pi i(\Delta_a-c/24)}
\mathrm{Tr}_{L_a}\!\left(
e^{2\pi i(L_0-\Delta_a)}\, q^{L_0-c/24}
\right)  \\[4pt]
&=
e^{2\pi i(\Delta_a-c/24)}
\mathrm{Tr}_{L_a}\!\left(
U_{\mathsf C}\, q^{L_0-c/24}
\right)  \\[4pt]
&=
e^{2\pi i(\Delta_a-c/24)}\,
\chi^{(\mathsf C)}_a(\tau).
\end{aligned}
\]
Thus the \(T\)-transform of the twisted affine character is the
\(\mathsf C\)-twined character, multiplied by the expected ground state phase.

\begin{table}
\centering
\resizebox{\textwidth}{!}{
\renewcommand{\arraystretch}{1.7}
\setlength{\tabcolsep}{10pt}
\begin{tabular}{|c|c|c|c|c|}
\hline
\dynname{\mathfrak{g}_k}
& \(( 1, 1)\)
& \(( 1, \mathsf C)\)
& \((\mathsf C, 1)\)
& \((\mathsf C,\mathsf C)\)
\\
\hline\hline
\dynname{A^{(1)}_{2n-1}}, \(n\ge 2\)
& \dynname{A^{(1)}_{2n-1}}
& \dynname{A^{(2)}_{2n-1}}
& \dynname{D^{(2)}_{n+1}}
& \dynname{A^{(2)}_{2n-1}}
\\
\diagAoneExt & \diagAoneExt & \diagAtwoNmaffExt & \diagDaffExt & \diagAtwoNmaffExt
\\[7pt]
\hline
\dynname{A^{(1)}_{2n}}, \(n\ge 1\)
& \dynname{A^{(1)}_{2n}}
& \dynname{A^{(2)}_{2n}}
& \dynname{A^{(2)}_{2n}}
& \dynname{A^{(2)}_{2n}}
\\
\diagAoneExt & \diagAoneExt & \diagAtwoNaffExt & \diagAtwoNaffExt & \diagAtwoNaffExt
\\[7pt]
\hline
\dynname{D^{(1)}_{n}}, \(n\ge 4\)
& \dynname{D^{(1)}_{n}}
& \dynname{D^{(2)}_{n}}
& \dynname{A^{(2)}_{2n-3}}
& \dynname{D^{(2)}_{n}}
\\
\diagDoneExt & \diagDoneExt & \diagDaffExt & \diagAtwoNmaffExt & \diagDaffExt
\\[7pt]
\hline
\dynname{E^{(1)}_{6}}
& \dynname{E^{(1)}_{6}}
& \dynname{E^{(2)}_{6}}
& \dynname{E^{(2)}_{6}}
& \dynname{E^{(2)}_{6}}
\\
\diagEoneExt & \diagEoneExt & \diagEtwoExt & \diagEtwoExt & \diagEtwoExt
\\[8pt]
\hline
\end{tabular}}
\caption{Affine algebras associated with the four \(T^2\) sectors. We use Kac's $\mathfrak g^{(r)}$ notation and suppress the level \(k\). The affine node is highlighted in {\color{blue}blue}.}
\label{tab:z2_outer_automorphism_sectors}
\end{table}

\subsection{Line Defects in Chern--Simons theory}
We have thus concluded that Chern--Simons theory with charge conjugation symmetry admits two distinguished classes of line defects:
\begin{itemize}
    \item ordinary Wilson lines \(W_R\), labeled by \(R\in \mathrm{Int}(\mathfrak g_k)\);
    \item \(\mathsf C\)-monodromy lines \(\mathsf M_a\), labeled by \(a\in \mathrm{Int}(\mathfrak g^{(2)}_k)\).
\end{itemize}
Table~\ref{tab:z2_outer_automorphism_sectors} summarizes the affine Lie
algebras governing each sector.

Together, these lines furnish the simple objects of a
\(\mathbb Z_2^{\mathsf C}\)-crossed braided tensor category~\cite{Turaev2000,Kirillov2004,Mueger2005,Barkeshli2019},
\begin{equation}
    \mathcal C^\times
    =
    \mathcal C_{\mathbf 1}
    \oplus
    \mathcal C_{\mathsf C}\, .
\end{equation}
The neutral component \(\mathcal C_{\mathbf 1}\) is the modular tensor category of ordinary Wilson lines \(W_R\). Equivalently,
\(\mathcal C_{\mathbf 1}\) is the category of integrable highest weight representations of the affine Lie algebra \(\mathfrak g_k\). The nontrivial component \(\mathcal C_{\mathsf C}\) is the \(\mathsf C\)-twisted defect sector; its simple objects are the monodromy lines \(\mathsf M_a\), labeled by integrable representations of the corresponding twisted affine Lie algebra \(\mathfrak g^{(2)}_k\).\footnote{A \(\Gamma\)-crossed braided extension of a modular tensor category is not
automatic. Once a braided \(\Gamma\)-action on \(\mathcal C\) is fixed,
symmetry fractionalization choices form, when unobstructed, a torsor over
\(H^2_\rho(\Gamma,\mathrm{Inv}(\mathcal C))\); the existence of the full
extension is further obstructed by a class in \(H^4(\Gamma,U(1))\), and the
remaining ambiguity includes stacking with \(2+1\)-dimensional bosonic
SPT phases, classified by \(H^3(\Gamma,U(1))\)~\cite{ENO2010}. In
Chern--Simons theory with charge conjugation symmetry, the monodromy
defects constructed here explicitly realize the
\(\mathbb Z_2^{\mathsf C}\)-twisted sector of such an extension.} 

Fusion in \(\mathcal C^\times\) respects the \(\mathbb Z_2\)-grading:
\begin{equation}
    \mathcal C_g\otimes \mathcal C_h
    \subset
    \mathcal C_{gh}\, .
\end{equation}
Thus, fusing an ordinary Wilson line with a \(\mathsf C\)-monodromy line gives an object in the \(\mathsf C\)-twisted sector, whereas the fusion of two \(\mathsf C\)-monodromy lines decomposes into objects in the neutral sector.

We have shown that the algebraic structure underlying the
\(\mathbb Z_2^{\mathsf C}\)-crossed braided tensor category of
Chern--Simons theory with charge conjugation symmetry is governed by the
interplay between untwisted and twisted affine Lie algebras. The untwisted
affine algebra \(\mathfrak g_k\) controls the ordinary Wilson lines and their
fusion, while the \(\mathsf C\)-twisted sector is naturally organized by
integrable representations of the corresponding twisted affine algebra
\(\mathfrak g^{(2)}_k\). In this way, the categorical data of the
charge conjugation extension 
--  including the action of \(\mathsf C\) on Wilson
lines, the spectrum of monodromy defects, and the modular transformations
between twisted sectors --  admit a concrete realization in the representation
theory of affine Lie algebras.

Physically, this is the categorical language of a topological phase of matter with a global symmetry.  A seminal insight of the condensed-matter and mathematical-physics communities is that gapped phases of matter enriched by a finite symmetry group~$\Gamma$ are classified, at the level of their anyonic content, by $\Gamma$-crossed braided tensor categories together with a compatible $\Gamma$-action~\cite{ENO2010,CGPW2016,Barkeshli2019}.  
   The   sector~$\mathcal{C}_{\mathbf{1}}$ describes the intrinsic topological order, while the   sector~$\mathcal{C}_g$ encodes the symmetry fractionalization and the defect content of the phase.  Gauging the   symmetry $\Gamma$ -- that is, summing over $g\in \Gamma$-twisted sectors on every cycle of the spacetime manifold -- defines a   new  modular tensor category.  This passage from symmetry-enriched to symmetry-gauged phases is one of the central tools for constructing and classifying topological orders.  Chern--Simons theory with $\mathsf C$-symmetry provides a particularly elegant and computable example of symmetry-enriched topological order. 

Arbitrary correlation functions of Wilson   and      defects in $G_k$ Chern--Simons theory can be computed from the data of the $\mathbb{Z}_2^{\mathsf C}$-crossed braided tensor category. Once this data is known, correlation functions are obtained by the same topological moves as for Wilson lines in ordinary Chern--Simons theory, now enriched by the possibility of dragging lines across symmetry branch cuts.  In this sense, the 
 $\mathbb{Z}_2^{\mathsf C}$-crossed braided tensor category 
 is the complete algebraic machinery behind all arbitrary correlators of  Wilson and monodromy defects.

\paragraph{$\bullet$ Modular S-matrix.}
The modular data has a simple geometric description.  The   $S$-matrix is obtained by evaluating Hopf links of simple objects, or equivalently by the modular $S$-transformation exchanging the two cycles of the torus. 

The modular data for the $\mathbb{Z}_2^C$-crossed braided tensor category 
$\mathcal C^\times$ relevant for Chern--Simons theory with $\mathsf C$-symmetry has a nice algebraic description. The $S$-matrix relevant for Wilson lines and ${\mathsf C}$-monodromy defects can be obtained by acting on the wavefunctions (characters) on the Hilbert spaces $\mathcal{H}^{(1,1)}$ and $\mathcal{H}^{(1,\mathsf{C})}$ respectively.  The $S$-matrix acting on Wilson lines, which create the states in $\mathcal{H}^{(1,1)}$,  is the usual $S$-matrix of the affine Lie algebra $\mathfrak{g}_k$. The $S$-matrix maps states of $\mathcal{H}^{(1,1)}$ into each other 
\beq
S:\mathcal{H}^{(1,1)}\rightarrow \mathcal{H}^{(1,1)}\,
\eeq
and is obtained by from the action on the characters of $\mathfrak{g}_k$
\begin{equation}
\chi_R(-1/\tau)=\sum_S S^{(1,1)}_{RS} \,\chi_S( \tau)\,.
\end{equation}

This $S$-matrix determines, in particular,  the   expectation value of a Wilson loop $W_R$ on an unknot. The normalized expectation value on $S^3$ is    the quantum dimension of the representation $R$~\cite{Witten:1988hf}: 
  \beq
\langle W_R\rangle_{\mathrm{norm}}={\frac{S^{(1,1)}_{R  0}} { S^{(1,1)}_{0  0}} }\,,
  \eeq
where $0$ denotes the trivial representation.

The modular \(S\)-matrix of the affine algebra \(\mathfrak{g}_k\)
can be written   as 
    a sum over the Weyl group  
of the finite horizontal Lie algebra \(\mathfrak g\) of \(\mathfrak{g}_k\)~\cite{kac_1990}\footnote{The bracket $\langle \cdot,\cdot \rangle$ is an
  invariant bilinear form normalized such that on the simple roots $\alpha_i$ of the horizontal algebra of the affine Lie algebra $\langle \alpha_i,\alpha_i\rangle =2 \tfrac{a_i^\vee}{a_i}$, where $a_i$ and $a_i^\vee$ are the marks and comarks associated to the (non-extended) $i$-th node of the affine Dynkin diagram. This implies that for simply-laced algebras, $Q^\vee=Q$.}
\begin{equation}
  S^{(1,1)}_{RS}
  =
  \frac{i^{|\Delta_+(\mathfrak g)|}}
  {(k+h^\vee)^{\mathrm{rank}(\mathfrak g)/2}\sqrt{|P/Q^\vee|}}
  \sum_{w\in W(\mathfrak g)}
  \epsilon(w)
  \exp\left(
  -\frac{2\pi i}{k+h^\vee}
  \left\langle w(\rho+R),\rho+S\right\rangle
  \right) .
  \label{Smatrix11}
  \end{equation}
where $\rho_{\mathfrak g}$ and $\cal W(\mathfrak g)$ is the Weyl vector and Weyl group  of $\mathfrak g$, $h^\vee$ is the dual Coxeter number of $\mathfrak g$, \(\Delta_+(\mathfrak g)\) is the set of positive roots, $P$ and $Q^\vee$ are the weight and co-root lattice of $\mathfrak g$, respectively. The $S$-matrix \eqref{Smatrix11} is symmetric and unitary \footnote{It also obeys
$\left(S^{(1,1)}\right)^2=C$, with 
$C_{RS}=\delta_{R,\overline{S}}\ 
$
where \(\overline{S}\) denotes the representation conjugate to \(S\). Equivalently,
\(S\otimes\overline{S}\) contains the vacuum representation \(\mathbf{1}\).}
\begin{equation}
 S^{(1,1)}  \left(S^{(1,1)}\right)^\dagger=\mathbf{1}\,.
\end{equation}
Determining the $S$-matrix for $\mathsf C$-monodromy defects is more subtle. Since the $S$-transfor\-mation maps 
\beq
S:\mathcal{H}^{(1,\mathsf{C})}\rightarrow \mathcal{H}^{(\mathsf{C},1)}\,,
\eeq
the $S$-matrix maps characters $ \chi_a$ of $\mathfrak{g}^{(2)}_k$ to characters $ \chi_\mu$ of $\breve{\mathfrak{g}}_k$. This defines another  $S$-matrix 
\begin{equation}
\chi_a(-1/\tau)=\sum_\mu S^{(\mathsf C,1)}_{a\mu} \,\breve{\chi}_\mu( \tau)\,.
\end{equation}

  This $S$-matrix determines, in particular, the   expectation value of a monodromy defect $\mathsf M_a$ on an unknot. The normalized expectation value on $S^3$ is    the quantum dimension of the representation   $a$ labeling the monodromy defect $\mathsf M_a$. This     is defined by  the following ratio of $S$-matrices (cf.~\cite{Barkeshli2019}): 
  \beq
\langle \mathsf M_a\rangle_{\mathrm{norm}}=\frac{S^{(\mathsf C,1)}_{a  0}} { S^{(1,1)}_{0  0} }\,,
  \eeq
  the one in numerator  acting on $\mathcal{H}^{(1,\mathsf C)}$ and the one on denominator on $\mathcal{H}^{(1,1)}$.
  
  A general property of \(\Gamma\)-crossed braided tensor categories is that all
grading components have the same total squared quantum
dimension~\cite{Barkeshli2019}. This implies that
\begin{equation}
    \sum_{R\in \mathrm{Int}(\mathfrak g_k)}
    \left(S^{(1,1)}_{R0}\right)^2
    =
    \sum_{a\in \mathrm{Int}(\mathfrak g^{(2)}_k)}
    \left(S^{(\mathsf C,1)}_{a0}\right)^2 .
\end{equation}
Equivalently, the total squared quantum dimension of the
\(\mathsf C\)-twisted sector \(\mathcal C_{\mathsf C}\), spanned by the
monodromy defects \(\mathsf M_a\), equals that of the neutral Wilson line
sector \(\mathcal C_{\mathbf 1}\).

This equality is a global constraint on the full collection of
\(\mathsf C\)-monodromy defects. A stronger statement about the individual
defects follows from the fact that \(\mathsf C\) acts nontrivially on the
Wilson line category, permuting representation labels as
\[
    W_R \longmapsto W_{\mathsf C(R)}\, .
\]
For a symmetry that nontrivially permutes simple objects of
\(\mathcal C_{\mathbf 1}\), as charge conjugation does, the corresponding symmetry defects are necessarily
non-Abelian~\cite{Barkeshli2019}. Hence the simple
\(\mathsf C\)-monodromy defects have quantum dimensions
\begin{equation}
    d_a
    =
    \frac{S^{(\mathsf C,1)}_{a0}}{S^{(1,1)}_{00}}
    >1,
    \qquad
    a\in \mathrm{Int}(\mathfrak g^{(2)}_k)\, .
\end{equation}

The modular $S$-matrix   $S^{(\mathsf C,1)}_{a  \mu}$ of the twisted affine algebra
$\mathfrak{g}^{(2)}_k$ can be written as  a   sum over the Weyl group of the finite horizontal subalgebra\footnote{Recall that $\mathring{\mathfrak g}=\mathfrak g_+$ for all twisted affine algebras except for $A_{2N}^{(2)}$, where 
$\mathring{\mathfrak g}={}^L\mathfrak g_+\simeq C_N$.}   of $\mathfrak{g}^{(2)}_k$~\cite{kac_1990}
\begin{equation}
 S^{(\mathsf C,1)}_{a  \mu}=i^{\,|\Delta_+(\mathring{\mathfrak{g}})|}{\cal M}(\mathfrak g, k) \sum_{w\in \cal W(\mathring{\mathfrak g})} \epsilon(w) \exp\left( -\frac{2\pi i } {k+ h^\vee} \langle w(\rho_{\mathring{\mathfrak g}}+a), \tau(\rho_{\breve{\mathfrak{g}}}+\mu\rangle \right)\,,
 \label{Smatrix1C}
\end{equation}
with  ${\cal M}(\mathfrak g, k)$ is a  real, positive nomalization factor and   where $\tau$ is the canonical
embedding of weight lattices
\begin{equation}
   \tau: P(\breve{\mathfrak{g}})\hookrightarrow
   P(\mathring{\mathfrak g})\,.
\end{equation}
 $\tau$  is the identity map when $\mathring{\mathfrak g}\simeq \breve{\mathfrak{g}}$. When 
 $\mathring{\mathfrak g}\ncong \breve{\mathfrak{g}}$, then the   Lie algebras are Langland duals to each other, that is ${}^L\mathring{\mathfrak g}\simeq \breve{\mathfrak{g}}$.

The modular $S$-matrix $S^{(\mathsf C,1)}_{a\mu}$, which is    real  and not symmetric,\footnote{It is symmetric for $A_{2N}^{(2)}$ because $a,\mu$ take values in the same algebra.}     obeys 
\begin{equation}
   S^{(\mathsf C,1)}  \left(S^{(\mathsf C,1)}\right)^\dagger= \left(S^{(\mathsf C,1)}\right)^\dagger  S^{(\mathsf C,1)}=  S^{(\mathsf C,1)}\left(S^{(\mathsf C,1)}\right)^T=\left(S^{(\mathsf C,1)}\right)^T S^{(\mathsf C,1)}=\mathbf{1}\,.
\end{equation}
This condition allows you to uniquely fix $\mathcal{M}(\mathfrak g, k)$ in $S^{(\mathsf C,1)}$. Explicit formulas for $S^{(\mathsf C,1)}_{a  \mu}$ will appear in Section~\ref{sec:four}. 

\paragraph{$\bullet$ Fusion coefficients.}

 The   line defects  in $G_k$ Chern--Simons theory    furnish a  
  $\mathbb{Z}_2^{\mathsf{C}}$-crossed braided tensor category 
\begin{equation}
    \mathcal{C}^{\times}
    =
    \mathcal{C}_{\mathbf{1}}
    \oplus
    \mathcal{C}_{\mathsf{C}}\,.
\end{equation}
The fusion product of Wilson lines and monodromy defects is compatible with
the \(\mathbb Z_2^{\mathsf C}\)-grading,
\[
\mathcal C_g\otimes \mathcal C_h \subset \mathcal C_{gh}.
\]
Explicitly, the fusion coefficients are given by Verlinde-type~\cite{Verlinde:1988sn} formulae
involving the ordinary modular matrix \(S^{(1,1)}\) and the
twisted modular matrix \(S^{(\mathsf C,1)}\). They are given by~\cite{Birke:1999ik,Deshpande:2019kwe,Dong:2023vok}: 
\begin{align}
R\otimes S &= \bigoplus_{T} N_{RS}{}^{T}\,T\,, &
N_{RS}{}^{T} &= \sum_{U\,\in\,\mathrm{Int}(\mathfrak g_k)}
\frac{S^{(1,1)}_{RU}\,S^{(1,1)}_{SU}\,
\overline{S^{(1,1)}_{TU}}}{S^{(1,1)}_{0U}}\,,
\label{eq:VerlindeWW}
\\[5pt]
R\otimes a &= \bigoplus_{b} N_{Ra}{}^{b}\,b\,, &
N_{Ra}{}^{b} &= \sum_{\mu\,\in\,\mathrm{Int}(\breve{\mathfrak{g}}_k)}
\frac{S^{(1,1)}_{R\,\iota(\mu)}\;S^{(\mathsf{C},1)}_{a\mu}\;
\overline{S^{(\mathsf{C},1)}_{b\mu}}}{S^{(1,1)}_{0\,\iota(\mu)}}\,,
\label{eq:VerlindeWD}
\\[5pt]
a\otimes b &= \bigoplus_{R} N_{ab}{}^{R}\,R\,, &
N_{ab}{}^{R} &= \sum_{\mu\,\in\,\mathrm{Int}(\breve{\mathfrak{g}}_k)}
\frac{S^{(\mathsf{C},1)}_{a\mu}\;S^{(\mathsf{C},1)}_{b\mu}\;
\overline{S^{(1,1)}_{R\,\iota(\mu)}}}{S^{(1,1)}_{0\,\iota(\mu)}}\,.
\label{eq:VerlindeDD}
\end{align}
The sum over $U$ in \eqref{eq:VerlindeWW}  is over the integrable representations of 
 the affine algebra $\mathfrak g_k$, while in \eqref{eq:VerlindeWD}--\eqref{eq:VerlindeDD} the sum is over $\mu$,  the integrable representations of  the orbit Lie algebra
$\breve{\mathfrak{g}}_k$.
 The map
\[
\iota:\mathrm{Int}(\breve{\mathfrak g}_k)
\longrightarrow
\mathrm{Int}(\mathfrak g_k)^{\mathsf C}
\]
is the standard level-preserving bijection from integrable representations
of the orbit Lie algebra to the \(\mathsf C\)-invariant integrable
representations of \(\mathfrak g_k\)~\cite{fuchs1995affine}. In affine Dynkin labels, \(\iota\) unfolds the orbit Lie algebra weight
\(\mu\) along the orbits of the outer automorphism, thereby producing the
corresponding outer automorphism invariant integrable weight of
\(\mathfrak g_k\). The explicit form of \(\iota(\mu)\) for each twisted
affine algebra is given in Table~\ref{tab:iota}.
\begin{table}[H]
\centering
\begin{tabular}{@{}llll@{}}
\toprule
\multicolumn{1}{c}{$\mathfrak g$} &
\multicolumn{1}{c}{$\breve{\mathfrak g}$} &
\multicolumn{1}{c}{$\mu$} &
\multicolumn{1}{c}{$\iota(\mu)$} \\
\midrule
$A_{2N-1}$ & $D_{N+1}^{(2)}$ &
$(b_0;\,b_1,\dots,b_{N-1},b_N)$ &
$(b_0;\,b_1,\dots,b_{N-1},\,b_N,\,b_{N-1},\dots,b_1)$ \\
$A_{2N}$ & $A_{2N}^{(2)}$ &
$(b_0;\,b_1,\dots,b_N)$ &
$(b_0;\,b_1,\dots,b_{N-1},\,b_N,b_N,\,b_{N-1},\dots,b_1)$ \\
$D_{N+1}$ & $A_{2N-1}^{(2)}$ &
$(b_0;\,b_1,\dots,b_{N-1},b_N)$ &
$(b_0;\,b_1,\dots,b_{N-1},b_N,b_N)$ \\
$D_4$ & $D_4^{(3)}$ &
$(b_0;\,b_1,b_2)$ &
$(b_0;\,b_1,b_2,b_1,b_1)$ \\
$E_6$ & $E_6^{(2)}$ &
$(b_0;\,b_1,b_2,b_3,b_6)$ &
$(b_0;\,b_1,b_2,b_3,b_2,b_1,\,b_6)$ \\
\bottomrule
\end{tabular}
\caption{Bijection between integrable representations of
$\breve{\mathfrak{g}}_k$ and the outer automorphism invariant
representations of $\mathfrak{g}_k$ for each simple $\mathfrak g$ with an outer
 automorphism.
Node numbering follows~\cite[Table~Aff~1]{kac_1990}; the semicolon
isolates the label of the affine node.}
\label{tab:iota}
\end{table}

\section{Chern--Simons theory and Holography}
\label{sec:suNdual}

Chern--Simons theory with gauge algebra $\mathfrak{g} = \mathfrak{su}(N), \mathfrak{so}(N),$ or $\mathfrak{sp}(N)$ admits an elegant large $N$ expansion. This naturally raises the question -- in the spirit of the't~Hooft expansion ~\cite{tHooft:1973alw} -- of what closed string theory provides the dual description of Chern--Simons theory.

The natural framework in which to address these questions is   A-model topological string theory~\cite{Witten:1988xj,Witten:1991mm}. Indeed, the open string A-model on a Calabi--Yau threefold with \(N\) Lagrangian branes supported on a Lagrangian three-cycle \(L\) is governed, on the brane worldvolume, by {\it analytically continued } Chern--Simons theory on \(L\), with gauge connection  suitably complexified~\cite{Witten:1992fb}.\footnote{Depending on the geometry, the Chern--Simons action can be corrected by worldsheet instantons~\cite{Witten:1992fb}.}  
 This construction mirrors the way in which the boundary conformal field theory in an AdS/CFT pair arises as the low energy limit of the open string theory living on a stack of D-branes in string theory \cite{Maldacena:1997re}.

A  holographic correspondence requires more than a matching of partition functions: it demands a precise dictionary between observables on both sides of the duality. In the present setting, this means that  all observables in Chern--Simons theory  -- Wilson loops and $\mathsf C$-monodromy defects -- must admit a dual     string description.  The search for the closed string dual description of monodromy defects is not merely a refinement of the correspondence, but a necessary ingredient in establishing the duality.

\subsection{String dual of partition function}
\label{partitionfunctiondual}
\label{sec:3.1}
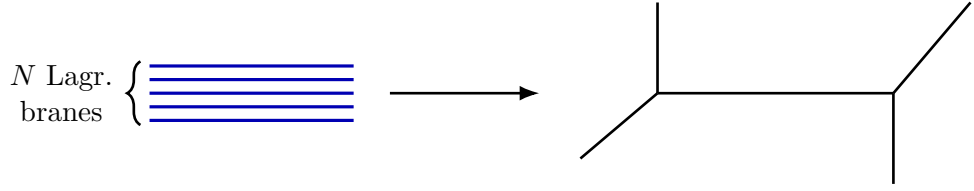
\begin{figure}
    \centering
\begin{tikzpicture}[
    >=Latex,
    line/.style={line width=1.0pt},
    brane/.style={line width=1.2pt,blue!70!black},
    oplane/.style={line width=1.2pt,red!75!black},
    xmark/.style={line width=1.35pt,red!80!black},
    every node/.style={font=\small}, scale =1.2
]
\foreach \y in {-0.30,-0.15,0,0.15,0.30} {
    \draw[brane] (-3.75,\y) -- (-1.5,\y);
}
\draw[decorate, decoration={brace, amplitude=5pt}, line width=1pt]
    (-3.85,-0.35) -- (-3.85,0.35)
    node[midway, left=6pt,align=center] {$N$ Lagr.\ \\branes};

\draw[->,line] (-1.10,0) -- (0.55,0);
\begin{scope}[shift={(1.85,0)}]
\coordinate (L) at (0,0);
\coordinate (R) at (2.60,0);
\coordinate (X) at ($(L)!0.5!(R)$);
\draw[line] (L) -- (R);
\draw[line] (L) -- ++(0,1.00);
\draw[line] (L) -- ++(-0.85,-0.72);
\draw[line] (R) -- ++(0.85,1.00);
\draw[line] (R) -- ++(0,-1.00);
\end{scope}
\end{tikzpicture}
    \caption{A stack of $N$ Lagrangian branes wrapping $S^3$ is replaced, after geometric transition by an exceptional $\mathbb{P}^1$ with K\"ahler parameter $t= Ng_s$ }
    \label{fig:resconifold}
\end{figure}

 We begin our search for the bulk description of monodromy defects by recalling the essential ingredients underlying the string dual   of   the partition function of \(SU(N)_k\) Chern--Simons theory on \(S^3\) \cite{Gopakumar:1998ki} and of its   Wilson line observables \(W_R\) \cite{Ooguri:1999bv,Gomis:2006sb}.\footnote{Important work  matching     knot and link invariants to open topological string amplitudes includes~\cite{Labastida:2000yw,Aganagic:2000gs,Aganagic:2001nx,Marino:2001re,Diaconescu:2002qa,Diaconescu:2011at}.} Analytically continued Chern--Simons theory on $S^3$ arises as the open string field theory on $N$ Lagrangian A-branes wrapping the zero-section $S^3\subset T^*S^3$. The closed string dual found by 
 Gopakumar--Vafa~\cite{Gopakumar:1998ki}, where the branes have been replaced by flux, 
 is the A-model topological string on the resolved conifold geometry, which henceforth we denote by $X$ (see Figure \ref{fig:resconifold}). For a review see~\cite{Marino:2005sj,Vafa:2025zah}.

In this paper, we exploit the fact that the gauge theory on the Lagrangian branes is
analytically continued Chern--Simons theory. This observation allows us to
substantially simplify the 't Hooft expansion of gauge theory correlation
functions.  The partition function of \(SU(N)_k\) Chern--Simons theory on \(S^3\) is the
vacuum matrix element of the modular \(S\)-matrix, given by\footnote{Here \(\rho\) is the Weyl vector,  \(R,S\) denote highest weights of integrable
representations of \( \mathfrak{su}(N)_k\) and the Weyl group $\mathcal{W}(\mathfrak{su}(N)) = S_N$}
\begin{equation}
S^{(1,1)}_{RS}
=
\frac{i^{\frac{N(N-1)}{2}}}
{\sqrt{N}\,(k+N)^{\frac{N-1}{2}}}
\sum_{w\in S_N}(-1)^w\,
\exp\!\left[
-\frac{2\pi i}{k+N}
\left\langle
w(R+\rho),\,S+\rho
\right\rangle
\right].
\end{equation}
Specializing to the vacuum representation, \(R=S=0\), gives
\begin{equation} 
Z_{SU(N)_k}(S^3)
=
S_{00}^{(1,1)}
=
\frac{1}{\sqrt{N({k+N})^{N-1}}}
\prod_{1\leq i<j\leq N}
2\sin\frac{\pi(j-i)}{k+N}.
\end{equation}
 Upon defining
\begin{equation}
   q=e^{\frac{2\pi i}{k+N}}=e^{-g_s}\,,
\end{equation}
the vacuum amplitude may be written, up to an overall normalization, as
\begin{equation}
\mathcal{Z}_{SU(N)_k}(S^3)
=
S_{00}^{(1,1)}
\doteq
\prod_{r=1}^{N-1}\left(1-q^r\right)^{N-r}.
\label{ZSU}
\end{equation}
Here \(\doteq\) denotes equality up to an overall prefactor, including
numerical constants and an overall power  of \(q\).\footnote{Such factors contribute to constant maps in the topological string free energy along with the Machmahon and the Euler function and do not affect the worldsheet instanton expansion} This suffices to compute the unambiguous part of A-model topological string amplitudes. We will use this notation
throughout. 
 A  detailed treatment of the omitted
normalization terms is given in Appendix \ref{app:constant-maps} (see also Appendix \ref{sec:SUNconstant}). 

In ordinary unitary Chern--Simons theory, the quantum parameter
\(q\) lies on the unit circle, since the level \(k\) is real
and quantized. In analytically continued Chern--Simons theory, however, \(k\)
is allowed to be complex, and \(q\) becomes a genuine complex parameter. This
allows us to work in a region with \(|q|<1\), where the amplitudes admit a
small \(q\) expansion.
This should be contrasted with the Gopakumar--Vafa expansion of
\(\mathcal Z_{SU(N)_k}(S^3)\)~\cite{Gopakumar:1998ki}, which is performed in
the large \(N\) 't Hooft limit with \(g_s\sim (k+N)^{-1}\) small, or
equivalently \(q=e^{-g_s}\to 1\).
 Taking the logarithm of \eqref{ZSU} and using the small \(q\) expansion
\begin{equation}
    \log(1-q^r)
    =
    -\sum_{n=1}^{\infty}\frac{q^{nr}}{n}\,,
\end{equation}
we can perform the sum over \(r\) explicitly, obtaining
\[
\log S_{00}^{(1,1)}
\doteq
-\sum_{n=1}^{\infty}\frac1n\,
\frac{(N-1)q^{n}-Nq^{2n}+q^{(N+1)n}}{(1-q^{n})^{2}}=
-\sum_{n=1}^{\infty}\frac1n\,
\frac{(N-1)-Nq^{n}+q^{Nn}}{[n]^2}\,,
\]
where
\beq
[n]= q^{n/2}-q^{-n/2}\,.
\eeq
The  sum over the first two terms can be reorganized as
\begin{align}
-\sum_{n=1}^{\infty}\frac{1}{n}\,
\frac{(N-1)-Nq^{n}}{[n]^2}
&=
\sum_{n=1}^{\infty}\frac{1}{n}
\left(
\frac{Nq^{n}}{q^{n}-1}
+
\frac{q^{n}}{(q^{n}-1)^2}
\right) \notag \\
&=
\log\!\left(\frac{M(q)}{\zeta(q)^N}\right),
\end{align}
where \(M(q)=\prod_{n\geq 1}(1-q^n)^{-n}\) is the MacMahon function and
\(\zeta(q)^{-1}=\prod_{n\geq 1}(1-q^n)\). Altogether, we find\footnote{We also define $\mathcal{F}^0_{X}(N) := \frac{M(q)}{\zeta(q)^N}$ and $\mathcal{F}^{\rm inst}_{X}(t) := -\sum_{n=1}^{\infty}
\frac{e^{-nt}}{n\,[n]^2}$, where $X$ is the resolved conifold}
\begin{equation}
\mathcal{Z}_{SU(N)_k}(S^3)
\doteq
\frac{M(q)}{\zeta(q)^N}
\exp\left(
-\sum_{n=1}^{\infty}
\frac{e^{-nt}}{n\,[n]^2}
\right),
\label{ZSUa}
\end{equation}
where
\begin{equation}
    q=e^{-g_s},
    \qquad
    q^N=e^{-t}.
\end{equation}
This is precisely the closed A-model topological string partition function on
the resolved conifold
\[
    X=\mathcal O(-1)\oplus \mathcal O(-1)\longrightarrow \mathbb P^1
\]
in the Gopakumar--Vafa parametrization, that is, in terms of BPS invariants \cite{Gopakumar:1998ii,Gopakumar:1998jq}.{\footnote{The experts might recognize the small $q$ expansion to be the one natural for the $B$-model: the Donaldson--Thomas partition function, which counts ideal sheaves on the resolved conifold. From this perspective the equivalence of the two expansions, is a non-trivial consequence of  the duality of the $A$ and $B$ model topological strings \cite{Iqbal:2003ds,Maulik:2003rzb,Nekrasov:2004js}. Yet, we want to  stress that it is a remarkable property of analytically continued Chern--Simons theory that  this equivalence can be illustrated without ever leaving the gauge theory side of the correspondance.}} Here \(g_s\) is the topological string
coupling and \(t\) is the K\"ahler parameter of the resolved conifold. The
exponential factor is the all genus contribution of nonconstant holomorphic
maps to  the target \(X\), while the prefactor (including the terms we have dropped in $\doteq$) encodes  the
constant map contributions, that we discuss separately in Appendix \ref{sec:SUNconstant}.

As stated above, our derivation is substantially different from the original Gopakumar--Vafa~\cite{Gopakumar:1998ki} (whose computation we review in Section~\ref{smallgs}), where they have proceeded by expanding $\log S_{00}$ at weak string coupling $\gs\ll1 $. Our expansion at $q\ll 1$, corresponds instead to the strong coupling $\gs \gg 1$. Thus, it is absolutely not obvious that the two are equivalent. In Appendix \ref{sucomparison}, we show that the two expansions match term-by-term once the $q$-series is re-expressed in terms of $\gs$. In particular, the Euler function $\zeta(e^{-\gs})$, which does not appear directly at small $\gs$,  plays an important role in matching the two, and should be regarded as an imprint of non-perturbative effects in $\gs$. We discuss this further in Section~\ref{constantmaps}.

Let us emphasize a somewhat striking point: nowhere in this derivation (and in any that will follow) have we used a large \(N\) approximation. From the Chern--Simons perspective, the resulting expansion is an exact,  finite \(N\) identity. What is special about the analytically continued theory is that the parameter \(q\)  is no longer constrained to the unit circle. One may therefore work in a chamber, for example \(|q|<1\), in which the relevant \(q\)-series are genuinely convergent rather than merely formal expansions. In this sense, the small \(q\) expansion is exact: it is obtained by analytically continuing the Chern--Simons answer away from the unitary locus and then expanding in a domain where the resulting series converges.   

\subsection[String dual of  Wilson loops]{\boldmath String dual of  Wilson loops}
We now turn to the dual string theory interpretation of Wilson loops \(W_R\).
The exact unnormalized expectation value of a Wilson loop supported on an
unknot and labeled by a representation \(R\) of \(SU(N)\) is
\begin{equation}
    \langle W_R\rangle
    =
    S^{(1,1)}_{R0}\, .
    \label{wilsonppp}
\end{equation}
Wilson lines admit  a complete string dual description~\cite{Ooguri:1999bv,Gomis:2006sb}. A Wilson line can be described either by a collection of branes on   $X$, a collection of antibranes on $X$, or by a pure closed string ``bubbling" Calabi--Yau geometry.\footnote{This is the Chern--Simons/topological string analogue of the
\(AdS_5\times S^5\) realization in terms of $D5$- branes  along $AdS_2\times S^4$ \cite{Yamaguchi:2006tq,Gomis:2006sb,Gomis:2006im}  
or $D3$- branes  along  $AdS_2\times S^2$
\cite{Drukker:2005kx,Gomis:2006sb,Gomis:2006im} 
  of  Wilson lines in
\({\cal N}=4\) super-Yang--Mills theory \cite{Witten:1998xy}.} The corresponding A-model partition functions can be computed exactly using the topological vertex  \cite{Aganagic:2003db},\footnote{Even in the regime where Calabi--Yau geometry or brane is not semiclassical.}
and exactly reproduce the small $q$ expansion of \eqref{wilsonppp}.
The exact correspondence for Wilson loops is summarized in   Table \ref{tab:summarySU(N)}.

\sbox{\MrowFigbox}{%
  \includegraphics[width=3.2cm]{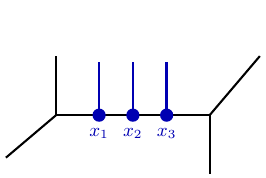}%
}

\sbox{\ProwFigbox}{%
  \includegraphics[width=3.2cm]{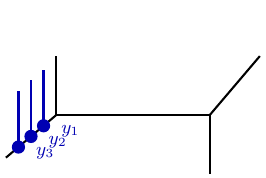}%
}

\sbox{\MrowTextbox}{%
  \begin{minipage}{6.4cm}\centering
  \vspace{0.2cm}
    \(M\) Branes on the inner edge\\
    \(x_i=g_s(R^T_i-i+M+\tfrac12)\)\\
    \(i=1,\ldots,M,\qquad \widehat t=t+Mg_s\)
      \vspace{0.2cm}
  \end{minipage}%
}

\sbox{\ProwTextbox}{%
  \begin{minipage}{6.4cm}\centering
    \vspace{0.2cm}
    \(P\)  Antibranes on outer non-compact edge\\
    \(y_i=g_s(R_i-i+P+\tfrac12)\)\\
    \(i=1,\ldots,P,\qquad \widehat t=t-Pg_s\)
      \vspace{0.2cm}
  \end{minipage}%
}

\setlength{\Mrowheight}{0pt}
\updrowheight{\Mrowheight}{\MrowFigbox}
\updrowheight{\Mrowheight}{\MrowTextbox}

\setlength{\Prowheight}{0pt}
\updrowheight{\Prowheight}{\ProwFigbox}
\updrowheight{\Prowheight}{\ProwTextbox}

\begin{table}[H]
  \centering
  \begin{tabular}{|>{\centering\arraybackslash}m{3.5cm}
                  | >{\centering\arraybackslash}m{7cm}
                  | >{\centering\arraybackslash}m{3.5cm}|}
                  \hline
    \(\langle W_R\rangle = S_{R0}\)
    & Branes A-model
    & Toric diagram\\
    \hline
    \hline
    Antisymmetric  \(\Lambda^n\)
    & \parbox{6.4cm}{\centering
      Brane on the inner compact edge\\
      \(x=(n+\tfrac12)g_s,\qquad \widehat t=t+g_s\)
    }
    & \includegraphics[width=3.2cm]{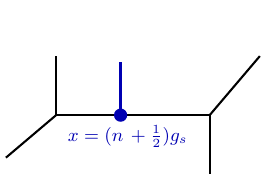}\\
    \hline
    Symmetric \(\mathrm{Sym}^l \)
    & \parbox{6.4cm}{\centering
       Antibrane on the outer non--compact edge\\
      \(y=(l+\tfrac12)g_s,\qquad \widehat t=t-g_s\)
        \vspace{0.2cm}
    }
    & \includegraphics[width=3.2cm]{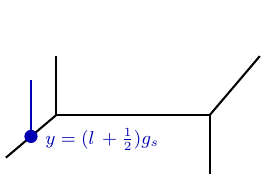}\\
    \hline \hspace{0.5cm}
    \multirow{2}{=}{%
      \begin{minipage}[c][0pt][c]{3.2cm}
        \centering
        \vspace{3cm}\hspace{1cm}
        \resizebox{\textwidth}{!}{%
          \begin{tikzpicture}[
            line width=1pt,
            transform shape,
            lbl/.style={inner sep=0.35pt, scale=1.7},
            clbl/.style={inner sep=0.2pt, scale=1.7}
          ]
            \def\rows{{6,5,3,3,2}}
            \foreach \i in {0,...,4} {
              \pgfmathsetmacro{\len}{\rows[\i]}
              \foreach \j in {1,...,\len} {
                \draw (\j-1,-\i) rectangle (\j,-\i-1);
              }
            }
            \node[lbl, right] at (6.35, -0.5) {$R_1$};
            \node[lbl, right] at (6.35, -1.5) {$R_2$};
            \node[lbl]           at (6.75, -2.5) {$\vdots$};
            \node[lbl, right] at (6.35, -4.5) {$R_P$};
            \node[clbl, below] at (0.5, -5.12) {$R_1^{T}$};
            \node[clbl, below] at (1.5, -5.12) {$R_2^{T}$};
            \node[clbl, below] at (2.5, -5.12) {$R_3^{T}$};
            \node[clbl, below] at (3.5, -5.12) {$R_4^{T}$};
            \node[clbl, below] at (4.5, -5.20) {$\cdots$};
            \node[clbl, below] at (5.5, -5.12) {$R_M^{T}$};
          \end{tikzpicture}%
        }%
      \end{minipage}%
    }
    &
    \Rowcell{\Mrowheight}{6.4cm}{\usebox{\MrowTextbox}} &
    \Rowcell{\Mrowheight}{3.2cm}{\usebox{\MrowFigbox}} \\
    \cline{2-3}
    &
    \Rowcell{\Prowheight}{6.4cm}{\usebox{\ProwTextbox}} &
    \Rowcell{\Prowheight}{3.2cm}{\usebox{\ProwFigbox}} \\
    \hline
  \end{tabular}
  \caption{Summary of the correspondence between \(SU(N)_k\) Chern--Simons Wilson line defects $W_R$
    and their A-model brane descriptions. A representation \(R\) of \(SU(N)\), specified
    by a Young tableau, admits two equivalent brane realizations in the resolved
    conifold: the brane description is determined by the column lengths
    \(R_i^T\), while the antibrane description is determined by the row lengths
    \(R_i\).
  }
  \label{tab:summarySU(N)}
\end{table}

\subsection[String duals of \texorpdfstring{$Spin/Sp$}{Spin/Sp} Chern--Simons]{\boldmath String duals to \texorpdfstring{$Spin/Sp$}{Spin/Sp} Chern--Simons}

As we shall see, the string dual description of monodromy defects in
\(SU(N)_k\) Chern--Simons theory is     formulated in terms of
unoriented topological string backgrounds. This should be distinguished from
the unoriented closed string duals of \(Spin(N)_k\) and \(Sp(N/2)_k\)
Chern--Simons theory on \(S^3\), proposed in~\cite{Sinha:2000ap}.  Remarkably, the two known orientifold projections of the resolved conifold ~\cite{Acharya:2002kv,Hori:2005bk,Krefl:2009md} find a dual gauge theory interpretation, one describes monodromy defects in $SU(N)_k$ Chern--Simons theory and the other is dual to \(Spin(N)_k\) and \(Sp(N/2)_k\)
Chern--Simons theory. 
In Appendix~\ref{App:su-so-sp-branes-geometry}, we review the duality
proposed in~\cite{Sinha:2000ap} and put forward brane and antibrane
descriptions of Wilson line defects in \(Spin(N)_k\) and \(Sp(N/2)_k\)
Chern--Simons theory. This includes, in particular, the dual description of
Wilson lines in spinorial representations of \(Spin(N)\).
 \smallskip

\section[\boldmath Monodromy defects    \texorpdfstring{$SU(N)_k$}{SU(N)k} Chern--Simons theory]{\boldmath Monodromy defects   \texorpdfstring{$SU(N)_k$}{SU(N)k} Chern--Simons theory}
\label{sec:four}
In Section~\ref{sec:dos} we have demonstrated  that $SU(N)_k$ Chern--Simons theory admits a collection of   monodromy defects $\mathsf{M}_a$, labeled by integrable representations of the ${\mathsf C}$-twisted affine algebra 
$\mathfrak{su}(N)^{(2)}_k$.

Recall from Section~\ref{mondefectrep} that the twisted affine algebra is completely 
determined by the outer automorphism group of $\mathfrak{g}$. The $\mathsf{C}$-even currents, 
taking values in $\mathfrak{g}_+$, are integrally moded, and the $\mathsf{C}$-odd currents, 
taking values in $\mathfrak{g}_-$, are half-integrally moded. The canonical 
$\mathfrak{su}(N)^{(2)}_k$ affine algebra is constructed from the following 
outer automorphisms   on the $\mathfrak{su}(N)$ Lie algebra generators:\footnote{The twisted affine algebra depends only on the outer automorphism class, not
on the choice of representative~\cite{kac_1990}.}
\begin{equation}
\begin{aligned}
    \mathfrak{su}(2N+1) &: \quad t \;\mapsto\; {-t^{\mathsf{T}}}
        && \Longrightarrow \quad \mathfrak{g}_+ = \mathfrak{so}(2N+1) \\[4pt]
    \mathfrak{su}(2N)   &: \quad t \;\mapsto\; {-\mathsf J\hspace{2pt}t^{\mathsf{T}} \mathsf J^{-1}}
        && \Longrightarrow \quad \mathfrak{g}_+ = \mathfrak{sp}(N) \,,
\end{aligned}
\label{outertwistedaa}
\end{equation}
where $t$ is a Hermitian matrix and $\mathsf J$ is the canonical antisymmetric matrix.  These moded currents generate a twisted affine Lie
algebra~\cite{kac_1990}. The associated generalized Cartan matrices are encoded by the
extended Dynkin diagrams \(A_{2N}^{(2)}\) and \(A_{2N-1}^{(2)}\) (see Table \ref{tab:twisted-affine}).

Monodromy defects in $SU(N)_k$ Chern--Simons theory, labeled by integrable representations of  $\mathfrak{su}(N)^{(2)}_k$,  are characterized by Dynkin labels $a_j$ of a representation $a$ of $\mathring{ \mathfrak{g}}=C_N$  obeying:
\begin{equation}
\begin{aligned}\label{integrablerep}
    \mathfrak{su}(2N+1)^{(2)}_k &: \quad 2a_1+2a_2+\cdots+2a_N \leq k, \\[4pt]
    \mathfrak{su}(2N)^{(2)}_k &: \quad a_1+2a_2+\cdots+2a_N \leq k .
\end{aligned}
\end{equation}
We collect the Dynkin labels of the representation into the vector
\(a=(a_1,a_2,\ldots,a_N)\).

As discussed in
Section~\ref{subsec:monodromy-defect-data},  correlators of monodromy defects $\mathsf M_a$
  are computed from the modular matrix
  \begin{equation}
      S^{(\mathsf C,1)}_{a\mu}\,,
  \end{equation}
where $a$ is an integrable representation of $\mathfrak{g}^{(2)}_k$ and $\mu$ is an integrable representation of the   
  orbit Lie algebra $\breve{\mathfrak{g}}_k$.
The orbit Lie algebra for  $\mathfrak{su}(2N+1)$ is $\breve{\mathfrak{g}}_k=\mathfrak{su}(2N+1)^{(2)}_k$ and for 
  $\mathfrak{su}(2N)$ is
$\breve{\mathfrak{g}}_k=\mathfrak{so}(2N+2)^{(2)}_k$. Their integrable weights are   characterized by Dynkin labels $\mu_j$ of a representation $\mu$ of $\breve{\mathfrak g}$, the horizontal subagebra of $\breve{\mathfrak{g}}_k$,  obeying:
\begin{equation}
\begin{aligned}
    \mathfrak{su}(2N+1)^{(2)}_k &: \quad 2\mu_1+2\mu_2+\cdots+2\mu_N \leq k, \\[4pt]
    \mathfrak{so}(2N+2)^{(2)}_k &: \quad 2\mu_1+2\mu_2+\cdots+2\mu_{N-1}+\mu_N \leq k .
\end{aligned}
\end{equation}
  Table \ref{tab:twisted-affine} summarizes the relevant twisted affine algebras \(\mathfrak{g}^{(2)}_k\), the corresponding orbit Lie   algebras \(\breve{\mathfrak g}_k\), together with their comarks and marks, and  their horizontal subalgebras. 
\begin{table}[t]
\centering
\resizebox{\textwidth}{!}{
\renewcommand{\arraystretch}{1.6}
\begin{tabular}{|c||c|c||c|c|}
\hline
\dynname{\mathfrak g} & \dynname{\mathfrak{g}^{(2)}_k} & \dynname{\mathring{ \mathfrak{g}}}
 & \dynname{\breve{\mathfrak g}_k } & \dynname{\breve{\mathfrak g}}\\[3pt]
\hline\hline
\dynname{A_{2N}} & \dynname{A_{2N}^{(2)}} & \dynname{C_N} & \dynname{A_{2N}^{(2)}} & \dynname{C_N}\\
\diagAtwoN & \diagAtwoNaff & \diagCN & \diagAtwoNaff & \diagCN\\[7pt]
\hline
\dynname{A_{2N-1}} & \dynname{A_{2N-1}^{(2)}} & \dynname{C_N} & \dynname{D_{N+1}^{(2)}} & \dynname{B_N}\\
\diagAtwoNm & \diagAtwoNmaff
& \diagCN & \diagDaff & \diagBN\\[7pt]
\hline
\end{tabular}}
\caption{Affine Lie algebra data. The numbers above are the comark $a_i^\vee$, the ones below are the marks $a_i$. We have highlighted in {\color{blue}blue}  the affine node. }
\label{tab:twisted-affine}
\end{table}
The $S$-matrix \(S_{a\mu}^{(\mathsf C,1)}\) in $SU(2N+1)_k$ Chern--Simons theory is given by~\cite{kac_1990}: 
\beq
S_{a\mu}^{(\mathsf C,1)} = \frac{i^{N^2}}{(k + 2N + 1)^{N/2}}\sum_{w\in {\cal W}(C_{N})} \epsilon(w) \exp{\left( -{\frac{2\pi i} {k+2N+1}} \langle w(\rho_C+a),  \rho_C+\mu\rangle\right)}
\eeq
where  
\begin{itemize}
\item $a$ is a highest weight of $C_{N}$  
\item $\mu$ is a highest weight of $C_{N}$

\end{itemize}

When \(\mu=0\),  the formula reduces to the \(C_N\) Weyl denominator formula, 
 that is,  a product over the positive roots of  $C_N$:
\beq\label{SmatSU2n+1}
S^{(\mathsf{C},1)}_{a0}
=
(k+2N+1)^{-N/2}\,
i^{N^2}\,
q^{-\langle\rho_C,\rho_C+a\rangle}
\prod_{\alpha\in\Delta_+(C_N)}
\left(1-q^{\langle\alpha,\rho_C+a\rangle}\right), 
\eeq
where 
\begin{equation}
q=\exp\left(\frac{2\pi i}{k+2N+1}\right)\,.
\end{equation}
The $S$-matrix \(S_{a\mu}^{(\mathsf C,1)}\) in $SU(2N)_k$ Chern--Simons theory is given  by~\cite{kac_1990}:\footnote{Note that this is not a symmetric matrix.} 
\beq
\begin{aligned}
S_{a\mu}^{(\mathsf C,1)}
 = \frac{1}{\sqrt{2}} \frac{i^{N^2}}{(k+2N)^{N/2}}\,
\,&\sum_{w\in {\cal W}(C_{N})} \epsilon(w) \exp{\left( -{\frac{2\pi i}{ k+2N}} \langle w(\rho_C+a),  \tau(\rho_B+\mu)\rangle\right)}
\end{aligned}
\eeq
where
\begin{itemize}
\item $a$ is a highest weight of $C_{N}$
\item $\mu$ is a highest weight of $B_{N}$ 
\end{itemize}

Specializing to \(\mu=0\), and using the fact that the Weyl groups of
\(B_N\) and \(C_N\) are   isomorphic, \(\mathcal W(C_N)\simeq
\mathcal W(B_N)\), we can rewrite the result in the form of the \(B_N\) Weyl
denominator, namely as a product over the positive roots of \(B_N\):
\beq
S_{a0}^{(\mathsf C,1)}
=
2^{-1/2}(k+2N)^{-N/2}\,
i^{N^2}\,
q^{-\langle\rho_B,\tau^{-1}(\rho_C+a)\rangle}
\prod_{\alpha\in\Delta_+(B_N)}
\left(1-q^{\langle\alpha,\tau^{-1}(\rho_C+a)\rangle}\right)\,,
\label{SmatSU2n}
\eeq
where
\begin{equation}
  q=\exp\left(\frac{2\pi i}{k+2N}\right)\,.
\end{equation}
$\tau^{-1}$ is the   embedding   
\begin{equation}
\tau^{-1}:\, P(C_N)\hookrightarrow P(B_N). 
\end{equation}
It acts on the orthogonal basis as
\begin{equation}
    \tau^{-1}(e_i)=e_i  \qquad i=1,\ldots,N\,.
\end{equation}
We can now proceed to finding the dual string interpretation of monodromy defects.\footnote{A  representation of these $S$-matrices as a determinant has appeared in \cite{Gaberdiel_2002}.} 

\subsection{The vacuum monodromy defect}
\label{vacuum}

For an untwisted affine Lie algebra \(\mathfrak g_k\),
\(S^{(1,1)}_{R0}\) is minimized, among all the integrable representations, by
the trivial representation \(R=0\).\footnote{If the affine Lie algebra has simple currents, then the simple current orbit of
the vacuum representation \(R=0\), equivalently its spectral-flow images, gives
the same value of \(S^{(1,1)}_{R0}\). These simple currents are associated with
the subgroup of automorphisms of the extended affine Dynkin diagram generated by
the special automorphisms that send the affine node \(0\) to another special
node.} This reflects the fact that inserting the
Wilson line \(W_R\) with \(R=0\) leaves the Chern--Simons partition function
unchanged.

The situation is different for monodromy defects. Inserting a defect
\(\mathsf M_a\) produces a genuinely new Chern--Simons amplitude. This remains
true   for any  object in \(\mathcal C_{\mathsf C}\), including the one labeled by the trivial
representation \(a=0\). For twisted affine Lie algebras \(\mathfrak g^{(2)}_k\), the
trivial representation \(a=0\) need not be the minimizer of
\(S^{(\mathsf C,1)}_{a0}\), a point we discuss now.

Motivated by the search for a dual string theory description of monodromy
defects, we record in Table~\ref{tab:minimizers} the integrable
representations \(a_*\) that minimize \(S^{(\mathsf C,1)}_{a0}\).\footnote{The
derivation is presented in Appendix~\ref{app:monodef}.} These representations
describe the lightest monodromy defect insertions, in the sense that they
produce the smallest modification of the Chern--Simons partition function. We
refer to the corresponding defects as vacuum monodromy defects. As shown in
Table~\ref{tab:minimizers}, their labels depend on the parities of \(k\) and
\(N\).

By contrast, monodromy defects labeled by other integrable representations may
be viewed as excited monodromy defects. In the dual string description, their
insertion is expected to correspond to a nontrivial modification of the closed
string background associated with the vacuum monodromy defects. As we shall
show, this expectation is borne out by the analysis below. 
\begin{table}[H]
\centering
{\renewcommand{\arraystretch}{1.3}
\begin{tabular}{|c|c|c|}
\hline
{Theory} & {Parity of $ k$} & {Vacuum monodromy defect} \\
\hline
\hline
$SU(2N)_k$ & $k$ odd
&
$(0,\ldots,0)$, $(k,0,\ldots,0)$
\\
\hline
$SU(2N)_k$ & $k$ even, $2N\leq k$
&
$(0,\ldots,0)$, $(k,0,\ldots,0)$
\\[1mm]
$SU(2N)_k$ & $2N=k+2$
&
$(0,\ldots,0)$, $(k,0,\ldots,0)$, $(0,\ldots,0,\tfrac k2)$
\\[1mm]
$SU(2N)_k$ & $k$ even, $2N>k+2$
&
$(0,\ldots,0,\tfrac k2)$
\\[.5mm]
\hline
\hline
$SU(2N+1)_k$ & $k$ even
&
$(0,\ldots,0,\tfrac k2)$
\\
$SU(2N+1)_k$ & $k$ odd
&
$(0,\ldots,0)$, $(0,\ldots,0,\tfrac{k-1}{2})$
\\[.5mm]
\hline
\end{tabular}}
\caption{Representations $a_*$ for vacuum monodromy defects in the $SU(N)_k$ Chern--Simons.}
\label{tab:minimizers}
\end{table}

 The degeneracy between the weights \((0,\ldots,0)\) and \((k,0,\ldots,0)\) in $SU(2N)_k$ Chern--Simons theory, valid for all levels \(k\), is a consequence of the simple current associated with the order two automorphism of the affine Dynkin diagram of \(A_{2N-1}^{(2)}\)
that exchanges the affine node \(0\) with node \(1\)  (see Table \ref{tab:twisted-affine}). The additional degeneracy with the $(0,\ldots, k/2)$ defect  for $k=2N-2$ is accidental, and not a consequence of a symmetry of $A_{2N-1}^{(2)}$. As $N$ is further increased, the monodromy 
 defect labeled by $(0,\ldots, k/2)$ becomes the vacuum monodromy defect in $A_{2N-1}^{(2)}$.

The case of \(A_{2N}^{(2)}\) is slightly subtle.  Although its Dynkin diagram
has no nontrivial symmetry, a twofold  degeneracy appears for odd \(k\).  The
origin of this degeneracy is not a diagram automorphism, but rather a
half-period reflection that becomes an allowed lattice symmetry precisely at
odd level.  In this sense the odd-level theory admits a simple current absent
at even level, leading to a nontrivial dependence of the spectrum on the
parity of \(k\).
Having identified the vacuum monodromy defects, we expand $S^{(\mathsf{C},1)}_{a_*0}$ in powers of small $q = e^{2\pi i/(k+N)}$   in analytically continued $SU(N)_k$ Chern--Simons theory, with a view toward finding their dual string theoretic description in Section~\ref{sec:4.2}. 
\subsubsection{\boldmath $ {SU(2N)}_k$,  odd level}

The vacuum monodromy defect is labeled by $a_*=0$.\footnote{Which is degenerate with the spectrally flowed representation $(k,0,\ldots,0)$.} Evaluating the associated $S$-matrix \eqref{SmatSU2n} we find:\footnote{As before omitting a normalization factor and overall power of $q$. See Appendix \ref{app:constant-maps} \eqref{eq:su2nodd-pref} for the omitted normalizations.} 
\begin{equation}
 S^{(\mathsf C,1)}_{00}\doteq
 \prod_{\alpha\in\Delta_+(B_N)}\!\big(1-q^{\langle \alpha,\tau^{-1}(\rho_C)\rangle}\big)
 =\prod_{1\le i<j\le N}\!\big(1-q^{z_i-z_j}\big)\big(1-q^{z_i+z_j}\big)
  \prod_{i=1}^{N}\big(1-q^{z_i}\big),
\label{eq:su2nodd-weyl}
\end{equation}
with \footnote{We use the normalization of the bilinear form $\langle e_i, e_j \rangle  = \delta_{ij}$}
\begin{equation}
    z_i=N+1-i=\langle e_i,\tau^{-1}(\rho_C)\rangle\,, 
\end{equation}
where $e_i$ are the orthogonal basis, in terms of which the positive roots are
\begin{equation}
\Delta_+(B_N)
=
\{\,e_i-e_j,\ e_i+e_j \mid 1\leq i<j\leq N\,\}
\cup
\{\,e_i \mid 1\leq i\leq N\,\},
\end{equation}
and the Weyl vector is
\begin{equation}
    \rho_C=
\sum_{i=1}^N (N-i+1)e_i\, .
\label{WeylCN}
\end{equation}

Taking the logarithm of $ S^{(\mathsf C,1)}_{00}$, expanding at small $q$ using $\log(1-q^{r})=-\sum_{n\ge1}q^{nr}/n$, and summing  over $i,j$, we arrive at
\begin{equation}
\log S^{(\mathsf C,1)}_{00}
\doteq
-\sum_{n=1}^{\infty}\frac1n\,
\frac{Nq^{n}-q^{2n}-Nq^{3n}+q^{(2N+2)n}}{(1-q^{n})^{2}\,(1+q^{n})}\,.
\label{eq:su2nodd-rational}
\end{equation}
Performing the sum over terms that have no $N$ in a $q$-exponent we find 
\begin{equation}
-\sum_{n=1}^{\infty}\frac1n\,
\frac{Nq^{n}-q^{2n}-Nq^{3n}}{(1-q^{n})^{2}(1+q^{n})}
=\log\frac{M(q)^{1/2}\,\zeta(q^{2})^{1/2}}{\zeta(q)^{\,N+1/2}}\,.
\label{eq:su2nodd-contact}
\end{equation}
Using  the   partial fraction
\[
\frac{1}{(1-q^n)^2(1+q^n)}
=
\frac12\frac{1}{(1-q^n)^2}
+
\frac12\frac{1}{1-q^{2n}}\,, 
\]
the remaining term in \eqref{eq:su2nodd-rational} can be written as
\begin{equation}
\begin{aligned}
-\sum_{n=1}^{\infty}
\frac{1}{n}
\frac{q^{(2N+2)n}}{(1-q^n)^2(1+q^n)}
&=
-\frac12\sum_{n=1}^{\infty}
\frac{1}{n}
\frac{q^{(2N+2)n}}{(1-q^n)^2}
-\frac12\sum_{n=1}^{\infty}
\frac{1}{n}
\frac{q^{(2N+2)n}}{1-q^{2n}} .
\end{aligned}
\label{intfoo}
\end{equation} 
Expressing
\beq
\frac{q^n}{(1-q^n)^2}=\frac{1}{[n]^2},
\qquad
\frac{q^{n/2}}{1-q^n}=-\frac{1}{[n]},
\eeq
where
\begin{equation}
   [n]=q^{n/2}-q^{-n/2}, 
\end{equation}
the first term in \eqref{intfoo} becomes
\beq
-\frac12\sum_{n=1}^{\infty}
\frac{q^{(2N+1)n}}{n[n]^2}.
\eeq
For the second term in \eqref{intfoo}, setting \(m=2n\), we get
\beq
-\frac12\sum_{n=1}^{\infty}
\frac{1}{n}
\frac{q^{(2N+2)n}}{1-q^{2n}}
=
-\sum_{\substack{m\geq 1\\ m\ \mathrm{even}}}
\frac{q^{(N+1)m}}{m(1-q^m)}
=
\sum_{\substack{m\geq 1\\ m\ \mathrm{even}}}
\frac{q^{(N+1/2)m}}{m[m]} .
\eeq
Therefore, we have found that
\beq
-\sum_{n=1}^{\infty}
\frac{1}{n}
\frac{q^{(2N+2)n}}{(1-q^n)^2(1+q^n)}
=
-\frac12\sum_{m=1}^{\infty}
\frac{q^{(2N+1)m}}{m[m]^2}
+
\sum_{\substack{m\geq 1\\ m\ \mathrm{even}}}
\frac{q^{(N+1/2)m}}{m[m]} .
\eeq
Finally, using \(e^{-t}=q^{2N+1}\), this becomes
\beq
-\sum_{n=1}^{\infty}
\frac{1}{n}
\frac{q^{(2N+2)n}}{(1-q^n)^2(1+q^n)}
=
-\frac12\sum_{m=1}^{\infty}
\frac{e^{-mt}}{m[m]^2}
+
\sum_{\substack{m\geq 1\\ m\ \mathrm{even}}}
\frac{e^{-mt/2}}{m[m]} .
\eeq
 Putting everything together, and anticipating the dual closed string interpretation to be discussed later, we write the expectation value of the vacuum monodromy defect as,
 \begin{subequations}
 \label{SU(2N)kodd}
\begin{equation}
\begin{aligned}
 \langle \mathsf M_0\rangle=\log S^{(\mathsf C,1)}_{00} &\doteq
  \mathcal F^0(N)+\mathcal F^{\rm inst}_{\Omega\sigma^{Sp}_+}(t)\,,
\end{aligned}
\end{equation}
where
\begin{equation}
\mathcal{F}^{\rm inst}_{\Omega\sigma^{Sp}_+}(t) =
 -\frac{1}{2}\sumall {\frac{e^{-mt}}{m[m]^2}}
 +\sumeven{\frac{e^{-mt/2}}{ m[m]}}\,,   
\end{equation}
and
\begin{equation}
  \mathcal{F}^0(N)=\frac{1}{2}\log\frac{M(q)\,\zeta(q^{2})}{\zeta(q)^{2N+1}}\,, 
\end{equation}
with
\begin{equation}
       e^{-t} =q^{2N+1}\,,\qquad q=e^{\frac{2\pi i}{k+2N}}=e^{-g_s}\,.
\end{equation}
\end{subequations}
\subsubsection{\boldmath $SU(2N)_k$,  even level}
For even $k$ and $k\geq 2N$ the vacuum monodromy defect is also labeled by $a_*=0$.\footnote{Which is degenerate with the spectrally flowed representation $(k,0,\ldots,0)$).} The small $q$ expansion  $ \langle \mathsf M_0\rangle$ is therefore identical to 
\eqref{SU(2N)kodd}.

For $2N>k+2$, the monodromy defect labeled by the $C_N$ representation $a_*=    (0,\ldots,k/2)$ has the smallest $S^{(\mathsf C,1)}_{a0}$, and is the lightest monodromy defect. For even \(k\), this representation lies on the boundary of the alcove of integrable highest weights of \(A_{2N-1}^{(2)}\).   A similar computation yields
\begin{equation}
 S^{(\mathsf C,1)}_{a_*0}\doteq
 \prod_{1\le i<j\le N}\!\big(1-q^{\widetilde z_i-\widetilde z_j}\big)
   \big(1-q^{\widetilde z_i+\widetilde z_j}\big)
 \prod_{i=1}^{N}\big(1+q^{\widetilde z_i}\big)\,,
\label{eq:su2neven-weyl}
\end{equation}
where now\footnote{{Since  $a_*=\tfrac{k}{2}\omega_N$, 
$z_i=\langle e_i,\tau^{-1}(\rho_C+a_*)\rangle=(N-i+1)+\tfrac{k}{2}$. Since
$q=e^{2\pi i/(k+2N)}$,  one has $q^{k/2}=-q^{-N}$, and  hence
\begin{equation}
  q^{z_i}\;=\;q^{\,N-i+1}\,q^{k/2}\;=\;-\,q^{\,1-i}\,.
\end{equation}
The sign propagates into each coroot factor with multiplicity:
\begin{align}
  e_i-e_j:&\quad (\omega_N,(e_i-e_j)^\vee)=0
    &&\Rightarrow\ q^{\,z_i-z_j}=q^{\,j-i}, &&(1-q^{\,j-i});\\
  e_i+e_j:&\quad (\omega_N,(e_i+e_j)^\vee)=2
    &&\Rightarrow\ q^{\,z_i+z_j}=q^{\,2-i-j}, &&(1-q^{\,2-i-j});\\
  2e_i:&\quad (\omega_N,(2e_i)^\vee)=(\omega_N,e_i)=1
    &&\Rightarrow\ q^{\,z_i}=-\,q^{\,1-i}, &&(1+q^{\,1-i}).
\end{align}and relabelling $i\mapsto N+1-i$,  we get  to the formula above with $\tilde z_i=N-i$.} }
\begin{equation}
  \widetilde z_i=N-i\,. 
\end{equation} 
Taking the logarithm, expanding at small $q$, and summing over the three families of positive roots  yields 
\begin{equation}
\log S^{(\mathsf C,1)}_{a_*0}
\doteq
-\sum_{n=1}^{\infty}\frac1n\,
\frac{(N-1)q^{n}-Nq^{3n}+q^{(2N+1)n}}{(1-q^{n})^{2}\,(1+q^{n})}\,.
\label{eq:su2neven-rationala}
\end{equation}
 Performing the sum over terms that have no $N$ in a $q$-exponent    we find 
\begin{equation}
-\sum_{n=1}^{\infty}\frac1n\,
\frac{(N-1)q^{n}-Nq^{3n}}{(1-q^{n})^{2}(1+q^{n})}
=\log\frac{M(q)^{1/2}}{\zeta(q)^{\,N-1/2}\,\zeta(q^{2})^{1/2}}\,.
\label{eq:su2neven-contact}
\end{equation}
The remaining term in \eqref{eq:su2neven-rationala}  carries an extra power of $q^{n}$ relative to previous  odd-level case. Using  the   partial fraction 
\begin{equation}
 \frac{q^n}{(1-q^n)^{2}(1+q^n)}=\frac12\frac{1}{(1-q^n)^{2}}-\frac12\frac{1}{1-q^{2n}}\,,  
 \label{minuspartialf}
\end{equation}
we find 
\beq
-\sum_{n=1}^{\infty}
\frac{1}{n}
\frac{q^{(2N+1)n}}{(1-q^n)^2(1+q^n)}
=
-\frac12\sum_{m=1}^{\infty}
\frac{q^{(2N-1)m}}{m[m]^2}
-
\sum_{\substack{m\geq 1\\ m\ \mathrm{even}}}
\frac{q^{(N-1/2)m}}{m[m]} .
\eeq
Finally, using \(e^{-t}=q^{2N-1}\), this becomes
\begin{equation}
  -\sum_{n=1}^{\infty}\frac1n\,\frac{q^{(2N+1)n}}{(1-q^{n})^{2}(1+q^{n})}
=-\frac12\sumall{ \frac{e^{-mt}}  {m[m]^{2}}}
-\sumeven{\frac{e^{-mt/2}} {m[m]}}\,. 
\end{equation}
Putting everything together,   the vacuum expectation value of the vacuum monodromy defect for $k$ even and  $k>2N-2$    is\footnote{We keep the same symbol  $\mathcal{F}^0$ as before as we believe no confusion can be made compared to the previous case, since we never make a discussion with distinct background in the same amplitude}
\begin{subequations}
\label{SU2N-keven-endpoint}
\begin{equation}
\begin{aligned}
 \langle \mathsf M_{a_*}\rangle=\log S^{(\mathsf C,1)}_{a_*0} &\doteq
  \mathcal F^0(N)+\mathcal F^{\rm inst}_{\Omega\sigma^{SO}_+}(t)\,,
\end{aligned}
\end{equation}
where
\begin{equation}
\mathcal{F}^{\rm inst}_{\Omega\sigma^{SO}_+}(t) =
 -\frac{1}{2}\sumall{\frac{e^{-mt}} {m[m]^2}}
 -\sumeven{\frac{e^{-mt/2}} {m[m]}}\,,   
\end{equation}
and
\begin{equation}
\mathcal{F}^0(N)=\frac{1}{2}\log\frac{M(q)}{\zeta(q)^{2N-1}\,\zeta(q^{2})}\,,  
\end{equation}
with
\begin{equation}
       e^{-t} =q^{2N-1}\,,\qquad q=e^{\frac{2\pi i}{k+2N}}=e^{-g_s}\,.
\end{equation}
\end{subequations}
\subsubsection{\boldmath $SU(2N+1) _k$,  even level}

The vacuum monodromy defect is always labeled by $a_*=(0,\cdots,0,\tfrac k2)$, a $C_N$ representation that lies on the boundary of the alcove of integrable highest weights of \(A_{2N}^{(2)}\). Evaluating the corresponding $S$-matrix    \eqref{SmatSU2n+1} we find:
\begin{equation} 
 S^{(\mathsf C,1)}_{a_*0}\doteq \prod_{\alpha\in\Delta_+(C_N)}
\left(1-q^{(\alpha,\rho_C+a)}\right)=
 \prod_{1\le i<j\le N}\!\big(1-q^{\widetilde z_i-\widetilde z_j}\big)\big(1-q^{\widetilde z_i+\widetilde z_j}\big)
  \prod_{i=1}^{N}\big(1-q^{2\widetilde z_i}\big)\,,
\end{equation}
where, now 
\begin{equation}
  \widetilde z_i=N+\tfrac12-i\,,
\end{equation}
and we have used that the positive roots in the orthogonal basis are 
\begin{equation}
\Delta_+(C_N)
=
\{\,e_i-e_j,\ e_i+e_j \mid 1\leq i<j\leq N\,\}
\cup
\{\,2e_i \mid 1\leq i\leq N\,\}\,,
\label{posrootsCN}
\end{equation}
and that   $q^{k+2N+1}=1$.

Taking the logarithm, expanding at small $q$, and summing  over the positive roots  yields
\begin{equation}
\log S^{(\mathsf C,1)}_{a_*0}
\doteq
-\sum_{n=1}^{\infty}\frac1n\,
\frac{Nq^{n}-q^{2n}-Nq^{3n}+q^{(2N+2)n}}{(1-q^{n})^{2}\,(1+q^{n})}\,.
\label{su2n+1even}
\end{equation}
This is \emph{identical} to the $SU(2N)$ odd-level rational function \eqref{eq:su2nodd-rational}. As we shall see, however, the same rational function can be reorganized in
 several ways. These different representations lead to different closed string interpretations, and hence to distinct dual closed string
backgrounds.

  The terms with no $N$ in a $q$-exponent in \eqref{su2n+1even} are the same as \eqref{eq:su2nodd-contact}. The remaining term in \eqref{su2n+1even} can be represented as a closed string background by writing $q^{(2N+2)n}=q^{(2N+1)n}q^{n}$ and using the minus partial fraction in \eqref{minuspartialf}. This yields  
\begin{equation}
-\sum_{n=1}^{\infty}
\frac{1}{n}
\frac{q^{(2N+2)n}}{(1-q^n)^2(1+q^n)}
=
-\frac12\sum_{m=1}^{\infty}
\frac{q^{2Nm}}{m[m]^2}
-
\sum_{\substack{m\geq 1\\ m\ \mathrm{even}}}
\frac{q^{Nm}}{m[m]} .
\end{equation}

Putting everything together,   the   expectation value of the vacuum monodromy defect  is
\begin{subequations}
\label{SU2Np1-keven-endpoint}

\begin{equation}
 \langle \mathsf M_{a_\ast}\rangle=\log S^{(\mathsf C,1)}_{a_\ast0} \doteq
  \mathcal F^0(N)+\mathcal F^{\rm inst}_{\Omega\sigma^{SO}_+}(t)\,,
\end{equation}
where
\begin{equation}
\mathcal{F}^{\rm inst}_{\Omega\sigma^{SO}_+}(t) =
 -\frac{1}{2}\sumall{\frac{e^{-mt}} {m[m]^2}}
 -\sumeven{\frac{e^{-mt/2}} {m[m]}}\,,   
\end{equation}
and
\begin{equation}  \mathcal{F}^0(N)=\log\frac{M(q)^{1/2}\zeta(q^{2})^{1/2}}{\zeta(q)^{\,N+1/2}}\,,  
\end{equation}
with
\begin{equation}
       e^{-t} =q^{2N}\,,\qquad q=e^{\frac{2\pi i}{k+2N+1}}=e^{-g_s}\,.
\end{equation}
\end{subequations}
We notice that the vacuum monodromy defect corresponding to the case of $SU(2N+1)_{k \; \rm even}$ has the same instanton partition function as the case of $SU(2N)_{k \; \rm even}$, but differs in the prefactor $\mathcal{F}^0$. In Appendix \ref{app:constant-maps}, we perform a small $g_s$ expansion and show that these two backgrounds have exactly the same perturbative expansion with the appropriate definition of $t$ and $g_s$ and only differ in terms non-perturbative in $g_s$. \footnote{This behavior is analogous to the difference we find for parition function of $Spin(2N)_k$ and $Spin(2N+1)_k$ Chern--Simons theory, see appendix \ref{App:su-so-sp-branes-geometry}} 
\subsubsection{\boldmath $SU(2N+1)_k$,  odd level}
The  trivial  representation  of $C_N$ (with $a=0$) describes the vacuum monodromy defect for odd level and rank. Evaluating the  $S$-matrix   \eqref{SmatSU2n+1} we find
\begin{equation}
 S^{(\mathsf C,1)}_{00}\doteq
 \prod_{1\le i<j\le N}\!\big(1-q^{z_i-z_j}\big)\big(1-q^{z_i+z_j}\big)
  \prod_{i=1}^{N}\big(1-q^{2z_i}\big)\,,
\end{equation}
with
\beq
z_i=N+1-i=\langle e_i, \rho_C\rangle\,,
\eeq
where we have used the positive roots \eqref{posrootsCN} and Weyl vector   \eqref{WeylCN} of $C_N$.

  Taking the logarithm, performing $q$-series expansion  and summing over $i,j$ we have
\begin{equation}
\log S^{(\mathsf C,1)}_{00}
\doteq
-\sum_{n=1}^{\infty}\frac1n\,
\frac{(N-1)q^{n}-Nq^{3n}+q^{(N+1)n}-q^{(N+3)n}+q^{(2N+3)n}}{(1-q^{n})^{2}\,(1+q^{n})}\,.
\label{su2n+1odd}
\end{equation}
The terms with no $N$ in a $q$-exponent in \eqref{su2n+1odd} sum up to  \eqref{eq:su2neven-contact}. 

Using that 
$q^{(N+1)n}-q^{(N+3)n}=q^{(N+1)n}(1-q^{2n})$, the first two terms with $N$ in a $q$-exponent in \eqref{su2n+1odd} can be combined into
\begin{equation}
-\sum_{n=1}^{\infty}\frac1n\,\frac{q^{(N+1)n}\big(1-q^{2n}\big)}{(1-q^{n})^{2}(1+q^{n})}
=-\sum_{n=1}^{\infty}\frac1n\,\frac{q^{(N+1)n}}{1-q^{n}}
=\sumall{\frac{q^{(N+1/2)m}}{m[m]}}\,.
\end{equation}
Writing $q^{(2N+3)n}=q^{(2N+2)n}q^{n}$ and using the minus partial fraction in \eqref{minuspartialf}, the remaining term in \eqref{su2n+1odd} reads
\begin{equation}
-\sum_{n=1}^{\infty}\frac1n\,\frac{q^{(2N+3)n}}{(1-q^{n})^{2}(1+q^{n})}
=-\frac12\sumall{\frac{q^{(2N+1)m}}{ m[m]^{2}}}
-\sumeven{\frac{q^{(N+1/2)m}} {m[m]}}\,.
\end{equation}

The  expectation value of the vacuum monodromy defect       is 
\begin{subequations}
\label{SU2Np1-kodd-endpoint}

\begin{equation}
 \langle \mathsf M_{0}\rangle=\log S^{(\mathsf C,1)}_{00} \doteq
  \mathcal F^0(N)+\mathcal F^{\rm inst}_{\Omega\sigma^{SO}_+ \oplus \mathcal B}(t)\,,
\end{equation}
where
\begin{equation}
\mathcal{F}^{\rm inst}_{\Omega\sigma^{SO}_+ \oplus \mathcal B}(t) =
-\frac{1}{2}\sumall{\frac{e^{-mt}} {m[m]^2}}
 -\sumeven{\frac{e^{-mt/2}} {m[m]}}
 +\sumall{\frac{e^{-mt/2}} {m[m]}}\,,   
\end{equation}
and
\begin{equation}  \mathcal{F}^0(N)=\log\frac{M(q)^{1/2}}{\zeta(q)^{\,N-1/2}\,\zeta(q^{2})^{1/2}}\,,  
\end{equation}
with
\begin{equation}
       e^{-t} =q^{2N+1}\,,\qquad q=e^{\frac{2\pi i}{k+2N+1}}=e^{-g_s}\,.
\end{equation}
\end{subequations}
\subsection[String theory dual of   vacuum monodromy defects]{\boldmath String theory dual of $SU(N)$ vacuum monodromy defects}
\label{sec:4.2}

In Section~\ref{vacuum} we have identified the vacuum monodromy defects in
\(SU(N)_k\) Chern--Simons theory and shown that their expectation values
admit a natural genus expansion in terms of the string coupling \(g_s\).  On the other hand, in Section~\ref{sec:suNdual}, we recalled the duality between
\(SU(N)_k\) Chern--Simons theory and the \(A\)-model topological string on the
resolved conifold geometry for the $S^3$ partition function and the Wilson line operators. 

We are thus led to ask the natural question how are monodromy defects in \(SU(N)_k\)
Chern--Simons theory represented in the dual string theory. More
concretely, what is the dual topological string description of these defects?
The very existence of such nontrivial observables provides a sharp stress-test
of the holographic duality: the duality should not only reproduce the
  string expansion of ordinary Chern--Simons observables, but also account
for the monodromy defects constructed in this paper.

Besides the explicit computations in 
 Section~\ref{vacuum}, where we have obtained, from the   modular $S$-matrix of the twisted affine Lie algebra,  the exact expectation value for the unknot vacuum monodromy defect  expanded as closed  a topological string  partition function, a simple argument motivates that the insertion of even the ``lightest" vacuum monodromy defect in Chern--Simons theory modifies  the {\it closed} string background away from the A-model on the resolved conifold.

In the presence of a monodromy defect, the components of the
Chern--Simons gauge field valued in the invariant subalgebra
\(\mathfrak g_+\subset \mathfrak{su}(N)\) have the ordinary propagator,
whereas the components valued in the \(\mathsf C\)-odd complement
\(\mathfrak g_-\) obey twisted boundary conditions. Thus the effect of the
monodromy defect is diagnosed by the number of twisted gauge field components,
namely by the dimension of \(\mathfrak g_-\simeq \mathfrak{su}(N)/\mathfrak g_+\).
Using \eqref{outertwistedaa}, this gives
\begin{equation}
     \dim \mathfrak g_-
     =
     \dim\!\left( \mathfrak{su}(N)/\mathfrak g_+ \right)
     =
     \begin{cases}
     \dim\!\left( \mathfrak{su}(N)/\mathfrak{sp}(N/2) \right)
     =
     \dfrac{N^2}{2}-\dfrac{N}{2}-1,
     & N~\text{even},\\[6pt]
     \dim\!\left( \mathfrak{su}(N)/\mathfrak{so}(N) \right)
     =
     \dfrac{N^2}{2}+\dfrac{N}{2}-1,
     & N~\text{odd}.
     \end{cases}
     \label{scaleNN}
\end{equation}

This elementary computation provides some important clues. First, the leading effect of
the monodromy defect is of order \(N^2/2\). Hence the defect changes the
genus zero, or sphere, contribution of the closed string free energy. The dual
closed string background therefore cannot be the same as the one associated
with Chern--Simons theory in the absence of the monodromy defect, namely the
\(A\)-model on the resolved conifold. Moreover, the fact that the leading
sphere contribution is reduced by a factor of two is naturally suggestive of an
orientifold background (cf. \cite{Sinha:2000ap}). 

Equation~\eqref{scaleNN} provides another indication that the vacuum monodromy
defects are dual to orientifold backgrounds. While the leading term scales as
\(N^2/2\), the subleading order \(N\) term changes sign between even and odd
\(N\). In the closed string expansion, an order \(N\) contribution is naturally
identified with the leading unoriented worldsheet, \(\mathbb{RP}^2\). Thus the even- and odd \(N\) cases are expected to be described by orientifold
backgrounds whose crosscap contributions differ by an overall sign. As we
review below, this sign is a discrete datum in the specification of the
orientifold background.

This discussion naturally leads us to consider orientifolds of the A-model on
the resolved conifold. An A-model  orientifold background is defined by combining
worldsheet parity, denoted by \(\Omega\), with an antiholomorphic involution
\(\sigma\) of the target Calabi--Yau geometry. The resulting A-model
orientifold action is generated by $\Omega \sigma$. For a fixed orientifold action \(\Omega\sigma\), there remains an additional
twofold choice of orientifold background. This discrete choice, described below, controls the sign of the corresponding
crosscap contribution. In the open string description, the same choice is
reflected in the action of the orientifold projection on the Chan--Paton factors.

We describe the resolved conifold $X$ as the Higgs branch of a two-dimensional
\(\mathcal N=(2,2)\) \(U(1)\) gauge theory with four chiral multiplets \(X_I\)
of charges
\[
(1,1,-1,-1).
\]
Thus the resolved conifold is obtained from
\begin{equation}
 |X_1|^2+|X_2|^2-|X_3|^2-|X_4|^2=\operatorname{Re}(t)
\end{equation}
by quotienting by the \(U(1)\) action. There are two natural antiholomorphic involutions of the resolved conifold\footnote{The maps \(\sigma^\pm\) are antiholomorphic since they act by complex
conjugation on the coordinates. Moreover, \((\sigma^+)^2=\mathrm{id}\), while
\((\sigma^-)^2=-\mathrm{id}\). Although \((\sigma^-)^2\) acts as the
simultaneous sign flip on the homogeneous coordinates, this sign flip is a
  gauge transformation in the GLSM description of the resolved
conifold. Hence it acts trivially on the quotient, and \(\sigma^-\) also
defines an antiholomorphic involution of the resolved conifold.}
\cite{Acharya:2002kv,Hori:2005bk,Krefl:2009md}
\begin{equation}
    \sigma^{\pm}:\ (X_1,X_2,X_3,X_4)
    \longrightarrow
    (\bar X_2,\pm \bar X_1,\bar X_4,\pm \bar X_3).
\end{equation}
If
\begin{equation}
    z=\frac{X_1}{X_2}
\end{equation}
is the affine coordinate on the exceptional \(\mathbb P^1\) in the resolved conifold, then these
involutions act as
\begin{equation}
      \sigma_\pm:\ z\longmapsto \pm \frac{1}{\bar z}.
\end{equation}

It follows that \(\sigma_-\) acts freely on the exceptional \(\mathbb P^1\),
and therefore the orientifold projection \(\Omega\sigma_-\) has no fixed locus
in the resolved conifold. In particular, this background contains no
orientifold plane. By contrast, \(\sigma_+\) fixes the real locus
\[
|z|^2=1,
\]
namely \(\mathbb{RP}^1\subset \mathbb P^1\). The corresponding fixed locus in
the resolved conifold is the Lagrangian submanifold
\[
\mathcal L\simeq \mathbb R^2\times \mathbb{RP}^1,
\]
and the orientifold projection by \(\Omega\sigma_+\) of the resolved conifold   contains an
orientifold plane supported on \(\mathcal L\).

It is instructive to analyze the action of $\Omega\sigma_\pm$  in the deformed conifold geometry,
\(T^*S^3\), where the original \(SU(N)_k\) Chern--Simons theory is realized by
\(N\) Lagrangian branes wrapping the zero-section \(S^3\). In terms of the
gauge invariant coordinates
\[
x=X_1X_3,\qquad
y=X_2X_4,\qquad
u=X_1X_4,\qquad
v=X_2X_3,
\]
the deformed conifold is
\begin{equation}
xy-uv=\mu,
\end{equation}
where \(\mu\) is the complex-structure deformation parameter. For real \(\mu\),
the two antiholomorphic involutions act as
\begin{equation}
\sigma_-:\quad
(x,y,u,v)\longmapsto
(\bar y,\bar x,-\bar v,-\bar u),
\end{equation}
and
\begin{equation}
\sigma_+:\quad
(x,y,u,v)\longmapsto
(\bar y,\bar x,\bar v,\bar u).
\end{equation}
Their fixed loci are markedly different. For \(\mu>0\),
\begin{equation}
\operatorname{Fix}(\sigma_-)
=
\left\{
y=\bar x,\quad v=-\bar u,\quad |x|^2+|u|^2=\mu
\right\}
\simeq S^3,
\end{equation}
whereas
\begin{equation}
\operatorname{Fix}(\sigma_+)
=
\left\{
y=\bar x,\quad v=\bar u,\quad |x|^2-|u|^2=\mu
\right\}
\simeq \mathbb R^2\times S^1.
\end{equation}

The orientifold projection \(\Omega\sigma_-\) therefore fixes the zero-section
\(S^3\) on which the  branes   and the Chern--Simons theory are supported.  \(\Omega\sigma_-\) adds an orientifold plane on that \(S^3\), that is, it does not introduce
a defect in the Chern--Simons theory. Instead, depending on the choice of
action on the Chan--Paton factors (see paragraph below), this orientifold projection realizes Chern--Simons
theory with orthogonal or symplectic gauge group~\cite{Sinha:2000ap}, i.e. 
$Spin(N)$ or $Sp(N/2)$, respectively. 
 In Appendix~\ref{App:su-so-sp-branes-geometry}, we review the string dual
description of the \(S^3\) partition functions and present the explicit
dictionary between Wilson line defects and brane or antibrane configurations in
the \(\Omega\sigma_-\) orientifold of the resolved conifold.

We denote by \(\Omega\sigma_{\pm}^{SO}\) and
\(\Omega\sigma_{\pm}^{Sp}\) the two possible choices of orientifold action
on the Chan--Paton factors. In the open string channel these two choices are
distinguished by the symmetry property of the Chan--Paton matrix implementing
the orientifold projection:
\begin{equation}
 \gamma_{\Omega\sigma_{\pm}^{SO}}^T
 =
 \gamma_{\Omega\sigma_{\pm}^{SO}},
 \qquad
 \gamma_{\Omega\sigma_{\pm}^{Sp}}^T
 =-
 \gamma_{\Omega\sigma_{\pm}^{Sp}} .
\end{equation}
Equivalently, in the closed string channel, the two choices differ by the
overall sign of the corresponding crosscap state: $\ket{C^
{SO}_\pm}=-
 \ket{C^
{Sp}_\pm}$. In
particular, the leading unoriented contribution to the closed string partition function,   from the \(\mathbb{RP}^2\) worldsheet, differs by an overall
sign between the \(\Omega\sigma_{(\cdot)}^{SO}\) and
\(\Omega\sigma_{(\cdot)}^{Sp}\) orientifold models. When fixed loci are present, the two projections give rise respectively to the
\(O^{-}\)- and \(O^{+}\)-planes. The superscript in \(O^{(\cdot)}\) indicates the sign of the
orientifold plane charge, equivalently the sign of the flux sourced by the
plane, which is \(-1\) for \(O^{-}\) and \(+1\) for \(O^{+}\).

\begin{figure}
\centering
\resizebox{0.8\textwidth}{!}{%
\begin{tikzpicture}[
    >=Latex,
    line/.style={line width=1.0pt},
    brane/.style={line width=1.2pt,blue!70!black},
    oplane/.style={line width=1.2pt,red},
    xmark/.style={line width=1.35pt,red!80!black},
    every node/.style={font=\small}
]

\foreach \y in {-0.30,-0.15,0,0.15,0.30} {
    \draw[brane] (-4.75,\y) -- (-1.75,\y);
}

\draw[oplane] (-3.25,-1) -- (-3.25,1) node[right] {$O^{\mp}$} ;

\draw[->,line] (-1.10,0) -- (0.55,0);

\begin{scope}[shift={(1.85,0)}]
\coordinate (L) at (0,0);
\coordinate (R) at (2.60,0);
\coordinate (X) at ($(L)!0.5!(R)$);

\draw[line] (L) -- (R);
\draw[line] (L) -- ++(0,1.00);
\draw[line] (L) -- ++(-0.85,-0.72);
\draw[line] (R) -- ++(0.85,1.00);
\draw[line] (R) -- ++(0,-1.00);

\draw[oplane,dashed] ($(X)+(0,-1)$) -- ($(X)+(0,1)$);

\fill[red] (X) circle (2.4pt);

\end{scope}
\end{tikzpicture}%
}
 \caption{Geometric transition for the \(\Omega\sigma_+^{SO/Sp}\)
orientifold. On the deformed conifold side, the orientifold plane \(O^{\mp}\)
intersects the stack of \(N\) Lagrangian branes along  
\(S^1\subset S^3\). After the transition, the branes are replaced by flux, and
the resolved conifold contains an orientifold plane  \(O^{\mp}\)  at the fixed locus of $\sigma_+$, shown as the dashed
red line.}
\label{omegasigmap}
\end{figure}
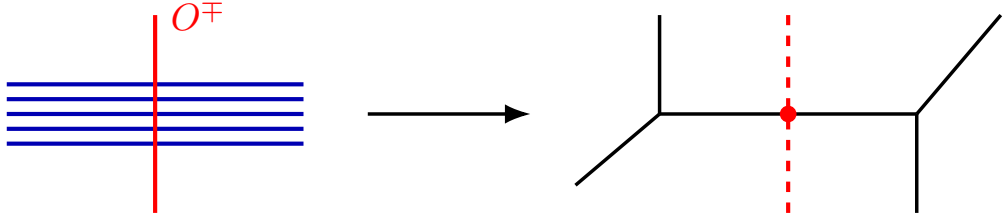

The orientifold projection \(\Omega\sigma_+^{SO/Sp}\), on the other hand, has fixed
locus \(\mathbb R^2\times S^1\) inside \(T^*S^3\). Its intersection with the
zero-section \(S^3\) is an \(S^1\), an unknot in $S^3$. Thus, from the point of view of the
\(SU(N)_k\) Chern--Simons theory on \(S^3\), this orientifold introduces a
codimension-two defect supported on a knot. This provides further evidence
that the closed string dual of   monodromy defects in $SU(N)_k$ Chern--Simons theory involves   an orientifold
background of the $\Omega\sigma_{+}^{{SO}/{Sp}}$ type. The brane configuration in \(T^*S^3\) and the toric diagram after the conifold transition are depicted in Figure \ref{omegasigmap}. 

A final important clue to find the precise closed string description of the vacuum monodromy defects is to look at what worldsheet maps contribute to the A-model partition function on the $\mathbb{RP}^2$ worldsheet.
The unoriented contribution is computed by summing over holomorphic maps $\phi$ from oriented double covers obeying the equivariance condition
\begin{equation}
\label{eq:orientifold-equivariant-unoriented}
\phi\circ \Omega
=
\sigma_\pm\circ \phi\,.
\end{equation}
If we denote by $w$ the coordinate on the $\mathbb{RP}^2$ worldsheet, the equivariance condition reads
\begin{equation}
  \Omega\sigma_\pm^{{SO}/{Sp}}:\qquad    z(-1/\bar w)=\pm{\frac{1} {\overline{z(w)}}}\,.
\end{equation}
Consider maps $z(w)=w^n$. Equivariance of the worldsheet maps imposes a parity selection rule on the
degree. For the two orientifold actions one finds~\cite{Acharya:2002kv}
\begin{equation}
    \begin{aligned}
    \Omega\sigma_-^{SO/Sp}:&\qquad n \ \text{odd},\\
    \Omega\sigma_+^{SO/Sp}:&\qquad n \ \text{even}.
    \end{aligned}
\end{equation}
Equivalently, the corresponding instanton sums are restricted to odd and even
degrees, respectively.

 Putting all these facts together, this implies that the worldsheet instanton part closed string partition function of an orientifold of the resolved conifold takes the following form:\footnote{Unoriented topological string amplitudes have been computed using localization
on moduli spaces of maps with involution, orientifold versions of the
topological vertex, and the real topological string/extended holomorphic
anomaly formalism, see e.g.
\cite{Bouchard:2004ie,
BouchardFloreaMarino2005,Walcher2007,
KreflWalcher2008,KreflWalcher2009,
KreflPasquettiWalcher2009,AganagicSchaeffer2012}.} 
\begin{equation}
\label{eq:free-orientifold-inst-free-energy}
\mathcal F^{\rm inst}_{\Omega \sigma_{{\color{green!65!black}\pm}}^{{\color{red}SO}/{\color{blue} Sp}}}(t)
=
-\underbrace{\frac12
\sum_{m\geq 1}
\frac{e^{-mt}}{m[m]^2}}_{\displaystyle\Sigma_{\rm or}\to X} \;\; 
{{\mathbin{\substack{{\color{red}\scalebox{1.05}{$-$}}\\[-1ex]{\color{blue}\scalebox{1.05}{$+$}}}}} }
\underbrace{\sum_{\substack{m\geq 1\\ m\ {\rm \color{green!65!black}even / odd}}}
\frac{e^{-\frac{mt}{2}}}{m[m]}}_{\displaystyle  \Sigma_{\rm un}\to X}\,.
\end{equation}
 \(t\) is the complexified K\"ahler parameter, measuring the area of the
\(\mathbb P^1\). Accordingly, \(e^{-t}\) is the instanton factor associated
with a worldsheet wrapping the \(\mathbb P^1\) once. The factor \(e^{-t/2}\) is   the
instanton factor for a crosscap worldsheet whose oriented double cover wraps
the \(\mathbb P^1\) once. Equivalently, the unoriented quotient carries half
the usual worldsheet action.
Therefore, the first term in \eqref{eq:free-orientifold-inst-free-energy} is   the sum over  maps from oriented worldsheets to the target space geometry $\phi: \Sigma_{\rm or}\to X$, while the second term corresponds to the contribution of the equivariant maps from unoriented worksheet $\phi: \Sigma_{\rm un} \to X$, and indeed only contains one power of the string coupling as illustrated in \eqref{eqnodd} and \eqref{eqneven}. 
The amplitude in \eqref{eq:free-orientifold-inst-free-energy} admits a natural
extension obtained by adding branes or antibranes to the orientifold
background. This extension will play a crucial role in identifying the string
duals of arbitrary monodromy defects.

The sign ${{\mathbin{\substack{{\color{red}\scalebox{1.05}{$-$}}\\[-1ex]{\color{blue}\scalebox{1.05}{$+$}}}}} }$ in \eqref{eq:free-orientifold-inst-free-energy} encodes the sign of    the crosscap state:    $\color{red}-$ in the case of $O^{\color{red}-}$ projection ($SO$ type), and $\color{blue}+$ for the $O^{\color{blue}+}$ projection ($Sp$ type). In the case of $\Omega\sigma_-^{(\cdot)}$ the former realizes $Spin(N)$ Chern--Simons while the latter   the $Sp(N/2)$ theory, and therefore, we keep denoting the choice of crosscap as $SO$/$Sp$ also for the $\Omega\sigma_+$ case of interest here. 

The general unoriented string free energy formula     \eqref{eq:free-orientifold-inst-free-energy} is exactly the one we have found by expanding the vacuum monodromy defects in Section~\ref{vacuum}, providing  a rather non-trivial test of our proposal.  
 
The closed string dual of the vacuum monodromy defects are precisely matched with the orientifold backgrounds in Table  \ref{tab:ts-backgrounds}:
\begin{table}[H]
\centering
{\renewcommand{\arraystretch}{1.2}
\begin{tabular}{|c|c|c|c|c|}
\hline
{Theory} & {Parity of $k$} & {Background} & Planes  & {Flux $e^{-t}$} \\
\hline
\hline
$SU(2N)_k$, & $k$ odd
&
$\Omega\sigma^{ Sp}_+$
& $\color{blue} O^+$&
$q^{2N+1} = q^{2N{\color{blue}+1}} $
\\
\hline
$SU(2N)_k$, & $k$ even, $2N\leq k$
&
$\Omega\sigma^{ Sp}_+$&
$\color{blue} O^+$
&
$q^{2N+1} = q^{2N{\color{blue}+1}}$
\\
$SU(2N)_k$, & $k$ even, $2N>k+2$
&
$\Omega\sigma^{ SO}_+$&
$\color{red} O^-$
&
$q^{2N-1} = q^{2N{\color{red}-1}}$
\\
\hline
$SU(2N+1)_k$, & $k$ even
&
$\Omega\sigma^{ SO}_+$&
$\color{red} O^-$
&
$q^{2N} = q^{2N+1{\color{red}-1}}$
\\
\hline
$SU(2N+1)_k$, & $k$ odd
&
$\Omega\sigma^{ SO}_+ \oplus \mathcal B$&
${\color{red} O^-} \oplus \color{green!70!black} \mathcal B $
&
$q^{2N+1} = q^{2N+1{\color{red}-1} {\color{green!70!black}+1}}$
\\
\hline
\end{tabular}}
\caption{Vacuum monodromy defects and corresponding dual  topological string backgrounds.  }
\label{tab:ts-backgrounds}
\end{table}
 As is clear from the table, the fluxes for $t$ work out perfectly to guarantee that the total flux supporting the $\mathbb{P}^1$ decreases by one unit of $g_s$ for the $\Omega\sigma^{SO}_+$ projection, and increases by one unit of $g_s$ for the $\Omega\sigma^{Sp}_+$ projection. As we shall see later, when we discuss non-vacuum monodromy defects, the total flux is always the total brane charge in the $T^*S^3$ realization of the monodromy defect: it includes the flux from the branes producing the Chern--Simons theory {\it and}   the flux due to the intersecting branes or antibranes and orientifolds that realize the monodromy defect.

\begin{figure}[H]
\centering
\resizebox{0.40\textwidth}{!}{%
\begin{tikzpicture}
  \toricCanvasOrientifold
  \toricConifoldBaseOrientifold
  \toricStuckBrane
\end{tikzpicture}%
}
\caption{$\Omega\sigma_+^{SO}\oplus \mathcal B$ orientifold of resolved conifold: there is a brane stuck at the
  midpoint of the internal edge, on top of the   $O$-plane
  (red dashed line).}
\label{fig:stuck-brane}
\end{figure}
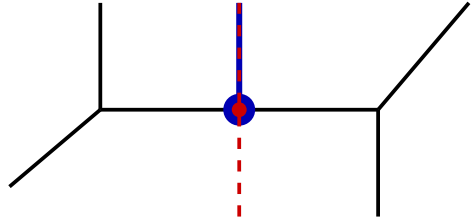

The case of $SU(2N+1)_k$  vacuum momodromy defect with $k$ odd has a peculiarity that deserves a comment.  First,   by analogy with the $SU(2N)_k$, odd level case, one  could have expected a background corresponding to the $\Omega\sigma^{Sp}_+$ projection. But, a symplectic projection with an odd $2N+1$ number of   branes is simply inconsistent. It is thus very natural that in the absence of the symplectic projection, the vacuum monodromy is in this case dual to an $\Omega\sigma_+^{SO}$ background with a stuck brane at the fixed locus of the projection, which we denote by $\Omega\sigma^{ SO}_+ \oplus \mathcal B$, as in Figure~\ref{fig:stuck-brane}. This is precisely what we find. Note,   that, indeed,  the flux works out perfectly:
 \begin{equation}
\qquad t = g_s(2N+1 \underbrace{{\color{red}- 1}}_{\color{red}O^-} \underbrace{\color{blue}+1}_{\color{blue} D})\,=g_s  \left(2N+1\right)\,.
 \end{equation}
 
\subsubsection{Constant maps}
\label{constantmaps}
We stress that the holographic dictionary the Chern--Simons observables and the topological A-model, is an equality between the unormalized modular $S$-matrix elements and the topological string partition function. In particular, as we explain in
Appendix~\ref{app:constant-maps} (see also Appendix~\ref{smallgs}), the overall normalization of the modular S-matrix $S_{00}$, which we have suppressed in the main text for
clarity, together with the prefactors \(M(q)\) and \(\zeta(q)\), contributes
only terms independent of \(e^{-t}\).  These terms contain physical content: they give the constant map
contribution to the topological string amplitude \cite{fulton1997notesstablemapsquantum,Bouchard:2004ie,Bouchard:2004yp, Sinha:2000ap}. Constant maps correspond to worldsheets collapsed to points in the target
space, \(\Sigma \to X\), where \(\Sigma\)
may be oriented or unoriented.  In the  $g_s$ Gopakumar--Vafa expansion, a worldsheet $\Sigma_{g,c}$ of genus $g$ and number of crosscap insertion $c$  contributes with the usual power \beq g_s^{-\chi\left(\Sigma_{g,c}\right)}, \qquad \chi\left(\Sigma_{g,c}\right) = 2- 2g - c\eeq, where $\chi$ is the Euler characteristic.  The constant map contributions therefore provide a
sharp diagnostic of the oriented/unoriented nature of the closed topological string partition function.\\
The MacMahon function has the $g_s$ expansion \eqref{macmahonexp}:  
\begin{equation}
\log M(q)=\frac{\zeta(3)}{\gs^2}+\frac1{12}\log\gs+\zeta'(-1)
+\sum_{g\ge2}\frac{B_{2g}B_{2g-2}}{2g(2g-2)(2g-2)!}\gs^{2g-2}
\end{equation}
and famously corresponds to a source for the constant maps terms.  Its appearance in the
small \(q\) expansion is therefore not surprising.  The Euler factor
\(\zeta(q)\), by contrast, deserves some comments, as it did not appear before, as far as we know,  in the context of the  genus  expansion. Its expansion at small $\gs$ is \footnote{Here
\(\zeta(q)=\prod_{n\geq 1}(1-q^n)^{-1}\) the inverse Euler function transforms under the modular transformation as,
\[
\zeta(e^{-g_s})
=
\sqrt{\frac{g_s}{2\pi}}\,
e^{\frac{\pi^2}{6g_s}-\frac{g_s}{24}}\,
\zeta(e^{-4\pi^2/g_s}) .
\]
Thus the non-perturbative factor is again the same function \(\zeta\), evaluated at
the dual variable \(e^{-4\pi^2/g_s}\).
}
\begin{equation}
\log\zeta(e^{-g_s})
=
\frac{\pi^2}{6g_s}
-\frac12\log\frac{2\pi}{g_s}
-\frac{g_s}{24}
+ \mathcal{O} \!\left(e^{-4\pi^2/g_s}\right).
\end{equation}
Thus its perturbative expansion around \(g_s=0\) terminates after the term
linear in \(g_s\), up to an exponentially small nonperturbative tail.  This
factor is essential for converting the small \(q\) expansion into the standard
small \(g_s\) genus expansion, as explained in
Appendix~\ref{smallgs}. The exponentially small terms \(e^{-4\pi^2/g_s}\) also have an interpretation.  They are the leading non-perturbative effects
dictated by the resurgent structure of the asymptotic expansion in $\gs$.\footnote{The small $\gs$ perturbative expansion, is as asymptotic  series in $\gs$ and requires a nonperturbative completion in order to be defined in the $\gs$ plane~\cite{Hatsuda:2013oxa, Alexandrov:2023wdj,Alim:2021mhp}. The nearest singularity in the Borel plane  occurs at precisely the action \(4\pi^2\), predicting
corrections of order \(e^{-4\pi^2/g_s}\).  These are exactly the exponentially
small terms in $\gs$ produced by the nonperturbative completion encoded in the inverse Euler
function \(\zeta(q)\). The analysis of nonperturbative corrections from Borel resummations of the small $\gs$ expansion was performed in \cite[Section 4.3]{Pasquetti:2010bps}, where they show that indeed the closest pole to the real axis  in the Borel plane (controlling the large genus asymptotic) predicts exactly the same exponential $e^{-4\pi^2/\gs}$ that we find (see equation $(4.39)$ there). }      After all prefactors are combined, the constant map contribution of
\(SU(N)_k\) Chern--Simons theory contains only even powers of \(g_s\):
\begin{equation}
\mathcal{F}^{\rm const}_{SU}
=
\underbrace{\frac{\#}{g_s^2}}_{\text{sphere}}
+
\underbrace{\#\, g_s^0}_{\text{torus}}
+
\underbrace{\#\, g_s^2+\cdots}_{g\geq 2}
=
\sum_{g\geq 0} c_g\, g_s^{2g-2}\,,
\end{equation}
as discussed in details  in Appendix  \ref{app:constant-maps} in equation \eqref{SUNconstant}. These, as expected, 
 correspond to maps from  oriented worldsheets, which have  for which $\chi \in 2\mathbb{Z}$.

We perform the same analysis for the partition functions of the
\(\Omega\sigma_\pm\) orientifold backgrounds in Appendix \ref{constSp}, \ref{spinconst}, and \ref{constmon}. This reveals that, in addition to the
usual oriented constant map contributions, there is an unoriented contribution
at the odd powers of \(g_s\):
\begin{align}
    \mathcal{F}^{\rm const}_{\Omega\sigma^{{\color{red}SO}/{\color{blue}Sp}}_-}
    &=
    \sum_{g\geq 0} c^-_g\, g_s^{2g-2}
    \;
    \boxed{
    {\mathbin{\substack{
    {\color{red}\scalebox{1.05}{$-$}}\\[-1ex]
    {\color{blue}\scalebox{1.05}{$+$}}
    }}}
    \frac{\pi^2}{8}\,\frac{1}{g_s}
    }\,,
    \\
    \mathcal{F}^{\rm const}_{\Omega\sigma^{{\color{red}SO}/{\color{blue}Sp}}_+}
    &=
    \sum_{g\geq 0} c^+_g\, g_s^{2g-2} \;
    \boxed{
    {\mathbin{\substack{
    {\color{red}\scalebox{1.05}{$-$}}\\[-1ex]
    {\color{blue}\scalebox{1.05}{$+$}}
    }}
    \frac{1}{g_s}
    \left(
        - \frac{t^2}{16}
        +
        \frac{\pi^2}{24}
    \right) 
    } {\mathbin{\substack{
    {\color{red}\scalebox{1.05}{$-$}}\\[-1ex]
    {\color{blue}\scalebox{1.05}{$+$}}
    }}} \frac{g_s}{48}} \, .
\end{align} as illustrated in \eqref{costSP} for $Sp(N)$, \eqref{constSO} for $Spin(N)$ and \eqref{constantmon} for the monodromy defects.
The \(g_s^{-1}\) contributions correspond to  equivariant constant maps from the one-crosscap worldsheet to the target,
\[
    \mathbb{RP}^2 \longrightarrow X\,,
    \qquad
    \chi(\mathbb{RP}^2)=1 .
\]
 While the one at order $\gs$ corresponds to equivariant constant maps from the three crosscap sphere (homeomorphic to the torus  with one crosscap $\Sigma_{1,1}$) $\Sigma_{0,3}\to X$, with $\chi\left(\Sigma_{0,3}\right) = -1$.\footnote{\label{fn:homeo} The \textit{unoriented} surface $\Sigma_{g,3+c}$ is homeomorphic to the \textit{unoriented} surface $\Sigma_{g+1,1+c}$, i.e, one can always trade 2 crosscap for an handle, as long as the surface remains unorientable. For e.g. 3 crosscap sphere $=\Sigma_{0,3}\cong \Sigma_{1,1}=$ one crosscap torus but, a normal torus $\Sigma_{1,0}=T^2 \not\cong \Sigma_{0,2}=$ Klein bottle as the torus is orientable.}
As expected, the sign of the unoriented contribution with odd number  of crosscaps insertions depend on the
\(SO/Sp\) choice of crosscap state~\cite{Bouchard:2004iu}.  There is, however, an important
distinction between the \(\Omega\sigma_-\) and \(\Omega\sigma_+\)
backgrounds.  In the \(\Omega\sigma_-\) case the \(t^2\) term is absent,
whereas it is present for \(\Omega\sigma_+\). As we discuss in Appendix  \ref{app:constant-maps} this is perfectly consistent with the absence of a fixed locus on the target in in the  $\Omega\sigma_-$ orientifold projection (see also \cite{Sinha:2000ap}), and it is therefore another confirmation our the holographic duality of monodromy defects.


This analysis also predicts that unoriented constant maps give no 
contributions at higher Euler number.  It would be interesting to understand this vanishing directly
from the mathematical theory of stable maps.

\subsection{String theory dual of excited monodromy defects}
\label{nontrivialmon}

We now turn to the string dual description of generic non-vacuum monodromy
defects. In \(SU(N)_k\) Chern--Simons theory, such excited monodromy defects
are labeled by non-vacuum integrable representations 
of the twisted affine algebra $A_N^{(2)}$. The vacuum expectation value of the excited defects 
is captured by the modular $S$-matrix entry $S^{(\mathsf{C},1)}_{a0}$ of $A_N^{(2)}$.
Excited monodromy defects can be physically thought of as bound states between a vacuum monodromy defect and a Wilson line, stacked on top of each other.  Remarkably, the spectrum of such bound states    are exactly in correspondence with the integrable representations of the twisted affine Lie algebra, as explained in Section~\ref{subsec:monodromy-defect-data}.

The analysis below provides a sharp   duality dictionary
between monodromy defects and their holographic string duals. As we will show,
excited monodromy defects have a remarkably simple interpretation in the dual
description: they are realized by adding specific collections of Lagrangian
branes or antibranes to the   \(\Omega\sigma^{SO/Sp}_+\)
orientifold background that describes the vacuum monodromy defect.

More precisely, a monodromy defect labeled by a Young tableau \(R\) with
\(M\) columns and \(P\) rows admits two equivalent descriptions. It can be
realized either by adding \(M\) branes on internal edges of the toric geometry,
or by adding \(P\) antibranes on external edges.\footnote{This is akin to the  dual topological string theory  description of ordinary $SU(N)$ and $Spin$/$Sp$ Wilson lines (reviewed in Appendix \ref{App:su-so-sp-branes-geometry}).}  In both descriptions the branes come with their orientifold images. The holographic map is summarized in Table \ref{tab:summarySUmono}.
{
\sbox{\MrowCYbox}{%
\begin{tikzpicture}[scale=0.8,baseline={(current bounding box.center)},
line/.style={line width=1.0pt},
brane/.style={line width=1.25pt,blue!70!black},
branepoint/.style={circle,fill=blue!70!black,inner sep=2.25pt},
redpoint/.style={circle,fill=red!80!black,inner sep=1.4pt},
axis/.style={line width=1.0pt,red!80!black,dashed},
every node/.style={font=\small}
]
\coordinate (L) at (0,0);
\coordinate (R) at (2.60,0);
\coordinate (C) at ($(L)!0.5!(R)$);
\draw[line] (L) -- (R);
\draw[line] (L) -- ++(0,1.00);
\draw[line] (L) -- ++(-0.85,-0.72);
\draw[line] (R) -- ++(0.85,1.00);
\draw[line] (R) -- ++(0,-1.00);
\draw[axis] ($(C)+(0,-1.05)$) -- ($(C)+(0,1.05)$);
\node[redpoint] at (C) {};
\foreach \t in {0.10, 0.20, 0.30} {
  \coordinate (Bpos) at ($(L)!\t!(R)$);
  \draw[brane] (Bpos) -- ++(0,0.84);
  \node[branepoint] at (Bpos) {};
}
\foreach \t in {0.70, 0.80, 0.90} {
  \coordinate (Bpos) at ($(L)!\t!(R)$);
  \draw[brane] (Bpos) -- ++(0,0.84);
  \node[branepoint] at (Bpos) {};
}
\end{tikzpicture}}

\sbox{\ProwCYbox}{%
\begin{tikzpicture}[scale=0.8,baseline={(current bounding box.center)},
line/.style={line width=1.0pt},
brane/.style={line width=1.25pt,blue!70!black},
branepoint/.style={circle,fill=blue!70!black,inner sep=2.25pt},
redpoint/.style={circle,fill=red!80!black,inner sep=1.4pt},
axis/.style={line width=1.0pt,red!80!black,dashed},
every node/.style={font=\small}
]
\coordinate (L) at (0,0);
\coordinate (R) at (2.60,0);
\coordinate (Lext) at ($(L)+(-0.85,-0.72)$);
\coordinate (Rext) at ($(R)+(0.85,1.00)$);
\coordinate (C) at ($(L)!0.5!(R)$);
\draw[line] (L) -- (R);
\draw[line] (L) -- ++(0,1.00);
\draw[line] (L) -- ++(-0.85,-0.72);
\draw[line] (R) -- ++(0.85,1.00);
\draw[line] (R) -- ++(0,-1.00);
\draw[axis] ($(C)+(0,-1.05)$) -- ($(C)+(0,1.05)$);
\node[redpoint] at (C) {};
\foreach \t in {0.25, 0.50, 0.75} {
  \coordinate (BposL) at ($(L)!\t!(Lext)$);
  \draw[brane] (BposL) -- ++(0,0.84);
  \node[branepoint] at (BposL) {};
}
\foreach \t in {0.25, 0.50, 0.75} {
  \coordinate (BposR) at ($(R)!\t!(Rext)$);
  \draw[brane] (BposR) -- ++(0,0.84);
  \node[branepoint] at (BposR) {};
}
\end{tikzpicture}}

\sbox{\MrowTextbox}{%
\begin{minipage}{6.4cm}\centering
  \(M\) branes on internal leg at
  \(x_i=\gs(R_i^T-i+M+\tfrac12)\), each with image at
  \(\widehat t-x_i\);\quad \(\widehat t=t+M\gs\)
\end{minipage}}

\sbox{\ProwTextbox}{%
\begin{minipage}{6.4cm}\centering\small
  \(P\) antibranes on external leg at
  \(y_i=\gs(R_i-i+P+\tfrac12)\), each with image at
  \(\widehat t+y_i\);\quad \(\widehat t=t-P\gs\)
\end{minipage}}

\setlength{\Mrowheight}{0pt}
\updrowheight{\Mrowheight}{\MrowCYbox}
\updrowheight{\Mrowheight}{\MrowTextbox}

\setlength{\Prowheight}{0pt}
\updrowheight{\Prowheight}{\ProwCYbox}
\updrowheight{\Prowheight}{\ProwTextbox}

\begin{table}[t]
  \centering
  \begin{tabular}{|>{\centering\arraybackslash}m{3.5cm}
                  | >{\centering\arraybackslash}m{7cm}
                  | >{\centering\arraybackslash}m{3.5cm}|}
                  \hline
                  \vspace{0.2cm}
    \(\langle \mathsf{M}_a\rangle = S^{(\mathsf{C},1)}_{a0}\) \vspace{0.2cm}
    &\vspace{0.2cm} Branes / image branes (A-model)
     \vspace{0.2cm}
    &
    \vspace{0.2cm}Toric diagram
    \vspace{0.2cm}\\
    \hline\hline
    \vspace{0.5cm}
    Antisymmetric \(\Lambda^\ell\)
    & \parbox{6.2cm}{\centering
\vspace{0.2cm}
      Brane on internal leg at \(x=\gs(\ell+\tfrac12)\), image at
      \(\widehat t-x\);\quad \(\widehat t=t+\gs\)
      \vspace{0.2cm}
    }
    & {\begin{tikzpicture}[scale=0.8,baseline={(current bounding box.center)},
line/.style={line width=1.0pt},
brane/.style={line width=1.25pt,blue!70!black},
branepoint/.style={circle,fill=blue!70!black,inner sep=2.25pt},
redpoint/.style={circle,fill=red!80!black,inner sep=1.4pt},
axis/.style={line width=1.0pt,red!80!black,dashed},
every node/.style={font=\small}
]
\coordinate (L) at (0,0);
\coordinate (R) at (2.60,0);
\coordinate (C) at ($(L)!0.5!(R)$);
\draw[line] (L) -- (R);
\draw[line] (L) -- ++(0,1.00);
\draw[line] (L) -- ++(-0.85,-0.72);
\draw[line] (R) -- ++(0.85,1.00);
\draw[line] (R) -- ++(0,-1.00);
\draw[axis] ($(C)+(0,-1.05)$) -- ($(C)+(0,1.05)$);
\node[redpoint] at (C) {};
\coordinate (Bpos) at ($(L)!0.30!(R)$);
\coordinate (IBpos) at ($(L)!0.70!(R)$);
\draw[brane] (Bpos) -- ++(0,0.84);
\node[branepoint] at (Bpos) {};
\draw[brane] (IBpos) -- ++(0,0.84);
\node[branepoint] at (IBpos) {};
\end{tikzpicture}}\\
    \hline
    \vspace{0.6cm}
    Symmetric \(\mathrm{Sym}^w\)
    & \parbox{6.2cm}{\centering
    \vspace{0.2cm}
       antibrane on external leg at \(y=\gs(w+\tfrac12)\), image at
      \(\widehat t+y\);\quad \(\widehat t=t-\gs\)
    }
    & {\begin{tikzpicture}[scale=0.8,baseline={(current bounding box.center)},
line/.style={line width=1.0pt},
brane/.style={line width=1.25pt,blue!70!black},
branepoint/.style={circle,fill=blue!70!black,inner sep=2.25pt},
redpoint/.style={circle,fill=red!80!black,inner sep=1.4pt},
axis/.style={line width=1.0pt,red!80!black,dashed},
every node/.style={font=\small}
]
\coordinate (L) at (0,0);
\coordinate (R) at (2.60,0);
\coordinate (Lext) at ($(L)+(-0.85,-0.72)$);
\coordinate (Rext) at ($(R)+(0.85,1.00)$);
\coordinate (C) at ($(L)!0.5!(R)$);
\draw[line] (L) -- (R);
\draw[line] (L) -- ++(0,1.00);
\draw[line] (L) -- ++(-0.85,-0.72);
\draw[line] (R) -- ++(0.85,1.00);
\draw[line] (R) -- ++(0,-1.00);
\draw[axis] ($(C)+(0,-1.05)$) -- ($(C)+(0,1.05)$);
\node[redpoint] at (C) {};
\coordinate (BposL) at ($(L)!0.50!(Lext)$);
\coordinate (BposR) at ($(R)!0.50!(Rext)$);
\draw[brane] (BposL) -- ++(0,0.84);
\node[branepoint] at (BposL) {};
\draw[brane] (BposR) -- ++(0,0.84);
\node[branepoint] at (BposR) {};
\end{tikzpicture}}\\
    \hline
    \multirow{2}{=}{%
      \begin{minipage}[c][0pt][c]{3.05cm}
        \centering
        \vspace{2cm}
        \resizebox{\linewidth}{!}{%
          \hspace{1cm}\begin{tikzpicture}[
            line width=1pt,
            transform shape,
            lbl/.style={inner sep=0.35pt, scale=1.7},
            clbl/.style={inner sep=0.2pt, scale=1.7}
          ]
            \def\rows{{6,5,3,3,2}}
            \foreach \i in {0,...,4} {
              \pgfmathsetmacro{\len}{\rows[\i]}
              \foreach \j in {1,...,\len} {
                \draw (\j-1,-\i) rectangle (\j,-\i-1);
              }
            }
            \node[lbl, right] at (6.35, -0.5) {$R_1$};
            \node[lbl, right] at (6.35, -1.5) {$R_2$};
            \node[lbl]           at (6.75, -2.5) {$\vdots$};
            \node[lbl, right] at (6.35, -4.5) {$R_P$};
            \node[clbl, below] at (0.5, -5.12) {$R_1^{T}$};
            \node[clbl, below] at (1.5, -5.12) {$R_2^{T}$};
            \node[clbl, below] at (2.5, -5.12) {$R_3^{T}$};
            \node[clbl, below] at (3.5, -5.12) {};
            \node[clbl, below] at (4.5, -5.20) {$\cdots$};
            \node[clbl, below] at (5.5, -5.12) {$R_M^{T}$};
          \end{tikzpicture}%
        }%
      \end{minipage}%
    }
    &
    \Rowcell{\Mrowheight}{6.4cm}{\usebox{\MrowTextbox}} &
    \Rowcell{\Mrowheight}{3.2cm}{\usebox{\MrowCYbox}} \\
    \cline{2-3}
    &
    \Rowcell{\Prowheight}{6.4cm}{\usebox{\ProwTextbox}} &
    \Rowcell{\Prowheight}{3.2cm}{\usebox{\ProwCYbox}} \\
    \hline
  \end{tabular}
  \caption{Summary of the correspondence between
    monodromy defects in \(SU(N)_k\) Chern--Simons theory and their
    orientifold A-model duals on the resolved conifold.  Blue segments
    are branes in the internal edge and antibranes in the external; each comes with an orientifold image
    reflected across the red dashed \(O\)-plane.  The positions are
    \(x_i=\gs(R^T_i-i+M+\tfrac12)\) and
    \(y_i=\gs(R_i-i+P+\tfrac12)\) for a partition \(R\) with \(M\)
    columns and \(P\) rows.  The closed background is an
    \(\Omega\sigma_+\) orientifold of the resolved conifold. Crosscap signs depends on the precise $\Omega\sigma_+^{SO/Sp}$ background.}
\label{tab:summarySUmono}
\end{table}

}

The logic of this section will be the same as before: we proceed by expanding the relevant modular $S$- matrix $S^{(\mathsf{C},1)}_{a0}$ in powers of $q$. All the details of the expansion are not presented, as they are all very similar to the ones we have seen before for the case of the vacuum monodromy defects. We instead focus on the geometric and physical interpretation of the various new terms that appear in the $q$-expansion, that are key to establish  the proposed duality.

\subsubsection{\boldmath Monodromy defects for \texorpdfstring{$SU(2N)_k$, $k$ odd}{SU(2N) k odd}}
\label{sec:su2Nodd}
The vacuum is the trivial label $a=0$ (degenerate with its simple current
partner $(k,0,\ldots,0)$). It reproduces   the following worldsheet instanton string amplitude \eqref{SU(2N)kodd}
\beq
\mathcal F^{\rm inst}_{\Omega\sigma^{Sp}_+}(t)
=-\frac12\sumall{\frac{e^{-mt}} {m[m]^2}}
 +\sumeven{\frac{e^{-mt/2}} {m[m]}}, \,\qquad  e^{-t}=q^{2N+1}\,.
\eeq
with the prefactor given by,
\[
  \mathcal{F}^0(N)=\frac{1}{2}\log\frac{M(q)\,\zeta(q^{2})}{\zeta(q)^{2N+1}}\,, 
\]
Excited  monodromy defects are labeled by integrable highest weight representation of the twisted affine algebra $A^{(2)}_{2N-1}$, described in  \eqref{integrablerep} (see Table \ref{tab:twisted-affine}). These can be  characterized by a representation $a$ of the horizontal $\mathring g = C_N$ algebra. Given a \(C_N\) representation \(a\) with Dynkin labels
\((a_1,\ldots,a_N)\), we associate to it a Young tableau with row lengths
\(R_i\). These row lengths are determined from the Dynkin labels as
$
    R_i=\sum_{j=i}^N a_j$, with  
    $i=1,\ldots,N$.
Equivalently, \(a_i=R_i-R_{i+1}\), with the convention \(R_{N+1}=0\).
 Therefore, a   non-vacuum monodromy defect   labeled by a representation $a$ can  be  described by a Young tableau composed of $P$ rows of length $R_i$ and $M$ columns  of length $R_i^T$:
\begin{align}
\begin{tikzpicture}[scale=0.7, baseline={(current bounding box.center)}]
  \def\rows{{6,5,3,3,2}}
  \foreach \i in {0,...,4} {
    \pgfmathsetmacro{\len}{\rows[\i]}
    \foreach \j in {1,...,\len} {
      \draw (\j-1,-\i) rectangle (\j,-\i-1);
    }
  }
  \node[right] at (6.4, -0.5) {$R_1$};
  \node[right] at (6.4, -1.5) {$R_2$};
  \node        at (6.9, -3) {$\vdots$};
  \node[right] at (6.4, -4.8) {$R_P$};
  \node[below] at (0.5, -5.15) {$R_1^{T}$};
  \node[below] at (1.5, -5.15) {$R_2^{T}$};
  \node[below] at (3.6, -5.25) {$\cdots$};
  \node[below] at (5.7, -5.15) {$R_M^{T}$};
\end{tikzpicture}
\label{fig:tabb}
      \end{align}

      \paragraph{Row description.}
We repeat the same steps of the  computation \eqref{eq:su2nodd-weyl}--\eqref{SU(2N)kodd}, but this time for a monodromy defect $\langle\mathsf M_{a}\rangle$   in a   representation $a$ of $C_N$, described by  a  tableau $R$ with rows
$R_1\ge\cdots\ge R_P>0$ (and $R_i=0$ for $i>P$).   Recall from \eqref{SmatSU2n} that the $\mu=0$ entry of the
twisted $S$-matrix is the $B_N$ Weyl product with the shifted coordinates
$z_i=\langle e_i,\tau^{-1}(\rho_C+a)\rangle$: 
\begin{equation}
S^{(\mathsf C,1)}_{a0}
\;\doteq\;
\prod_{1\le i<j\le N}\!\big(1-q^{z_i-z_j}\big)\big(1-q^{z_i+z_j}\big)
  \prod_{i=1}^{N}\big(1-q^{z_i}\big),
\qquad
z_i=R_i+N+1-i .
\label{eq:su2nodd-weyl-R}
\end{equation}
Because $R_i=0$ for $i>P$, the collection of $z_i$ separate into a ``background"
block and a block that is modified by $R_i$, which we denote as the ``brane" block
\begin{equation}
\{z_i\}_{i=1}^{N}
=\underbrace{\{1,2,\ldots,N-P\}}_{\text{Background }}
\;\sqcup\;
\underbrace{z_i=\{R_i+N+1-i\}_{i=1}^{P}}_{\text{Brane}},
\qquad z_i>N-P\ \ (i\le P),
\end{equation}
and \eqref{eq:su2nodd-weyl-R} factorises accordingly into background-background, background-brane and brane-brane terms. The product where $1\leq i,j \leq N-P$, the  background-background contribution, results in      exactly identical computation as we did for the vacuum monodromy defect, but with shifted flux $N\to N-P$, while the brane-brane terms and mixed background-brane terms produce new contributions, whose interpretation we  unearth. The result is the following: 
 \begin{equation}
 \begin{split}
\log S^{(\mathsf C,1)}_{a0}
\;\doteq\;
&\overbrace{\mathcal F^0(N-P)
+ \mathcal F^{\rm inst}_{\Omega\sigma^{Sp}_+}\left( \widehat t \, \right)}^{\rm Background}\quad +\quad \sum_{i=1}^{P}\overbrace{\sumall{\frac{e^{-my_i}-^{-m(\that+y_i)}}{m[m]}}}^{{\color{blue} \rm Open \, string}}+\\[+4pt]
 \qquad &-\sum_{1\le i<j\le P}\Big[\underbrace{\sumall{\frac{e^{-m(y_i-y_j)}}{m}}}_{\color{orange!65!black}\rm Annulus\, \, \bar{B}-\bar{B}}+\underbrace{\sumall{\frac{e^{-m(\widehat{t}+y_i-y_j)}} {m}}}_{\color{teal!65!black}\rm Annulus\, \, \bar{B}-Image \,\bar{B}}\Big]\,,
\label{eq:SU2N-odd-rows}
\end{split}
\end{equation}
where \[
 e^{-\that}=q^{2N+1-2P}, \qquad \quad  y_i = \gs\Big(R_i-i+P+\tfrac12\Big),\qquad i=1,\ldots,P \leq N
\]
\begin{figure}
  \centering
  \begin{subfigure}[t]{0.32\textwidth}
    \centering
   \includegraphics[width = \linewidth]{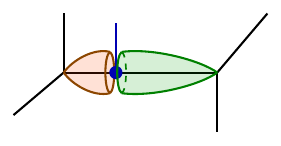}
    \caption{Brane in internal leg}
    \label{fig:disk-a}
  \end{subfigure}
  \hfill
  \begin{subfigure}[t]{0.32\textwidth}
    \centering
    \includegraphics[width = \linewidth]{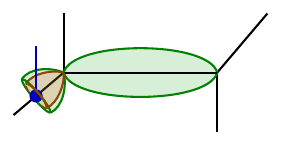}
    \caption{Brane in external leg}
    \label{fig:disk-b}
  \end{subfigure}
    \hfill
    \begin{subfigure}[t]{0.32\textwidth}
    \centering
    \includegraphics[width = \linewidth]{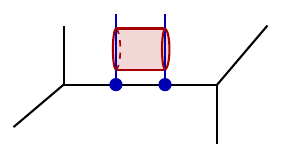}
    \caption{Annulus between branes}
    \label{fig:annulus}
  \end{subfigure}
\caption{Open string partition function for a brane in the internal \ref{fig:disk-a} and external leg \ref{fig:disk-b}. In {\color{orange!55!black} red} the strings wrap a distance $x_i$, and correspond to the first term in \eqref{disks}, while in {\color{green!60!black} green} the strings cover the distance $\widehat{t} \mp  x_i$ corresponding to the second term in \eqref{disks}. In \ref{fig:annulus} an annulus stretching between two parallel branes.  }  \label{fig:disksall}
\end{figure}

\begin{figure}[ht]
    \centering
    \includegraphics[width=0.5\linewidth]{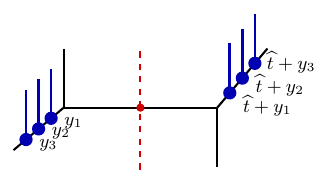}
    \caption{Antibranes ending on the external edge at position $y_i = \gs\Big(R_i-i+P+\tfrac12\Big)$ of the $\Omega\sigma_+$ geometry. These are dual to generic representations described as $P$ rows of length $R_i$. 
}
    \label{fig:rowsdual}
\end{figure}

This expression has an elegant interpretation in the  A-model as the topological string amplitude of   $P$ Lagrangian antibranes on a external edge of the  $\Omega\sigma^{Sp}_+$ orientifold of the resolved conifold, as in   Figure \ref{fig:rowsdual}. To illustrate this, let us give a detailed description of all the terms in \eqref{eq:SU2N-odd-rows}. 

\begin{itemize}
    \item \underline{\textbf{Background}.}  The first term in \eqref{eq:SU2N-odd-rows} is the closed string instanton partition function for the $\Omega\sigma_+^{Sp}$ background but with shifted flux: 
    \beq 
 {\frac{\widehat{t}}\gs} = 2N \underbrace{+1}_{O^+}  \underbrace{-2P}_{P\,  \rm antibranes}\,.
    \eeq
The shifted flux is consistent with a stack of $P$   antibranes/image pairs: each of them decreases the total flux by exactly 1 unit of $g_s$, and in the orientifold projection, each brane at $y_i$ is paired with its image at $\widehat{t_i}+y_i$  as in  Figure \ref{fig:rowsdual} and \ref{fig:image}. Hence the total flux is lowered by $2P$ units of $g_s$.(analogously $N\to N-P$).
\begin{figure}
  \centering
  \begin{subfigure}[t]{0.32\textwidth}
    \centering
   \includegraphics[width = 1.2\linewidth]{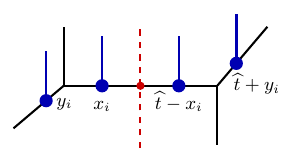}
    \caption{$\Omega\sigma_+$ covering space: brane and its image}
    \label{fig:image}
  \end{subfigure}
  \hfill
  \begin{subfigure}[t]{0.32\textwidth}
    \centering
    \includegraphics[width = 1.2\linewidth]{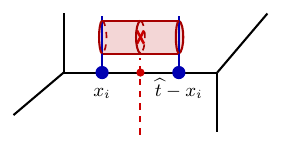}
    \caption{M\"obius strip amplitude between a brane and its image in covering space}
    \label{fig:mobius}
  \end{subfigure}
    \hfill
    \begin{subfigure}[t]{0.32\textwidth}
    \centering
    \includegraphics[width = 1.2\linewidth]{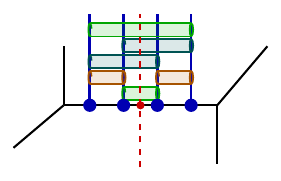}
    \caption{Annuli between branes and image branes}
\label{fig:annuli}
  \end{subfigure}
\caption{Branes and their image in the double cover of the $\Omega\sigma_+$ projection. In Figure \ref{fig:image}, a brane and its plane reflection image w.r.t.\ the $O$ plane. In Figure \ref{fig:mobius} the M\"obius strip amplitude (overlap with the crosscap state). In Figure \ref{fig:annuli}, the various annuli configurations for 2 branes and their paired images. In \textcolor{orange!65!black} {orange} the equal magnitude annuli $[\mathcal B$]-[$\mathcal B$] and {[Image $\mathcal B$]} - {[Image $\mathcal B$] }in \eqref{B-B}; in {\color{teal!65!black}   teal} the annuli [Image $\mathcal B$]-[$\mathcal B$] and [$\mathcal B$]-[Image $\mathcal B$]; in {\color{green!65!black} green} the annuli between a brane and its own image in \eqref{annulusitself}. }
  \label{fig:disks}
\end{figure}

\item \underline{\color{blue}\textbf{Open string.}} Each of the summand referred to as ``{\color{blue}Open string}'' in \eqref{eq:SU2N-odd-rows} is an open topological string partition function associated to strings ending on the brane at position $y_i$. To see this, let us first show this for the case of the oriented string theory. The open topological string amplitude for a single {\color{red}brane} in the internal leg or an {\color{blue}antibrane} in the external leg at position $p_i$ of the resolved conifold is given by:
\beq 
\label{disks}   
\sum_{m\geq 1} \frac{e^{-m p_i}}{m[m]} \;{\mathbin{\substack{{\color{red}\scalebox{1.05}{$+$}}\\[-1ex]{\color{blue}\scalebox{1.05}{$-$}}}}} \;\sum_{m\geq 1} \frac{e^{-m (\widehat{t}  \, {\mathbin{\substack{{\color{red}\scalebox{.9}{$-$}}\\[-0.9ex]{\color{blue}\scalebox{.9}{$+$}}}}}\, p_i)}  }{m[m]}, \qquad \widehat{t} = t\,  {\mathbin{\substack{{\color{red}\scalebox{1.05}{$+$}}\\[-1ex]{\color{blue}\scalebox{1.05}{$-$}}}}}  \, g_s\,.
\eeq
They correspond to the resummation of the all-genus, one-boundary contributions from open strings ending on the (anti-)brane in the resolved conifold geometry as in Figure \ref{fig:disk-a} (internal) and \ref{fig:disk-b} (external). The exponentials $e^{-m p_i}$ are just the exponentials of the worldsheet areas corresponding to non-constant holomorphic maps. For an internal brane at $p_i = x_i$, these worldsheet span an area $x_i$ and $\widehat{t}- x_i$ as in Figure \ref{fig:disk-a}, while for a brane in the external edge at $p_i=y_i$ the areas are $y_i$ and $\widehat t+y_i$, as shown  in Figure \ref{fig:disk-b}. The relative $-$ sign for the external case, is to be attributed to the nodal point of the worldsheet.

In the covering space of the $\Omega\sigma_+^{Sp}$ orientifold, each of the contributions from a brane at $y_i$   is halved (double cover), but paired with its equal in magnitude coming from the image at $\widehat{t}+y_i$. Hence for $P$ external brane we exactly land on the term in the second bracket of \eqref{eq:SU2N-odd-rows}. 
\item \underline{\textbf{Annuli}}. Let us now describe the annuli terms in \eqref{eq:SU2N-odd-rows} in {\color{orange!65!black}orange} and {\color{teal!65!black}teal} there. These are annuli amplitudes for string stretching between the distinct branes in the geometry.  An annulus stretching between two (anti)branes at $p_i$ and  $p_j$ contributes:
\beq
\D(p_i-p_j)\;=\;\sumall{\frac{e^{-m(p_i-p_j)}} {m}}\;=\;-\log\!\left(1-e^{-(p_i-p_j)}\right),
\label{annulusdef}
\eeq

All possible annuli between the branes, between the image branes, and between the branes
and their images contribute to the topological string amplitude. However, in the covering
space for \(\Omega\sigma\) backgrounds, the contribution from these oriented annuli is
halved, \( \mathcal{A}\to \frac{1}{2}\mathcal{A}\).  The annuli stretching between each pair {\color{orange!65!black}[$\mathcal{B}$]-[$\mathcal{B}$]}  exactly equal the ones between their  images {\color{orange!65!black}[Image $\mathcal{B}$]--[Image $\mathcal{B}$]} (these are also in  {\color{orange!65!black}orange } in Figure \ref{fig:annuli}). Same happens for the {\color{teal!65!black}[Image $\mathcal{B}$]--[$\mathcal{B}$]} and the  {\color{teal!65!black}[$\mathcal{B}$]--[Image $\mathcal{B}$]} annuli: 
\begin{align}\label{B-B}
\underbrace{-\frac{1}{2}\sum_{m\geq 1} \frac{e^{-m (x_i - x_j)}}{m}}_{\rm \color{orange!65!black}\text{[$\mathcal B$]-[$\mathcal B$]}} - \underbrace{\frac{1}{2}\sum_{m\geq 1} \frac{e^{-m (\widehat t - x_j - \widehat t + x_i)}}{m}}_{\rm \color{orange!65!black}\text{[Image  $\mathcal B$]-[Image  $\mathcal B$]}} &= -\mathcal{A}(x_i-x_j)\\ \label{IB-IB}
\underbrace{-\frac{1}{2}\sum_{m\geq 1} \frac{e^{-m (\widehat t - x_i - x_j)}}{m}}_{\rm \color{teal!65!black}\text{Image $\mathcal B$-$\mathcal B$}} \underbrace{-\frac{1}{2}\sum_{m\geq 1} \frac{e^{-m (\widehat t - x_j -  x_i)}}{m}}_{\rm \color{teal!65!black}\text{[$\mathcal B$]- [Image  $\mathcal B$]}} &= -\mathcal{A}(\widehat t - x_i-x_j)
\end{align}
The same formulas apply to the case of an antibranes at $y_i$ and $y_j$ and their corresponding images at $\that + y_i$ and $\that + y_j$ with the same overall sign due to $\braket{\bar {\mathcal B}}{\bar{ \mathcal B}} = \braket{\mathcal B}{\mathcal B}$, while annuli between a brane and an antibrane or between their images have a relative sign flip $\braket{\mathcal B}{\bar{\mathcal B}} = -\braket{\mathcal B}{\mathcal B}$. This is represented in Figure \ref{fig:annuli}.\\
In \eqref{eq:SU2N-odd-rows} the annuli terms correspond exactly to the annuli stretching between all distinct pairs of antibranes at $y_i$ and $y_j$ and their images at $\widehat{t}+y_i$ and $\widehat{t}+y_j$, except the annuli between a brane and its own image. This contribution is absent, as is clear from the absence of the \(i=j\) terms in the
sum in \eqref{eq:SU2N-odd-rows}. The reason is subtle and it is due to the specific $\Omega\sigma^{Sp}_+$ orientifold projection, and it is therefore a strong test of our duality. The annulus between a (the \textcolor{green!65!black} {green} annuli in Figure \ref{fig:annuli}) a {\color {red}brane} (minus sign), or {\color{blue} antibrane} (plus sign) and its own image contributes:
\beq  \label{annulusitself}
-\frac{1}{2}\sum_{m\geq 1} \frac{e^{-m (\widehat t {\mathbin{\substack{{\color{red}\scalebox{.8}{$-$}}\\[-0.75ex]{\color{blue}\scalebox{.8}{$+$}}}}} 2p_i)}}{m} = -\frac{1}{2} \mathcal{A}(\widehat t {\mathbin{\substack{{\color{red}\scalebox{.9}{$-$}}\\[-0.9ex]{\color{blue}\scalebox{.9}{$+$}}}}} 2p_i)
\eeq
In the unoriented case, there is also an M\"obius strip amplitude,  the overlap of a brane at $p_i$ with the crosscap state (see Figure \ref{fig:mobius}). These unoriented contributions in the $\Omega\sigma^{{\color{red}SO/Sp}}_{\color{blue}\pm}$ projection have the amplitude \footnote{By $\braket{B}{C}$ we indicate the corresponding contribution to the partition function  $e^{-\braket{B}{C}}$}:
 \beq
 \label{eq:mobius}-\braket{\color{green!50!black}\mathcal B/\bar{\mathcal B}}{C_{{\color{blue}\pm}}^{\color{red}SO/Sp}}  = -\underbrace{\color{red}(\mp)}_{SO/Sp} \underbrace{\color{green!50!black}(\pm)}_{\mathcal B /\bar{ \mathcal B}}
\sum_{m\, \rm\color{blue} even/odd} \frac{e^{- m(\widehat t -2p_i)/2}}{m}
\eeq
There are two sources of relative signs: the crosscap state $\braket{\mathcal B}{C_\pm^{SO}}$ in the $\Omega\sigma^{SO}_\pm$ ($O^-$ plane) has the opposite sign to the one in the $\Omega\sigma^{Sp}_\pm$ projection ($O^+$ plane), and the usual relative brane/antibrane sign. Furthermore, the orientifold projection determines whether the allowed degrees of the equivariant maps are even or odd.\\
In this case, we have $\braket{\bar{\mathcal B}}{C_{{-}}^{Sp}}$ which combines with the annulus to give:
\beq 
-\overbrace{\frac{1}{2} \sum_{m\geq 1 } \frac{e^{-m (\widehat t + 2y_i )}}{m}}^{\rm{ \color{green!65!black} Annulus}} - \overbrace{\underbrace{(-1)}_{\bar{\mathcal B}} \underbrace{(+1)}_{ Sp} \sum_{m\, \rm even}  \frac{e^{-m (\widehat t + 2y_i )/2}}{m}}^{\rm{\color{red} \text{M\"obius}} } = 0 ,
 \eeq
Therefore, the self-image annulus cancels with the M\"obius contribution and indeed does not appear in \eqref{eq:SU2N-odd-rows}. 
\end{itemize}

This detailed discussion elucidates how the various terms in \eqref{eq:SU2N-odd-rows} derived from the twisted   $S$-matrix perfectly match their dual description in topological string theory as external antibranes see Figure \ref{fig:rowsdual}. 
For later convenience, we package the answer as: 
\beq 
\eqref{eq:SU2N-odd-rows}= \log S^{(\mathsf C,1)}_{a0}
\;\doteq\;
\mathcal F^0(N-P)
+\mathcal F^{\rm inst}_{\Omega\sigma^{Sp}_+}\left( \widehat t \, \right)
+\SymP(y;\that), \qquad e^{-\that}=q^{2N+1-2P}
\eeq
where:
\beq \label{antibranesblock}\SymP(y;\that)
=\sum_{i=1}^{P}\underbrace{\sumall{e^{-my_i}-\frac{e^{-m(\that+y_i)}} {m[m]}}}_{\color{blue} \rm Open \, string}
 -\sum_{1\le i<j\le P}\Big[\underbrace{\D(y_i-y_j)}_{\color{teal!65!black}\rm Annulus\, \, \bar{B}-\bar{B}}+\underbrace{\D(\that+y_i-y_j)}_{\color{teal!65!black}\rm Annulus\, \, \bar{B}-Image \,\bar{B}}\Big].
\eeq
is the  $P$ antibrane amplitude excluding the self-annuli term.

\paragraph{Column description.}
Dually, the same defect $\mathsf{M}_{a}$ can be described by a Young tableau with $M$ columns of length $R_i^T$ as in \eqref{fig:tabb}. 

\begin{figure}
    \centering
    \includegraphics[width=0.5\linewidth]{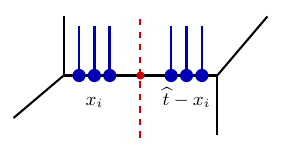}
    \caption{Branes ending on the internal edge at position $x_i = \gs\Big(R^T_i-i+M+\tfrac12\Big)$ of the $\Omega\sigma_+$ geometry. These are dual to generic representations described as $P$ rows of length $R_i$ }
    \label{fig:columnsdual}
\end{figure}

In the column representation of Young tableau corresponding to $a$,  columns behave like \emph{holes}:  all but $M$ sites 
$\{1,2,\ldots,N+M\}$ are filled as:
\begin{equation}
\{1,2,\ldots,N+M\}
=\underbrace{\{z_i=R_i+N+1-i\}_{i=1}^{N}}_{\text{Background}}
\;\sqcup\;
\underbrace{\{h_i=N+i-R^{T}_i\}_{i=1}^{M}}_{\text{Holes}}\,.
\end{equation}
The computation proceeds as before, the result is:
\begin{equation}
\begin{aligned}
\log S^{(\mathsf C,1)}_{R,0}
&\;\doteq\;
\mathcal F^0(N-M)
+\mathcal F^{\rm inst}_{\Omega\sigma^{Sp}_+}\left( \widehat t \, \right)
+\AntiP(x;\that)-
\underbrace{\color{blue}\sum_{i=1}^{M}\D(\that-2x_i)}_{\rm{\color{green!65!black} Annulus} \oplus \color{red}\text{M\"obius} }, \quad e^{-\that}=q^{2N+1+2M}
\label{eq:SU2N-odd-cols}
\end{aligned}
\end{equation}
where, as in \eqref{antibranesblock} we have packaged the terms: \beq\AntiP(x;\that)
 =\sum_{i=1}^{M}\underbrace{\sumall{\frac{e^{-mx_i}+e^{-m(\that-x_i)}} {m[m]}}}_{\color{blue} \rm Open \, string}
 -\sum_{1\le i<j\le M}\Big[\underbrace{\D(x_i-x_j)}_{\color{orange!65!black}\rm Annulus\, \, B-B}+\underbrace{\D(\that-x_i-x_j)}_{\color{orange!65!black}\rm Annulus\, \, B-Image\,  B}\Big],
 \eeq
into a brane block excluding the self-annuli, and 
\beq
x_i = \gs\Big(R^{T}_i-i+M+\tfrac12\Big),\qquad i=1,\ldots,M\,.
\eeq
This we recognize as the topological string amplitude for $P$ branes at position $x_i$ (and their image at $\widehat{t}-x_i$ ) in the $\Omega\sigma^{Sp}_+$ background as in Figure \ref{fig:columnsdual}. Indeed, the flux: \beq
\widehat{t}/\gs = 2N \underbrace{+1}_{O^+} + \underbrace{2M}_{M \rm  branes}\,,
\eeq
as well the open string amplitude (cf. with \eqref{disks}) and the annuli (cf. with \eqref{B-B} and \eqref{IB-IB}) perfectly match the description as $P$ internal edge branes in $\Omega\sigma^{Sp}_+$. 
The only new term is  the last one in {\color{blue} blue} in \eqref{eq:SU2N-odd-cols}. This we can identify as the sum of the annulus between each brane and its image, and the unoriented M\"obius amplitudes of branes and images: 
\beq 
-\underbrace{\frac{1}{2}\sum_{m\geq1}\frac{e^{-m(\that-2x_i)}}{m}}_{\text{\color{green!65!black}Annulus}}
-\underbrace{\underbrace{(+1)}_{\mathcal{B}}\underbrace{(+1)}_{Sp}
\sum_{m\,{\rm even}}\frac{e^{-m(\that-2x_i)/2}}{m}}_{\text{\color{red}M\"obius}}
\Bigg] = -
\underbrace{\color{blue}\sum_{i=1}^{M}\D(\that-2x_i)}_{\rm{\color{green!65!black} Annulus} \oplus \color{red}\text{M\"obius} }.
\eeq
The two contributions now add, rather than cancel, because the change
\(\mathcal{B}\to\bar{\mathcal{B}}\), changing their relative sign.
This combined term multiplies the partition function by,
\[
\exp[-\mathcal A(\that-2x_i)]=1-e^{-(\that-2x_i)}
\]
for each brane-image pair.
This factor vanishes as $x_i\to \that/2$, implying that brane and its image cannot
meet at the fixed locus of $\Omega\sigma^{Sp}_+$. This is consistent with the lack of a representation that inserts a stuck brane at the fixed locus of this symplectic projection.

\subsubsection{\boldmath  Monodromy defects for \texorpdfstring{$SU(2N)_k$, $k$ even}{SU(2N) k even}}
For $2N>k+2$, as discussed around Table~\ref{tab:minimizers}, for even level the
trivial label is \emph{not} the vacuum: the minimizer of
$S^{(\mathsf C,1)}_{a0}$ sits at the opposite endpoint 
$a_\ast=(0,\ldots,0,\tfrac k2)$. 
The instanton partition function for this vacuum monodromy defect is,
\beq
\mathcal F^{\rm inst}_{\Omega\sigma^{SO}_+}(e^{-t})
=-\frac12\sumall{\frac{e^{-mt}}{m[m]^2}}
 -\sumeven{\frac{e^{-mt/2}} {m[m]}} ,\qquad e^{-t}=q^{2N-1}\,.
\eeq
and the prefactor given by, 
\[
\mathcal{F}^0(N)=\frac{1}{2}\log\frac{M(q)}{\zeta(q)^{2N-1}\,\zeta(q^{2})}\,, 
\]
It is convenient to parametrize the representations $a$   by complementary partitions $r$ measured from the vacuum   $a_\ast$:
\begin{equation}
r_i = \frac{k}{2} - R_{N+1-i}, \qquad i = 1,2,...,N
\label{eq:endpoint-dictionary}
\end{equation}
where $R_i = \sum_{j=i}^{N} a_j$ are the partitions of a representation $a$. Using the complementary Young tableau $r$ \footnote{Note that \(r_N\) is not positive for every representation \(a\). However, the simple current automorphism \(a_0\leftrightarrow a_1\) sends \(r_N\to -r_N\). Therefore, without loss of generality, we can work with the representative for which \(r_N\ge0\), or equivalently with the partition
\[
(r_1,r_2,\ldots,|r_N|).
\]
which is weakly decreasing. } and expanding the modular  $S$-matrix leads to the same formula as \eqref{eq:su2nodd-weyl-R}, but with $\widetilde x_j=r_j+N-j$.
\paragraph{Column description.}
By plugging this into the modular $S$-matrix we find:
\begin{equation}
\log S^{(\mathsf C,1)}_{a0}
\;\doteq\;
\mathcal F^0(N-M)
+\mathcal F^{\rm inst}_{\Omega\sigma^{SO}_+}\left( \widehat t \, \right)
+\AntiP(x;\that), \qquad e^{-\that}=q^{2N-1+2M}\,,
\label{eq:SU2N-even-cols}
\end{equation}
where, for an $r$ with $M$ columns, the internal branes sit at
\[
x_i= g_s(r_i^T - i  + M  + \tfrac{1}{2})
\]
The annulus between the brane at $x_i$ and its own image cancels against the
M\"obius strip amplitude,
\beq
-\overbrace{\frac{1}{2} \sumall \frac{e^{-m (\widehat t - 2x_i )}}{m}}^{\rm{ \color{green!65!black} Annulus}} - \overbrace{\underbrace{(+1)}_{\mathcal B} \underbrace{(-1)}_{ SO} \sumeven  \frac{e^{-m (\widehat t - 2x_i )/2}}{m}}^{\rm{\color{red} \text{M\"obius}} } = 0 .
\eeq
Therefore, we see this is again perfectly consistent with the proposed duality.
This cancellation allows for the brane and its image to sit on top of the fixed locus of $\Omega \sigma^{SO}_{+}$.\\
Consider the brane at $x_1$,
\beq
x_1=\frac{\that}{2}-\big(N-r^{T}_1\big)\gs ,
\label{eq:su2neven-x1}
\eeq
this reaches $\that/2$ when $r$ has a full first column, $r^{T}_1=N$.  This is a  brane  at the orientifold fixed locus, coincident with its
own image.  
More generally the number of full-height columns of $r$ is its last Dynkin
label $a^{r}_N=r_N$, and these are dual to 
branes at positions 
\beq
x_i=\frac{\that}{2}-(i-1)\gs,\qquad i=1,\ldots,r_N ,
\eeq
independently of the rest of $r$. We also note, that the stuck locus is empty at the fixed points of the simple current automorphism of $A^{(2)}_{2N-1}$ (see Table \ref{tab:twisted-affine} )
 $a_0=a_1$.

\paragraph{Row description.} The dual external antibrane description for a representation with $P$ rows is, 
\begin{equation}
\log S^{(\mathsf C,1)}_{a0}
\;\doteq\;
\mathcal F^0(N-P)
+\mathcal F^{\rm inst}_{\Omega\sigma^{SO}_+}(\widehat t\, )
+\SymP(y;\widehat t )
\;-\;\underbrace{\color{blue}\sum_{i=1}^{P}\D(\that+2y_i)}_{\rm{\color{green!65!black} Annulus} \oplus \color{red}\text{M\"obius} }\,,
\label{eq:SU2N-even-rows}
\end{equation}
where
\beq
y_i=\gs(r_i-i+P+\tfrac12)\,,
\eeq
where    $e^{-\that}=q^{2N-1-2P}$.
Here the overlap of an antibrane with the crosscap state carries the
opposite sign, so the self-image annulus and the M\"obius strip add up
instead of cancelling.
\beq
-\overbrace{\frac{1}{2} \sum_{m\geq 1 } \frac{e^{-m (\widehat t + 2y_i )}}{m}}^{\rm{ \color{green!65!black} Annulus}} - \overbrace{\underbrace{(-1)}_{\bar{\mathcal B}} \underbrace{(-1)}_{{SO}} \sum_{m\, \rm even}  \frac{e^{-m (\widehat t + 2y_i )/2}}{m}}^{\rm{\color{red} \text{M\"obius}} } = -{\color{blue}\mathcal{A} (\widehat t + 2y_i) } ,
\eeq
We exclude the case of $P=N$ as the flux in the internal edge becomes negative.

\paragraph{\boldmath Case $2N \leq k$.}
For $2N\leq k$ the trivial rep $(0,\cdots,0)$ is the vacuum, and the background is $\Omega\sigma_+^{Sp}$. The non-vacuum representations are labelled by the usual Young tableau $R$ measured from the trivial corner, as in for $SU(2N)_{k \, \rm odd}$, with the same topological string amplitudes as Section~\ref{sec:su2Nodd}.  Thus, we will not repeat it here again.
\footnote{At the wall $2N=k+2$ the trivial, the
simple current and the endpoint corners are threefold degenerate; this
is the level-rank self-dual point, where
$e^{-t}=q^{2N-1} = e^{-2\pi i (2N-1)/(2N +2N-2)} =-1$.}

\subsubsection{\boldmath Monodromy defects for \texorpdfstring{$SU(2N+1)_k$, $k$ even}{SU(2N+1) k even}}

The vacuum is the endpoint $a_\ast=(0,\ldots,0,\tfrac k2)$, with
flux $e^{-t}=q^{2N}$ and pure orientifold background
\[
\mathcal F^{\rm inst}_{\Omega\sigma^{SO}_+}(t)
=-\frac12\sumall{\frac{e^{-mt}} {m[m]^2}}
 -\sumeven{\frac{e^{-mt/2}} {m[m]}} ,
\]
with no stuck brane. With the prefactor given by, 
\[ \mathcal{F}^0(N)=\log\frac{M(q)^{1/2}\zeta(q^{2})^{1/2}}{\zeta(q)^{\,N+1/2}}\,,  
\]
Parameterizing the representation as \eqref{eq:endpoint-dictionary} measured from $a_\ast$,\footnote{For any representation $a$ of $A^{(2)}_{2N}$, $r_i$ are automatically weakly decreasing and non-negative, therefore can be thought of as a Young tableau. Unlike the \(SU(2N)_k\) case, there is no simple current automorphism to choose a representative.}
defects are encoded in the complement partitions $r$ and the positions $x_i$ and $y_i$ below are built from column lengths $r^T_i$ and row lengths $r_i$ exactly as above.

\paragraph{Column description.}
For $r$ with $M$ columns, we have the internal edge brane description with the flux is shifted to
$e^{-\that}=q^{2N+2M}$ and
\begin{equation}
\log S^{(\mathsf C,1)}_{a0}
\;\doteq\;
\mathcal F^0(N-M)
+\mathcal F^{\rm inst}_{\Omega\sigma^{SO}_+}\left( \widehat t \, \right)
+\AntiP(x;\that).
\label{eq:SU2Np1-even-cols}
\end{equation}
As in the $SU(2N)_{k\,\rm even}$ case, the annulus between a brane and its own image cancels against their M\"obius strip amplitude \footnote{The representation $a=(0^{N})$ has  description as the background $\Omega \sigma^{SO}_+ \oplus \mathcal{B}$},
\beq
-\overbrace{\frac{1}{2} \sum_{m\geq 1 } \frac{e^{-m (\widehat t - 2x_i )}}{m}}^{\rm{ \color{green!65!black} Annulus}} - \overbrace{\underbrace{(+1)}_{\mathcal{B}} \underbrace{(-1)}_{SO} \sum_{m\, \rm even}  \frac{e^{-m (\widehat t - 2x_i )/2}}{m}}^{\rm{\color{red} \text{M\"obius}} } = 0 .
\eeq
\paragraph{Row description.}
For $r$ with $P$ rows, we have the external edge antibrane description with the flux $e^{-\that}=q^{2N-2P}$ and
\begin{equation}
\log S^{(\mathsf C,1)}_{a0}
\;\doteq\;
\mathcal F^0(N-P)
+\mathcal F_{\Omega\sigma^{SO}_+}\left( \widehat t \, \right)
+\SymP(y;\that)
\;-\;\underbrace{\color{blue}\sum_{i=1}^{P}\D(\that+2y_i)}_{\rm{\color{green!65!black} Annulus} \oplus \color{red}\text{M\"obius} }.
\label{eq:SU2Np1-even-rows}
\end{equation}
For the external antibranes the crosscap overlap carries the opposite
sign, so the self-image annulus and the M\"obius strip add up,
\beq
-\overbrace{\frac{1}{2} \sum_{m\geq 1 } \frac{e^{-m (\widehat t + 2y_i )}}{m}}^{\rm{ \color{green!65!black} Annulus}} - \overbrace{\underbrace{(-1)}_{\bar{\mathcal B}} \underbrace{(-1)}_{SO} \sum_{m\, \rm even}  \frac{e^{-m (\widehat t + 2y_i )/2}}{m}}^{\rm{\color{red} \text{M\"obius}} } = -{\color{blue}\mathcal{A} (\widehat t + 2y_i) } ,
\eeq
leaving the image annulus $\D(\that+2y_i)$.  Everything is completely
analogous to the $SU(2N)_{k\,\rm even}$ case.\footnote{Again we exclude the case of $P = N$, as the flux in the internal edge becomes zero}

\subsubsection{\boldmath Monodromy defects for \texorpdfstring{$SU(2N+1)_k$, $k$ odd}{SU(2N+1) k odd}}
The vacuum is the trivial label, with flux $e^{-t}=q^{2N+1}$ and
background $\Omega\sigma^{SO}_+\oplus \mathcal{B}$: an $O^-$-plane
transverse to the edge with a single brane stuck at its fixed locus,
\[
\mathcal F^{\rm inst}_{\Omega\sigma^{SO}_+\oplus \mathcal B}(t)
=-\frac12\sumall{\frac{e^{-mt}} {m[m]^2}}
 -\sumeven{\frac{e^{-mt/2}}{ m[m]}}
 +\sumall{\frac{e^{-mt/2}} {m[m]}} .
\]
With the prefactor, 
\[
\mathcal{F}^0(N)=\log\frac{M(q)^{1/2}}{\zeta(q)^{\,N-1/2}\,\zeta(q^{2})^{1/2}}\,,  
\]
Excited monodromy defects are labelled by usual  partitions $R$ measured from the
trivial corner.

\paragraph{Column description.}
For $R$ given by $M$ columns the flux is shifted to
$e^{-\that}=q^{2N+1+2M}$ and
\begin{align}
\log S^{(\mathsf C,1)}_{R,0}
\;\doteq\;
\mathcal F^0(N-M)
+\mathcal F_{\Omega\sigma^{SO}_+\oplus \mathcal B}\left( \widehat t \, \right)
+\AntiP(x;\that)
\;-\;\underbrace{\color{blue}\sum_{i=1}^{M}\D\!\left(\tfrac{\that}{2}-x_i\right)}_{\rm{\color{orange!65!black} Annulus}\,[\mathrm{Stuck}\,B]\text{--}[B]}.
\label{eq:SU2Np1-odd-cols}
\end{align}
In the equation above, there is no annulus between the brane at $x_i$ and its image since it exactly cancels with the Mobi\"us strip contribution: 
\beq 
-\overbrace{\frac{1}{2} \sum_{m\geq 1 } \frac{e^{-m (\widehat t - 2x_i )}}{m}}^{\rm{ \color{green!65!black} Annulus}} - \overbrace{\underbrace{(+1)}_{\mathcal{B}} \underbrace{(-1)}_{SO} \sum_{m\, \rm even}  \frac{e^{-m (\widehat t - 2x_i )/2}}{m}}^{\rm{\color{red} \text{M\"obius}} } = 0 .
 \eeq
We also find that there are no representation that places more branes at the fixed locus, this is supported by the fact that the contribution $\mathcal{A}(\tfrac{t}{2} - x)$ makes the amplitude vanish as $x \rightarrow \tfrac{t}{2}$.

\paragraph{Row description.}
For $R$ given by $P$ rows the flux is $e^{-\that}=q^{2N+1-2P}$ and
\begin{equation}
\log S^{(\mathsf C,1)}_{R,0}
\;\doteq\;
\mathcal F^0(N-P)
+\mathcal F^{\rm inst}_{\Omega\sigma^{SO}_+\oplus \mathcal B}\left( \widehat t \, \right)
+\SymP(y;\that)
\;-\;\underbrace{\color{blue}\sum_{i=1}^{P}\D(\that+2y_i)}_{\rm{\color{green!65!black} Annulus} \oplus \color{red}\text{M\"obius} }
\;+\;\underbrace{\color{magenta!65!black}\sum_{i=1}^{P}\D\!\left(\tfrac{\that}{2}+y_i\right)}_{\rm{\color{orange!65!black} Annulus}\,[\mathrm{Stuck}\,B]\text{--}[\bar B]}.
\label{eq:SU2Np1-odd-rows}
\end{equation}
This time, the contribution between a antibrane and the crosscap state  comes with opposite sign and therefore: \beq 
-\overbrace{\frac{1}{2} \sum_{m\geq 1 } \frac{e^{-m (\widehat t + 2y_i )}}{m}}^{\rm{ \color{green!65!black} Annulus}} -\overbrace{\underbrace{(-1)}_{\bar{\mathcal{B}}} \underbrace{(-1)}_{SO} \sum_{m\, \rm even}  \frac{e^{-m (\widehat t + 2y_i )/2}}{m}}^{\rm{\color{red} \text{M\"obius}} } = -{\color{blue}\mathcal{A} (\widehat t + 2y_i) } .
 \eeq
The {\color{magenta!65!black}last term} is the annulus between each pair of antibranes and image antibranes and the stuck brane at $\tfrac{\widehat{t}}{2}$, $\braket{\bar{\mathcal B}}{\mathcal B} = -\braket{\mathcal B}{\mathcal B}$, therefore has a relative sign with the usual annulus contribution.
\beq  
-\underbrace{(-1)}_{\braket{\bar {\mathcal B}}{\mathcal B}}\frac{1}{2}\underbrace{\sum_{m\geq 1 } \frac{e^{-m (\widehat t/2  + y_i )}}{m} }_{\rm{\color{green!65!black} Annulus\, [Stuck \, \mathcal B ]-[\bar{\mathcal B}]}} - 
\underbrace{(-1)}_{\braket{\bar{\mathcal  B}}{\mathcal B}}\underbrace{\frac{1}{2} \sum_{m\geq 1 } \frac{e^{-m (\widehat t  + y_i - \widehat t/2 )}}{m} }_{\rm{\color{green!65!black} Annulus\, [Stuck \, \mathcal B]-[Image \, \bar{\mathcal B}]}} = {\color{magenta!65!black}\mathcal{A} (\widehat t/2 + y_i) }
 \eeq
 \begin{table}[t]
  \centering
  \renewcommand{\arraystretch}{1.45}
  \begin{tabular}{|c|c|c|c|}
  \hline
  Theory & Background & $M$ branes int. & $P$ antibranes ext.\\
  \hline
  \small$SU(2N)_{\scriptstyle k>N+1, \rm even }$
  & $\Omega\sigma^{SO}_+$
  & $\AntiP$
  & $\SymP-\sum_i\D(\that{+}2y_i)$
  \\
  \small$SU(2N)_{k\, \rm odd}$
  & $\Omega\sigma^{Sp}_+$
  & $\AntiP-\sum_i\D(\that{-}2x_i)$
  & $\SymP$
  \\
  \small$SU(2N{+}1)_{k\,  \rm even}$
  & $\Omega\sigma^{SO}_+$
  & $\AntiP$
  & $\SymP-\sum_i\D(\that{+}2y_i)$
  \\
  \small $SU(2N{+}1)_{k\,  \rm odd}$
  & $\Omega\sigma^{SO}_+{\oplus}\,\mathcal B$
  & $\AntiP-\sum_i\D(\tfrac{\that}{2}{-}x_i)$
  & $\SymP-\sum_i\D(\that{+}2y_i)+\sum_i\D(\tfrac{\that}{2}{+}y_i)$
  \\
  \hline
  \end{tabular}
  \caption{Brane content of the four $SU$ monodromy sectors.  $K=M$
  (columns, $+$ sign in the flux) or $K=P$ (rows, $-$ sign).  Each entry
  is to be added to the corresponding background free energy
  $\mathcal F\left( \widehat t \, \right)$ and constant map terms.  In the $k$-even sectors the labels are measured
  from the endpoint vacuum via \eqref{eq:endpoint-dictionary}.}
  \label{tab:SUmono-branes}
  \end{table}

\subsubsection{\boldmath Summary and comparison with $\Omega\sigma_-$ projection}
These results provide a strong confirmation of our proposed duality. In particular all the cancellations between M\"obius and annulus that we have observed by directly expanding the modular $S$-matrix (summarized in Table \ref{tab:SUmono-branes}), perfectly match the 
signs of the crosscaps in the $\Omega\sigma_+$ projection. 
It is instructive to compare these results to the general representations in $Spin$/$Sp$ theories. These, as explained in Appendix \ref{App:su-so-sp-branes-geometry}, correspond to branes in the $\Omega\sigma^{SO/Sp}_-$ projection as we illustrate in Table \ref{tab:Binternal-four}. In the $\Omega\sigma_-$ case, there is no M\"obius annulus cancellations (like in the $\Omega\sigma_+$ case), as the unoriented contribution are from equivariant maps of degree odd.
 \begin{table}[ht]
 { \centering\resizebox{\textwidth}{!}{%
  {\renewcommand{\arraystretch}{1.9}
  \begin{NiceTabular}{|>{\centering\arraybackslash}m{2.4cm}|>{\centering\arraybackslash}m{2.5cm}|>{\centering\arraybackslash}m{2.1cm}>{\centering\arraybackslash}m{3.2cm}>{\centering\arraybackslash}m{5.8cm}|}
    \hline
    Branes & Background & \multicolumn{3}{c|}{Contribution} \\
    \cline{3-5}
    & & Universal & Annulus & M\"obius \\
    \hline\hline
    \Block{2-1}{Branes internal leg}
    & $\Omega\sigma^{{ \color{blue}SO}/{\color{red}Sp}}_+$
    & $\mathcal{B}$
    & $\displaystyle
    -{\frac{1}{2} \sum_{m\geq 1 } \frac{e^{-m (\widehat t - 2x_i )}}{m}}$
    & $\displaystyle
    ({\mathbin{\substack{{\color{red}\scalebox{1.05}{$+$}}\\[-1ex]{\color{blue}\scalebox{1.05}{$-$}}}}}1)(-1)  {\sum_{{\rm \textbf{even}} } \frac{e^{-m (\widehat t - 2x_i )/2}}{m}}$
    \bigstrut \\
    & $\Omega\sigma^{ {{ \color{blue}SO}/{\color{red}Sp}}}_-$
    & $\mathcal{B}$
    & $\displaystyle
    -\frac{1}{2} \sum_{m\geq 1 } \frac{e^{-m (\widehat t - 2x_i )}}{m}$
    & $\displaystyle
    ({\mathbin{\substack{{\color{red}\scalebox{1.05}{$+$}}\\[-1ex]{\color{blue}\scalebox{1.05}{$-$}}}}}1)(-1){\sum_{\rm \textbf{odd}}  \frac{e^{-m (\widehat t - 2x_i )/2}}{m}}$
    \bigstrut \\
    \hline\hline
    \Block{2-1}{ Antibranes external leg}
    & $\Omega\sigma^{{ \color{blue}SO}/{\color{red}Sp}}_+$
    & $\overline{\mathcal{B}}$
    & $\displaystyle
    -{\frac{1}{2} \sum_{m\geq 1 } \frac{e^{-m (\widehat t + 2y_i )}}{m}}$
    & $\displaystyle
({\mathbin{\substack{{\color{blue}\scalebox{1.05}{$+$}}\\[-1ex]{\color{red}\scalebox{1.05}{$-$}}}}}1)(-1)  {\sum_{{\rm \textbf{even}} } \frac{e^{-m (\widehat t + 2y_i )/2}}{m}}$
    \bigstrut \\
    & $\Omega\sigma^{ {{ \color{blue}SO}/{\color{red}Sp}}}_-$
    & $\overline{\mathcal{B}}$
    & $\displaystyle
    -\frac{1}{2} \sum_{m\geq 1 } \frac{e^{-m (\widehat t + 2y_i )}}{m}$
    & $\displaystyle
    ({\mathbin{\substack{{\color{blue}\scalebox{1.05}{$+$}}\\[-1ex]{\color{red}\scalebox{1.05}{$-$}}}}}1)(-1){\sum_{\rm \textbf{odd}}  \frac{e^{-m (\widehat t + 2y_i )/2}}{m}}$
    \bigstrut \\
  \hline
    \end{NiceTabular}}}}
  \caption{ Comparison between $\Omega\sigma_+$ and $\Omega\sigma_-$ background. The latter are dual to $Spin$/$Sp$ Chern--Simons theory, cf.\ Appendix \ref{App:su-so-sp-branes-geometry} . }
   \label{tab:Binternal-four}
   \end{table}

\subsection[Topological vertex for \texorpdfstring{$\Omega\sigma_+$}{Omegasigma+}orientifold projection ]{\boldmath Topological vertex for \texorpdfstring{$\Omega\sigma_+$}{Omegasigma+} orientifold projection}
The topological vertex \cite{Aganagic:2003db} is a crucial computational tool providing a solution for closed and open topological strings on toric Calabi--Yau geometries.
The elementary building block is the vertex $C_{R_1R_2R_3}$ associated with a $\mathbb{C}^3$ patch (the trivalent vertex of the toric web); combinatorially it is the generating function of plane partitions -- the three-dimensional analogue of Young diagrams -- counting boxes stacked in a corner of $\mathbb{R}^3_{\ge0}$ with the three asymptotic faces fixed by the representations $R_1,R_2,R_3$.
The resolved conifold is assembled by gluing two such vertices along a shared internal edge. Gluing amounts to summing over all partitions (Young diagrams) flowing on that edge, weighted by the edge propagator $(-Q)^{|R|}$, with $Q=e^{-t}$ the K\"ahler parameter. The partition function  is the instance in which neither external nor internal edges are decorated by a nontrivial representation, so that
\beq
\mathcal Z^{\rm inst}_{X} \;=\; \sum_R (-Q)^{|R|}\, C_{00R}\,C_{00R^{T}}
\;=\; \prod_{i,j\ge 1}\bigl(1-Q\,q^{\,i+j-1}\bigr),
\label{eq:Zcon}
\eeq
the last equality following from the dual Cauchy identity with $C_{00R}=s_R(q^\rho)$, $(q^\rho)_i=q^{\,i-1/2}$. Decorating the external legs reconstructs the amplitudes with branes ending on the geometry.

The vertex has been extended to a real version in the presence of an antiholomorphic involution of the target ~\cite{Krefl:2009md,Krefl:2009mw}, dual to orientifold planes (and branes) in the geometry. For the conifold the relevant involutions act on the base $\mathbb{P}^1$; the freely-acting one, $z\to-1/\bar z$, has quotient $\mathbb{RP}^2$ and acts on the internal edge as the transpose $R\to R^{T}$, so that the surviving configurations are self-conjugate. Its real vertex is the $Spin$/$Sp$ amplitude
\beq
\mathcal Z^{\rm inst}_{\Omega\sigma^{SO/Sp}_-} \;=\; \sum_{R=R^{T}} (-1)^{(|R|\mp r_R)/2}\, Q^{|R|/2}\, C_{00R}
\;=\; \prod_{i\ge1}\bigl(1\pm\sqrt{Q}\,q^{\,i-1/2}\bigr)\prod_{i<j}\bigl(1-Q\,q^{\,i+j-1}\bigr),
\label{eq:realvertex}
\eeq
where $r_R$ is the number of boxes on the diagonal of $R$ and the upper (lower) sign gives $SO$ ($Sp$). Yet the vertex technology reproducing the monodromy-defect dual that is  the geometry with a transverse $O$-plane carrying no extra stacked brane, was not developed there. Here we fill this gap and provide a set of vertex rules for the $\Omega\sigma_+$ geometry. The monodromy-defect dual is the same involution $z\to 1/\bar z$  but stripped of that brane, and it is a genuinely different amplitude. In Frobenius coordinates $R=(\alpha_1\cdots\alpha_d\,|\,\beta_1\cdots\beta_d)$ this restricts the edge representation to the shifted self-conjugate families $\alpha_i=\beta_i+1$ ($Sp$) or $\beta_i=\alpha_i+1$ ($SO$) having even number of boxes. The $\Omega\sigma_+$ vertex rule is therefore
\beq
\;\mathcal Z^{\rm inst}_{\Omega\sigma^{Sp}_+}=\!\!\sum_{R:\,\alpha_i=\beta_i+1}\!\!(-1)^{|R|/2}\,Q^{|R|/2}\,C_{00R}\,,
\qquad
\mathcal{Z}^{\rm inst}_{\Omega\sigma^{SO}_+}=\!\!\sum_{R:\,\beta_i=\alpha_i+1}\!\!(-1)^{|R|/2}\,Q^{|R|/2}\,C_{00R}
\label{eq:sigmaplusvertex}
\eeq
the weight $(-1)^{|R|/2}Q^{|R|/2}=\sqrt{(-Q)^{|R|}}$ being the square root of the conifold edge propagator in \eqref{eq:Zcon} (both families have $|R|$ even, so the weight is real).

\paragraph{Partition function.}
We evaluate \eqref{eq:sigmaplusvertex} in closed form. Using $C_{00R}=s_R(q^\rho)$ and the homogeneity of the Schur function, $Q^{|R|/2}s_R(q^\rho)=s_R(x)$ with
\beq
x_i \;\equiv\; \sqrt{Q}\,q^{\,i-1/2}\,, \qquad x_i x_j \;=\; Q\,q^{\,i+j-1}\,,
\eeq
so that \eqref{eq:sigmaplusvertex} is a signed sum of Schur functions over the shifted self-conjugate families. These are resummed by the (signed) Littlewood identities ~\cite{Littlewood:1950,Macdonald:1995},
\beq
\sum_{R:\,\alpha_i=\beta_i+1}(-1)^{|R|/2}\,s_R(x)=\prod_{i\le j}\bigl(1-x_i x_j\bigr)\,,
\qquad
\sum_{R:\,\beta_i=\alpha_i+1}(-1)^{|R|/2}\,s_R(x)=\prod_{i< j}\bigl(1-x_i x_j\bigr)\,,
\label{eq:littlewood}
\eeq
Equivalently, isolating the unpaired mode of the fixed $S^1$ exhibits the defect amplitude as the real vertex \eqref{eq:realvertex} dressed by the boundary factor of that mode,
\beq
\mathcal Z^{\rm inst}_{\Omega\sigma^{Sp}_+}\;=\;\Bigl[\prod_{i\ge1}\bigl(1-\sqrt{Q}\,q^{\,i-1/2}\bigr)\Bigr]\, \mathcal Z^{\rm inst}_{\Omega\sigma^{SO}_-}\,,
\qquad
\mathcal Z_{\Omega\sigma^{SO}_+}\;=\;\Bigl[\prod_{i\ge1}\bigl(1+\sqrt{Q}\,q^{\,i-1/2}\bigr)\Bigr]^{-1} \mathcal Z_{\Omega\sigma^{Sp}_-}\,,
\label{eq:boundaryfactor}
\eeq
the prefactor collecting the unpaired mode and its oscillator descendants on the fixed $S^1$. Inserting $x_i x_j=Q\,q^{\,i+j-1}$ into \eqref{eq:littlewood} gives the closed forms
\beq
\;
\mathcal Z_{\Omega\sigma^{Sp}_+}=\prod_{i\le j}\bigl(1-Q\,q^{\,i+j-1}\bigr)\,,
\qquad
\mathcal Z_{\Omega\sigma^{SO}_+}=\prod_{i< j}\bigl(1-Q\,q^{\,i+j-1}\bigr)\,.\;
\label{eq:sigmaplusZ}
\eeq
The free energy follows by taking the logarithm of \eqref{eq:sigmaplusZ} and resumming
\beq
\log \mathcal Z^{\rm inst}_{\Omega\sigma^{SO/Sp}_+} =\mathcal{F}^{\rm inst}_{\Omega\sigma^{SO/Sp}_+}
\;=\; -\frac12\sum_{n\ge1}\frac{Q^{\,n}}{n\,[n]^2}\;\mp\!\!\sum_{ n\, \text{even}}\!\frac{Q^{\,n/2}}{n\,[n]} =  -\frac12\sum_{n\ge1}\frac{e^{-n t }}{n\,[n]^2}\;\mp\!\!\sum_{ n\, \text{even}}\!\frac{e^{-n t/2}}{n\,[n]} 
\label{eq:sigmaplusF}
\eeq
i.e.\ exactly one half of the oriented conifold free energy together with the \emph{even}-winding contributions of the fixed locus, matching exactly \eqref{eq:free-orientifold-inst-free-energy}.

 \medskip\medskip
\section*{Acknowledgments}

The authors  would like to thank  M. Cheng, D.-E. Diaconescu,  D. Gaiotto, M. Mari\~no, C. Vafa and  J. Walcher    for   discussions. 
Research at Perimeter Institute is supported in part by the Government of Canada through the Department of Innovation, Science and Economic Development Canada and by the Province of Ontario through the Ministry of Colleges and Universities.

\vfill\eject

\appendix 

\section{\boldmath Holography for \texorpdfstring{$SU(N)$, $Spin(N)$ and $Sp(N)$ Chern--Simons }{SU(N), Spin(N) and Sp(N) Chern--Simons}}
\label{App:su-so-sp-branes-geometry}
In this Appendix we will review the dual for different Wilson lines $W_R$ in $SU$/$Spin $/$Sp$ Chern--Simons theory in terms of branes/ antibranes. This has been studied in the literature, this section will be a mix of review of already established results in $SU$ Chern--Simons theory and some new results in $Spin$/$Sp$ which will give more insights into the kind of geometries discussed in the main text, regarding the monodromy defects.  

\subsection[\texorpdfstring{$SU(N)_k$}{SU(N)k} Chern--Simons theory]{\boldmath \texorpdfstring{$SU(N)_k$}{SU(N)k} Chern--Simons theory}
\label{sec:su-branes-geometry}

\begin{figure}[H]
\centering
\resizebox{0.46\textwidth}{!}{%
\begin{tikzpicture}[
    web/.style={line width=1.15pt,line cap=round,line join=round},
    leadingfill/.style={fill=orange!65!red,fill opacity=0.13},
    leadingedge/.style={line width=1.00pt,orange!70!red,line cap=round,line join=round},
    subfill/.style={fill=blue!55!cyan,fill opacity=0.09},
    subedge/.style={line width=1.00pt,blue!65!black,line cap=round,line join=round},
    boundary/.style={line width=0.95pt,black,line cap=round},
    >=Latex
]
\path[use as bounding box] (-0.88,-0.86) rectangle (3.35,1.68);

\coordinate (L) at (0,0);
\coordinate (R) at (2.45,0);
\coordinate (Ll) at ($(L)+(-0.78,-0.78)$);
\coordinate (Ru) at ($(R)+(0.78,0.92)$);
\coordinate (Rv) at ($(R)+(0,-0.86)$);

\path[subfill]
  (L)
  .. controls (-0.24,0.27) and (-0.27,0.88) .. (-0.22,1.20)
  arc[start angle=180,end angle=360,x radius=0.22,y radius=0.07]
  .. controls (0.27,0.88) and (0.24,0.27) .. (L)
  -- cycle;
\path[subfill] (L) to[bend left=27] (R) to[bend left=27] (L);

\path[leadingfill]
  (L)
  .. controls (-0.34,0.27) and (-0.38,0.96) .. (-0.31,1.36)
  arc[start angle=180,end angle=360,x radius=0.31,y radius=0.10]
  .. controls (0.38,0.96) and (0.34,0.27) .. (L)
  -- cycle;

\draw[web] (L) -- (R);
\draw[web] (L) -- (Ll);
\draw[web] (R) -- (Ru);
\draw[web] (R) -- (Rv);

\draw[subedge] (L) .. controls (-0.24,0.27) and (-0.27,0.88) .. (-0.22,1.20);
\draw[subedge] (0,1.20) ellipse (0.22 and 0.07);
\draw[subedge] (L) .. controls (0.24,0.27) and (0.27,0.88) .. (0.22,1.20);
\draw[subedge] (L) to[bend left=27] (R);
\draw[subedge] (R) to[bend left=27] (L);

\draw[leadingedge] (L) .. controls (-0.34,0.27) and (-0.38,0.96) .. (-0.31,1.36);
\draw[leadingedge] (L) .. controls (0.34,0.27) and (0.38,0.96) .. (0.31,1.36);
\draw[leadingedge] (0,1.36) ellipse (0.31 and 0.10);
\draw[boundary] (L) -- (0,1.08);
\end{tikzpicture}%
}
\caption{Leading and subleading disk saddles for the fundamental Wilson line of \(SU(N)_k\).}
\label{fig:su-fundamental-disk-classes}
\end{figure}
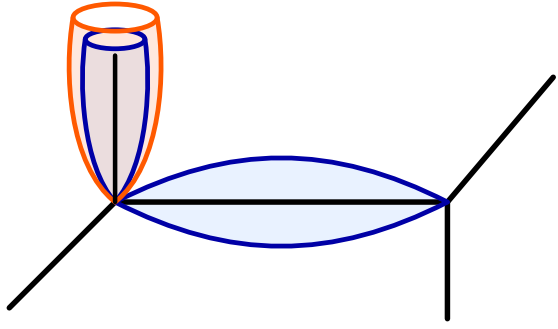

The brane/antibrane dictionary for Wilson lines in $SU(N)_k$ Chern--Simons theory was established  in~\cite{Gomis:2006sb}. Here is a quick summary of those results,

\paragraph{The Fundamental Wilson line.}
Before introducing the branes, we briefly review, following~\cite{Dymarsky:2006ve}, how a Wilson line in the vector representation of \(SU(N)_k\) can be described as a non-compact disc amplitudes. Consider the quantum dimension
\beq
\langle W_{\Box}\rangle_{\textrm{norm}}
=\frac{S_{0\Box}}{S_{00}}
=\frac{q^{N/2}-q^{-N/2}}{q^{1/2}-q^{-1/2}}\, .
\eeq
Using \(q=e^{-g_s}\) and the resolved-conifold K\"ahler parameter
\[
t=N g_s,
\]
the same expression becomes
\beq
\langle W_{\Box}\rangle_{\rm norm   }
=
\frac{e^{\frac{t}{2}}-e^{-\frac{t}{2}}}
{2\sinh{\tfrac{g_s}{2}}}
=
\frac{e^{t/2}-e^{-t/2}}{g_s}
\left(1-\frac{g_s^2}{24}+\mathcal O(g_s^4)\right).
\eeq
The two exponentials in the numerator have a direct worldsheet interpretation in terms of disks ending on the unknot $\alpha$; see Figure~\ref{fig:su-fundamental-disk-classes}.  The leading saddle is the disk with regularised AdS$_2$ area \(-t/2\), whose contribution is \(e^{-\mathrm{Area}}=e^{t/2}\). The subleading saddle is obtained by adding the area of the compact \(\mathbb P^1\) to this disk, shifting the area by \(+t\). Therefore
\[
e^{-\mathrm{Area}}=e^{-(-t/2+t)}=e^{-t/2}.
\]
The relative minus sign between the two contributions originates from the
node of the holomorphic disk map. Signs of this kind are familiar from
equivariant localization, where they arise from the orientation of the
fixed locus; physically, they should be traced to the one-loop
determinants of fluctuations around the corresponding instanton saddle.

\paragraph{Branes and  Antibranes for the Wilson line.}

Alternatively, Wilson lines  $W_R(\alpha)$ supported by an unknot $\alpha\subset S^3$ can be described by non--compact Lagrangian branes ending on the conormal of $\alpha$. This can be obtained by performing the small $q$ expansion of the unnormalized Wilson line expectation value, 
\[
\langle W_R \rangle = S_{0R}
\]
Here, we only describe the instanton part of the free energy, \(\mathcal{F}^{\rm inst}\). The other pieces coming from the Weyl-product formula which contribute to the nonperturbative parts of the amplitude takes the expected form of \(\mathcal{F}^{0}_{X}(N-M)\) for a geometry with \(M\) branes, or \(\mathcal{F}^{0}_{X}(N-P)\) for a geometry with \(P\) antibranes where,

\begin{equation}
  \mathcal{F}^0_{X}(N) = \frac{M(q)}{\zeta(q)^N}  \qquad  \mathcal{F}_{X}^{\rm inst}(t)= - \sum_{m\ge1} \frac{e^{-mt}}{m[m]^2} 
\end{equation}

\begin{figure}[H]
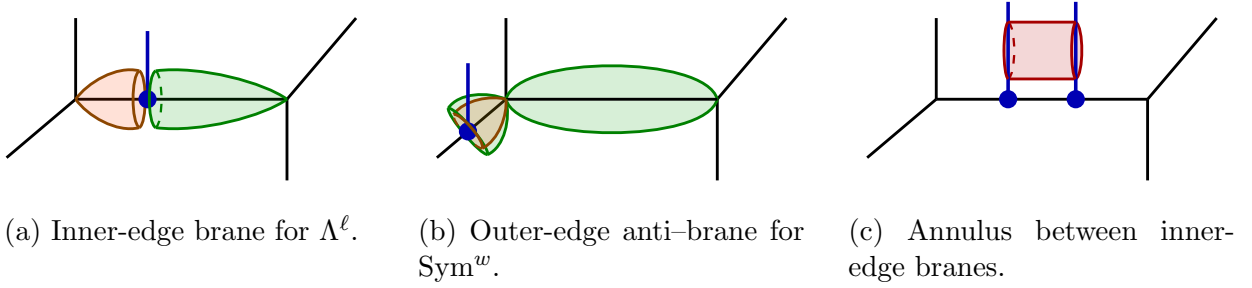

\centering
\begin{subfigure}[t]{0.31\textwidth}
\centering
\includegraphics[width=\linewidth]{Figures/disk.pdf}
\caption{Inner-edge brane for $\Lambda^\ell$.}
\label{fig:su-antisymmetric-inner-brane-panel}
\end{subfigure}
\hfill
\begin{subfigure}[t]{0.31\textwidth}
\centering
\includegraphics[width=\linewidth]{Figures/disk2.pdf}
\caption{Outer-edge anti--brane for $\mathrm{Sym}^w$.}
\label{fig:su-symmetric-outer-antibrane-panel}
\end{subfigure}
\hfill
\begin{subfigure}[t]{0.31\textwidth}
\centering
\includegraphics[width=\linewidth]{Figures/annulus.pdf}
\caption{Annulus between inner-edge branes.}
\label{fig:su-internal-brane-annulus-panel}
\end{subfigure}
\caption{The basic brane, antibrane, and annulus contributions}
\label{fig:su-antisymmetric-inner-brane}
\end{figure}

\paragraph{\boldmath Antisymmetric - $\Lambda^\ell$.}
\[
\lambda_{\Lambda^\ell}
=
e_1+\cdots+e_\ell
=
(\underbrace{1,\ldots,1}_{\ell},0,\ldots,0).
\]
The antisymmetric representation is realised by a single D--brane on the inner compact edge, see Figure \ref{fig:su-antisymmetric-inner-brane}.  The brane shifts the flux by one unit,
\[
\widehat t=g_s(N+1),
\qquad
x=g_s\!\left(\ell+\tfrac12\right),
\]
and gives the instanton free energy
\begin{equation}
\mathcal{F}_{X}^{\rm inst}(\widehat t\, )
+
\sum_{m\ge1}\frac{e^{-mx}+e^{-m(\widehat t-x)}}{m[m]} .
\end{equation}

\paragraph{\boldmath Symmetric - $\mathrm{Sym}^w$.}
\[
\lambda_{\mathrm{Sym}^w}
=
w\,e_1
=
(w,0,\ldots,0).
\]
The symmetric representation is equivalently realised by a single anti--brane on the outer non--compact edge.
\[
\widehat t=g_s(N-1),
\qquad
y=g_s\!\left(w+\tfrac12\right),
\]
and the instanton free energy is
\begin{equation}
\mathcal{F}_{X}^{\rm inst}(\widehat t\, )
+
\sum_{m\ge1}\frac{e^{-my}-e^{-m(\widehat t+y)}}{m[m]} .
\end{equation}

\paragraph{General representation.}
Let \(R\) be a Young tableau with row lengths \(R_i\), column lengths \(R^T_i\), \(M\) columns, and \(P\) rows.  The Wilson line can be either written as a collection of $M$ D--branes on the inner compact edge or a collection of $P$ anti--branes on the outer non--compact edge.

In terms of the branes, we have 
\[
\begin{aligned}
\widehat t &= g_s(N+M),\\
x_i &= g_s\!\left(R^T_i-i+M+\tfrac12\right),
\qquad i=1,\ldots,M .
\end{aligned}
\]
\begin{equation}
\begin{aligned}
\mathcal{F}_{X}^{\rm inst}(\widehat t\, )
+
\sum_{i=1}^{M}
\sum_{m\ge1}
\frac{e^{-mx_i}+e^{-m(\widehat t-x_i)}}{m[m]}
-
\sum_{1\le i<j\le M}\mathcal{A}(x_i-x_j).
\end{aligned}
\end{equation}
The last term is the annulus between distinct D--branes.\footnote{Where \[
\mathcal{A}(u)=\sum_{m\ge1}\frac{e^{-mu}}{m} 
\] is the annulus contribution between two branes at distance $u$.} 
In terms of the anti--branes, we have 
\[
\begin{aligned}
\widehat t &= g_s(N-P),\\
y_i &= g_s\!\left(R_i-i+P+\tfrac12\right),
\qquad i=1,\ldots,P .
\end{aligned}
\]
The corresponding instanton free energy is
\begin{equation}
\begin{aligned}
\mathcal{F}_{X}^{\rm inst}(\widehat t\, )
+
\sum_{i=1}^{P}
\sum_{m\ge1}
\frac{e^{-my_i}-e^{-m(\widehat t+y_i)}}{m[m]}
-
\sum_{1\le i<j\le P}\mathcal{A}(y_i-y_j).
\end{aligned}
\end{equation}

\subsection[\texorpdfstring{$Spin(N)_k$}{Spin(N)k} and \texorpdfstring{$Sp(N)_k$}{Sp(N)k} Chern--Simons theory]{\boldmath $Spin(N)_k$ and $Sp(N/2)_k$ Chern--Simons theory}
\label{sec:so-branes-geometry}
\begin{figure}
  \centering
\resizebox{0.8\textwidth}{!}{%
  \begin{tikzpicture}[
      >=Latex,
      line/.style={line width=1.0pt},
      brane/.style={line width=1.2pt,blue!70!black},
      oplane/.style={line width=1.2pt,red},
      xmark/.style={line width=1.35pt,red},
      every node/.style={font=\small}
  ]
  
  \foreach \y in {-0.30,-0.15,0,0.15,0.30} {
      \draw[brane] (-4.75,\y) -- (-1.75,\y);
  }
  \draw[decorate, decoration={brace, amplitude=5pt}, line width=1pt]
      (-4.85,-0.35) -- (-4.85,0.35)
      node[midway, left=6pt,align=center] {$2N$ Lagr.\ \\branes};
  \draw[oplane] (-4.75,0.52) -- (-1.75,0.52);
  \node[above left, red] at (-4.75,0.52) {$O^{\mp}$};
  
  \draw[->,line] (-1.10,0) -- (0.55,0);
  
  \begin{scope}[shift={(1.85,0)}]
  \coordinate (L) at (0,0);
  \coordinate (R) at (2.60,0);
  \coordinate (X) at ($(L)!0.5!(R)$);
  
  \draw[line] (L) -- (R);
  \draw[line] (L) -- ++(0,1.00);
  \draw[line] (L) -- ++(-0.85,-0.72);
  \draw[line] (R) -- ++(0.85,1.00);
  \draw[line] (R) -- ++(0,-1.00);
  
  \draw[xmark] ($(X)+(-0.10,-0.10)$) -- ($(X)+(0.10,0.10)$);
  \draw[xmark] ($(X)+(-0.10,0.10)$) -- ($(X)+(0.10,-0.10)$);
  \end{scope}
  \end{tikzpicture}%
  }
  \caption{Large $N$ geometric transition from $Spin$ ($O^-$ plane) and $Sp$ ($O^+$ plane) Chern--Simons theory to the orientifolded resolved geometry}
  \label{fig:so-sp-large-n-transition}
  \end{figure}
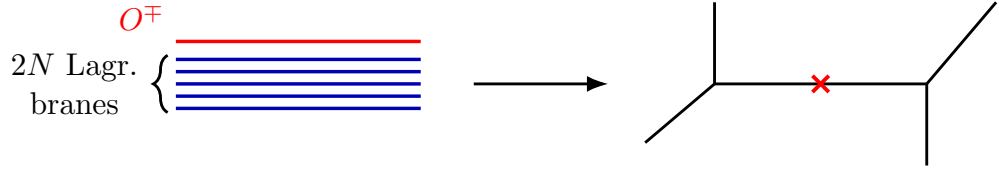

It was shown in \cite{Sinha:2000ap}
that $Spin/Sp$ Chern--Simons theory is dual to the A-model topological string on an orientifold of the resolved conifold. This is the orientifold projection in our setup called, $\Omega\sigma^{SO/Sp}_-$. Before the back reaction in the deformed conifold side, we have an O-plane parallel to the original stack of \(N\) branes; see Figure~\ref{fig:so-sp-large-n-transition}. The orientifold projection acts on the worldvolume gauge fields as
\begin{equation}
  \label{eq:sosp-equiv}
  A_\mu(x)
  =
  -\gamma_{\Omega\sigma_-^{SO/Sp}}\,A_\mu(x)^T\,
  \gamma_{\Omega\sigma_-^{SO/Sp}}^{-1}
\end{equation}
Requiring this projection to square to the identity gives
\[
\gamma_{\Omega\sigma_-^{SO}}^T=\gamma_{\Omega\sigma_-^{SO}},
\qquad \text{or} \quad \quad
\gamma_{\Omega\sigma_-^{Sp}}^T=-\gamma_{\Omega\sigma_-^{Sp}} .
\]
Up to a choice of basis, these are the two inequivalent Chan--Paton choices:
the orthogonal projection
\(\gamma_{\Omega\sigma_-^{SO}}=\mathbbm{1}_N\) and the symplectic projection
\(\gamma_{\Omega\sigma_-^{Sp}}=\mathsf J\). \footnote{Here \(\mathsf J\) denotes the standard symplectic matrix,
\[
\mathsf J=\begin{pmatrix}
0 & \mathbbm{1}_N\\ 
-\mathbbm{1}_N & 0
\end{pmatrix}.
\]
} These two choices give, respectively, $Spin$ and $Sp$ Chern--Simons theory on the worldvolume of the branes. We now derive the corresponding A-model string partition function via the small $q$ expansion \ref{sec:suNdual}.

\paragraph{\boldmath $Spin(2N)_k$.}
Consider the modular \(S\)-matrix explicitly for
\(Spin(2N)_k\). 
\begin{equation}
S_{RS}
=
\frac{1}{2} \frac{i^{N(N-1)}}{(k+2N-2)^{N/2}}
\sum_{w\in \mathcal{W}(D_N)}
\epsilon(w)\,
\exp\!\left[
-{\frac{2\pi i}{k+2N-2}}
\left\langle w(R+\rho_D),S+\rho_D\right\rangle
\right].
\label{eq:sosp-Smatrix-DN-full}
\end{equation}
Setting \(R=S=0\) gives the \(S^3\) partition function. Using the positive roots of $D_N$
\[
\Delta_+(D_N)
=
\{\,e_i-e_j,\;e_i+e_j \mid 1\le i<j\le N\,\}
\]
and with appropriate normalization factors ignored we have,
 \footnote{In this derivation, we dont consider the normalization pieces. A discussion about the full $S$-matrix can be found in  Appendix \ref{app:constant-maps}.} 
\[
S_{00}
\doteq
\prod_{1\le i<j\le N}
\left(1-q^{z_i-z_j}\right)
\left(1-q^{z_i+z_j}\right),
\]
with
\[
z_i=N-i=\langle e_i,\rho_D\rangle,
\qquad
\rho_D=\sum_{i=1}^{N}(N-i)e_i
\]
Taking the log and performing the small $q$ expansion, we get the A-model on $\Omega\sigma^{ SO}_-$ orientifold of the resolved conifold.

\begin{equation}
  \begin{aligned}
  \log S_{00}
  &\doteq
  \mathcal{F}^{0}_{\Omega\sigma^{SO}_{-}}(N)
  +
  \mathcal{F}_{\Omega\sigma^{ SO}_{-}}^{\rm inst}(t),
  \\[4pt]
  \mathcal{F}_{\Omega\sigma^{SO}_{-}}^{\rm inst}(t)
  &=
  -\frac12
  \sum_{m\geq 1}
  \frac{e^{-mt}}{m[m]^2}
  -
  \sum_{\substack{m\geq 1\\ m\ \mathrm{odd}}}
  \frac{e^{-mt/2}}{m[m]},
  \\[4pt]
  \mathcal{F}^{0}(N)
  &=
  \frac12
  \log
  \frac{M(q)\,\zeta(q^2)}
  {\zeta(q)^{2N+1}}.
  \end{aligned}
  \end{equation}
with
\[
q=e^{\frac{2\pi i}{k+2N-2}} = e^{-g_s}\qquad t = g_s(2N-1)
\]

 The free energy has half the oriented-string contribution and contributions from single crosscap, with only odd multicoverings as discussed in Section \ref{sec:4.2}. The expansion in $\gs$ of the odd multicovering contains only odd powers of $\gs$, as shown in \eqref{eqnodd}, consistent with maps from unoriented worldsheets.

\paragraph{$\boldsymbol{Spin(2N+1)_k}$.} 
Now consider the case of $Spin(2N+1)_k$, 

\begin{equation}
S^{Spin(2N+1)_k}_{RS}
=
 \frac{1}{2}\frac{i^{N^2}}{(k+2N-1)^{N/2}}
\sum_{w\in \mathcal{W}(B_N)}
\epsilon(w)\,
\exp\!\left[
-\frac{2\pi i}{k+2N-1}
\left\langle w(R+\rho_B),S+\rho_B\right\rangle
\right].
\end{equation}
With the positive roots,
 \[
\Delta_+(B_N)
=
\{\,e_i-e_j,\ e_i+e_j \mid 1\le i<j\le N\,\}
\cup
\{\,e_i\mid 1\le i\le N\,\}.
\]
We have
\[
S_{00}
\doteq
\prod_{1\le i<j\le N}
\left(1-q^{z_i-z_j}\right)
\left(1-q^{z_i+z_j}\right)
\prod_{i=1}^{N}
\left(1-q^{z_i}\right),
\]
with
\[
z_i
=
N-i+\frac12
=
\langle e_i,\rho_B\rangle,
\qquad
\rho_B
=
\sum_{i=1}^{N}
\left(N-i+\frac12\right)e_i .
\]
A very similar computation for this different gauge group gives, rather surprisingly, the same instanton part of the partition function, with only the flux shifted from \(2N{\color{red} {-1}}\) to \((2N+1){\color{ red}{-1}}\) \( =2N\).
\begin{equation}
\begin{split}
\log S_{00}
&\doteq
\mathcal{F}^{0}_{\Omega\sigma^{SO}_{-}}(N)
+
\mathcal{F}_{\Omega\sigma^{SO}_{-}}^{\rm inst}(t),
\\[4pt]
\mathcal{F}_{\Omega\sigma^{SO}_{-}}^{\rm inst}(t)
&=
-\frac12
\sum_{m\geq 1}
\frac{e^{-mt}}{m[m]^2}
-
\sum_{\substack{m\geq 1\\ m\ \mathrm{odd}}}
\frac{e^{-mt/2}}{m[m]},
\\[4pt]
\mathcal{F}^{0}(N)
&=
\frac12
\log
\frac{M(q)}
{\zeta(q^2)\,\zeta(q)^{2N-3}\,\zeta(q^{1/2})^2}.
\end{split}
\end{equation}

\[
q=e^{\frac{2\pi i}{k+2N-1}}=e^{-g_s},
\qquad
t=2Ng_s .
\]
Even though the instanton-dependent part of the partition function is the same for
\(Spin(2N)_k\) and \(Spin(2N+1)_k\), the prefactor
\(\mathcal F^{0}\) is different. In Appendix~\ref{app:constant-maps}, we perform the
expansion in small-\(g_s\)  and find that the perturbative expansions of the combination
\(\mathcal F^{0}\) together with the normalization terms agree in the two cases, as was already observed in \cite{Sinha:2000ap} with  appropriate definition of $g_s$ and $t$. However, from the small-\(q\) expansion,
we find that the difference between the \(Spin(2N)_k\) and \(Spin(2N+1)_k\) answers is
entirely due to nonperturbative terms contained in the prefactor \(\mathcal F^{0}\).\footnote{This very analogous to the topological string parition function for monodromy defects corresponding to $SU(2N)_{k\; \rm even}$ and $SU(2N+1)_{k \; \rm even}$ in Section \ref{vacuum}.}

 \paragraph{$\boldsymbol{Sp(N)_k}$.}For the case of $Sp(N)_k$ consider the $S$-matrix,\footnote{
  Here we use the convention \(Sp(N)\equiv USp(2N)\), so that \(Sp(N)\) has rank \(N\) and fundamental representation of dimension \(2N\). We also use a normalization in which the long roots have length squared \((2e_i,2e_i)=4\), rather than the standard normalization \((2e_i,2e_i)=2\). This convention is useful for a direct comparison with the \(Spin(N)_k\) cases.
  } 
\begin{equation}
  S^{Sp(N)_k}_{RS}
  =
  \frac{1}{2^{N/2}}\frac{i^{N^2}}{(k+N+1)^{N/2}}
  \sum_{w\in \mathcal{W}(C_N)}
  \epsilon(w)\,
  \exp\!\left[
  -\frac{\pi i}{k+N+1}
  \left\langle w(R+\rho_C),S+\rho_C\right\rangle
  \right].
  \end{equation}
with the positive roots
\[
\Delta_+(C_N)
=
\{\,e_i-e_j,\ e_i+e_j \mid 1\le i<j\le N\,\}
\cup
\{\,2e_i\mid 1\le i\le N\,\}.
\]
We have the Weyl product,
\[
S_{00}
\doteq
\prod_{1\le i<j\le N}
\left(1-q^{z_i-z_j}\right)
\left(1-q^{z_i+z_j}\right)
\prod_{i=1}^{N}
\left(1-q^{2z_i}\right),
\]
with
\[
z_i=N+1-i=\langle e_i,\rho_C\rangle,
\qquad
\rho_C=\sum_{i=1}^{N}(N+1-i)e_i .
\]
Performing the small $q$ expansion we get the A-model free energy on the $\Omega \sigma_-^{Sp}$ orientifold projection of the resolved conifold,
\begin{equation}
\begin{aligned}
\log S_{00}
&\doteq
\mathcal F^0_{\Omega\sigma^{Sp}_{-}}(N)
+
\mathcal F_{\Omega\sigma^{Sp}_{-}}^{\rm inst}(t),
\\[4pt]
\mathcal F_{\Omega\sigma^{Sp}_{-}}^{\rm inst}(t)
&=
-\frac12
\sum_{m\ge1}
\frac{e^{-mt}}{m[m]^2}
+
\sum_{\substack{m\ge1\\ m\ {\rm odd}}}
\frac{e^{-mt/2}}{m[m]},
\\[4pt]
\mathcal F^0(N)
&=
\frac12
\log
\frac{M(q)}
{\zeta(q)^{2N-1}\zeta(q^2)} .
\end{aligned}
\end{equation}

\[
q=e^{\frac{\pi i}{k+N+1}}=e^{-g_s},
\qquad
t=(2N+1)g_s .
\] 
This differs from the $Spin(N)_k$ expressions in exactly the two ways fixed by the charge of the O-plane. In the Sp projection the crosscap state has the opposite sign, 
\[
| C_-^{Sp} \rangle = -| C_-^{SO} \rangle
\]
therefore the odd-cover crosscap term appears with the opposite sign. The same choice of $O$-plane charge also reverses the shift of the K\"ahler flux to $+g_s$ compared to $-g_s$ for the $SO$ projection.

\subsubsection[Branes/Antibrane for $Spin(2N)_k$ Wilson
lines]{\boldmath Branes/Antibrane for $Spin(2N)_k$ Wilson
lines}

\begin{figure}[H]
\centering
\resizebox{0.40\textwidth}{!}{%
\begin{tikzpicture}[
    web/.style={line width=1.15pt,line cap=round,line join=round},
    mobiusfill/.style={fill=blue!55!cyan,fill opacity=0.10},
    mobiusedge/.style={line width=1.05pt,blue!65!black,line cap=round,line join=round},
    crossmark/.style={line width=1.20pt,red!75!black,line cap=round},
    boundary/.style={line width=0.95pt,black,line cap=round}
]
\path[use as bounding box] (-0.88,-0.86) rectangle (3.12,1.54);

\coordinate (L) at (0,0);
\coordinate (R) at (2.45,0);
\coordinate (M) at (1.225,0);
\coordinate (Mt) at ($(M)+(0,0.21)$);
\coordinate (Mb) at ($(M)+(0,-0.21)$);
\coordinate (Ll) at ($(L)+(-0.78,-0.78)$);
\coordinate (Ru) at ($(R)+(0.78,0.92)$);
\coordinate (Rv) at ($(R)+(0,-0.86)$);

\path[mobiusfill]
  (L)
  .. controls (-0.24,0.27) and (-0.27,0.88) .. (-0.22,1.20)
  arc[start angle=180,end angle=360,x radius=0.22,y radius=0.07]
  .. controls (0.27,0.88) and (0.24,0.27) .. (L)
  -- cycle;
\path[mobiusfill]
  (L)
  .. controls (0.35,0.24) and (0.82,0.24) .. (Mt)
  arc[start angle=90,end angle=-90,radius=0.21]
  .. controls (0.82,-0.24) and (0.35,-0.24) .. (L)
  -- cycle;

\draw[web] (L) -- (R);
\draw[web] (L) -- (Ll);
\draw[web] (R) -- (Ru);
\draw[web] (R) -- (Rv);

\draw[mobiusedge] (L) .. controls (-0.24,0.27) and (-0.27,0.88) .. (-0.22,1.20);
\draw[mobiusedge] (0,1.20) ellipse (0.22 and 0.07);
\draw[mobiusedge] (L) .. controls (0.24,0.27) and (0.27,0.88) .. (0.22,1.20);
\draw[boundary] (L) -- (0,1.08);

\draw[mobiusedge] (L) .. controls (0.35,0.24) and (0.82,0.24) .. (Mt);
\draw[mobiusedge] (L) .. controls (0.35,-0.24) and (0.82,-0.24) .. (Mb);
\draw[mobiusedge] (M) circle (0.21);
\draw[crossmark] ($(M)+(-0.15,-0.15)$) -- ($(M)+(0.15,0.15)$);
\draw[crossmark] ($(M)+(-0.15,0.15)$) -- ($(M)+(0.15,-0.15)$);
\end{tikzpicture}%
}
\caption{The M\"obius saddle for the fundamental Wilson line of \(Spin(2N)_k\) }
\label{fig:so-fundamental-mobius-saddle}
\end{figure}
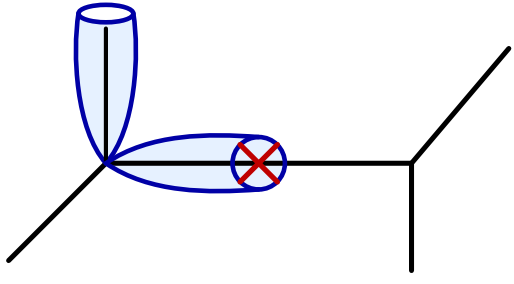

\paragraph{The Fundamental Wilson line.}
Similar to the \(SU(N)_k\) case, we can give a worldsheet interpretation to the normalized Wilson line in the vector representation of \(Spin(2N)_k\). Consider its quantum dimension.
\begin{equation}
\langle W_{\Box}\rangle_{\rm norm}
=
\frac{S_{0\Box}}{S_{00}}
=
1+\frac{q^{\frac{2N-1}{2}}-q^{-\frac{2N-1}{2}}}
{q^{1/2}-q^{-1/2}} .
\end{equation}
With the resolved-conifold K\"ahler parameter
\[
t=g_s(2N-1),
\]
this separates into an oriented disk contribution and a one-crosscap contribution,
\begin{equation}
\langle W_{\Box}\rangle_{\rm norm}
=
\underbrace{
\frac{e^{\frac{t}{2}}-e^{-\frac{t}{2}}}
{2\sinh{\tfrac{g_s}{2}}}
}_{\text{oriented}}
+
\underbrace{1}_{\text{unoriented}} .
\end{equation}
The oriented term is the same that appears for the fundamental Wilson line in the \(SU(N)_k\) theory, with \(t=g_s(2N-1)\).  There is a new unoriented contribution corresponding to the one-crosscap, or M\"obius saddle.\footnote{There are no further higher-genus unoriented contributions, nor any multicover sectors} This has, in addition to the regularized \(AdS_2\) contribution, an extra contribution equal to half the area of the compact \(\mathbb{P}^1\); see Figure~\ref{fig:so-fundamental-mobius-saddle}.
Hence
\[
\mathrm{Area}=-\frac{t}{2}+\frac{t}{2}=0,
\qquad
e^{-\mathrm{Area}}=1 .
\]
For this \(SO\) projection, the M\"obius contribution enters with a ``+" sign.\footnote{For \(Sp(N)_k\),
\[
\langle W_{\Box}\rangle_{\rm norm}
=
\frac{q^{\frac{2N+1}{2}}-q^{-\frac{2N+1}{2}}}
{q^{1/2}-q^{-1/2}}
-1 .
\]
The M\"obius term changes sign, as expected from the crosscap sign flip
\(\Omega\sigma_-^{SO}\to\Omega\sigma_-^{ Sp}\).}

\paragraph{Branes and  Antibrane}
Alternatively, just as in the \(SU(N)_k\) case, working with the unnormalized Wilson line, we can give a brane/antibrane description for the modular $S$-matrix entry. 
\[
\langle W_{R} \rangle=S_{R0}\,.
\]
Again, we specify only the instanton part of the free energy, this can be supplemented with $\mathcal{F}^0(N-M)$ for a geometry with $M$ brane/image pairs or $\mathcal{F}^0(N-P)$ for a geometry with $P$ antibrane/image pairs  where,
\begin{equation}
  \mathcal{F}^{0}(N)
  =
  \frac12
  \log
  \frac{M(q)\,\zeta(q^2)}
  {\zeta(q)^{2N+1}}. 
\end{equation}

\paragraph{\boldmath Antisymmetric - $\Lambda^\ell$}
\[
\lambda_{\Lambda^\ell}
=
e_1+\cdots+e_\ell
=
(\underbrace{1,\ldots,1}_{\ell},0,\ldots,0)\,.
\]
\[
\widehat t = g_s(2N+1)
\qquad
x = g_s(l+\tfrac{1}{2})
\]

\begin{equation}
\mathcal{F}_{\Omega\sigma^{SO}_-}^{\rm inst}(\widehat t\, )
+\underbrace{\sum_{m\ge1}\frac{1}{m}\,
\frac{e^{-mx}+e^{-m(\widehat t-x)}}{[m]}}_{\textcolor{blue}{\text{open string}}}
-\underbrace{\frac12\sum_{m\ge1}\frac{e^{-m(\widehat t-2x)}}{m}}_{\textcolor{green!65!black}{\text{self-image annulus}}}
+\underbrace{\sum_{\substack{m\ge1\\ m\ \mathrm{odd}}}
\frac{e^{-m(\widehat t/2-x)}}{m}}_{\textcolor{red}{\text{M\"obius}}}
\end{equation}

For the antisymmetric representation we notice the the flux shift from \(g_s(2N-1) \) to \( g_s(2N+1)\). This signals that the representation \(\Lambda^\ell\) adds a brane together with its orientifold image. The first  term is the oriented open string contribution with a single boundary from the brane and its image. The second term is half of the self-image annulus with modulus
\(
\widehat t-2x
\)
as in \eqref{annulusitself}. The last is the odd-cover M\"obius contribution with the sign
\[
- \langle \mathcal{B} | C_-^{ SO} \rangle =-(+1)(-1)=+1.
\]
Note that in the sum over the odd windings fo the $\Omega\sigma_-$ M\"obius amplitude, is consistent to what found in Section~\ref{nontrivialmon} (see discussion around \eqref{eq:mobius}) for $\Omega\sigma_+$ where instead the sum is over the even windings.
\paragraph{\boldmath Symmetric traceless - $\mathrm{Sym}^w_0$}
\[
\lambda_{\mathrm{Sym}^w_0.}
=
w e_1
=
(w,0,\ldots,0)
\]
\[
\widehat t = g_s(2N-3)
\qquad
y = g_s(w+\tfrac{1}{2})
\]

\begin{equation}
\mathcal{F}_{\Omega\sigma^{SO}_-}^{\rm inst}(\widehat t\, )
+
\sum_{m\ge1}\frac{1}{m}\,
\frac{e^{-my}-e^{-m(\widehat t+y)}}{[m]}
-\frac12
\sum_{m\ge1}
\frac{e^{-m(\widehat t+2y)}}{m}
-
\sum_{\substack{m\ge1\\ m\ \mathrm{odd}}}
\frac{e^{-m(y+\widehat t/2)}}{m}
\end{equation}
Analogously, the symmetric traceless Wilson line adds an antibrane and its orientifold image on the external leg, but now with the mobius contribution of opposite sign.
\[
- \langle \bar{\mathcal{B}} | C_-^{SO} \rangle =-(-1)(-1)=-1.
\]

\paragraph{\boldmath Spinors - $S_\pm$.}
\[
\lambda_{S_+}
=
\frac 1 2 (1,\ldots,1,1)
\qquad
\lambda_{S_-}
= \frac 12 (1,\ldots,1,-1)
\]
\[
\widehat t = g_s(2N)
\]

\begin{equation}
\begin{aligned}
\mathcal{F}_{\Omega\sigma^{SO}_-}^{\rm inst}(\widehat t\, )
+
\sum_{m\ge1}
\frac{1}{m}
\frac{e^{-m\widehat t/2}}{[m]}
\end{aligned}
\end{equation}
The case of the spinor Wilson lines is quite different from the previous two cases.\footnote{Note that both \(S_+\) and \(S_-\) give the same topological string amplitude. This is consistent with the charge conjugation symmetry of $Spin(2N)$ Chern--Simons theory.} In this case the Wilson line inserts a single brane at the ``stuck'' locus, without an image brane.\footnote{By the stuck locus we mean the \(S^1\subset S^2\) which, under the antipodal identification \(S^2\to \mathbb{RP}^2\), is mapped to itself and descends to an \(\mathbb{RP}^1\subset \mathbb{RP}^2\), which is to be contrasted with the fixed locus $S^1 \subset S^2$  in the case of $\Omega \sigma_+$.}

This increases the flux by only \(+g_s\), compared to the \(\pm 2g_s\) shift in the antisymmetric and symmetric-traceless cases. The prefactor \(\mathcal F^0\) also does not follow the usual shift rule for ordinary branes or antibranes. Instead, the spinor Wilson lines contains a new contribution
\begin{equation}
\mathcal F^0(N)
+
\log\frac{\zeta(q)}{\zeta(q^2)} .
\end{equation}

\begin{figure}[H]
\centering
\resizebox{0.92\textwidth}{!}{%
\begin{tikzpicture}[
    line/.style={line width=1.0pt},
    brane/.style={line width=1.25pt,blue!70!black},
    branepoint/.style={circle,fill=blue!70!black,inner sep=2.25pt},
    xmark/.style={line width=1.25pt,red!80!black},
    every node/.style={font=\small}
]

\begin{scope}[shift={(0,0)}]
\coordinate (L) at (0,0);
\coordinate (R) at (2.60,0);
\coordinate (X) at ($(L)!0.5!(R)$);
\coordinate (A1) at ($(L)!0.34!(R)$);
\coordinate (A2) at ($(L)!0.66!(R)$);

\draw[line] (L) -- (R);
\draw[line] (L) -- ++(0,1.00);
\draw[line] (L) -- ++(-0.85,-0.72);
\draw[line] (R) -- ++(0.85,1.00);
\draw[line] (R) -- ++(0,-1.00);

\draw[brane] (A1) -- ++(0,0.84);
\draw[brane] (A2) -- ++(0,-0.84);
\node[branepoint] at (A1) {};
\node[branepoint] at (A2) {};

\draw[xmark] ($(X)+(-0.10,-0.10)$) -- ($(X)+(0.10,0.10)$);
\draw[xmark] ($(X)+(-0.10,0.10)$) -- ($(X)+(0.10,-0.10)$);

\node at (1.30,-1.35) {$\Lambda^\ell$};
\end{scope}

\begin{scope}[shift={(4.15,0)}]
\coordinate (L) at (0,0);
\coordinate (R) at (2.60,0);
\coordinate (X) at ($(L)!0.5!(R)$);
\coordinate (SL) at ($(L)!0.55!($(L)+(-0.85,-0.72)$)$);
\coordinate (SR) at ($(R)!0.55!($(R)+(0.85,1.00)$)$);

\draw[line] (L) -- (R);
\draw[line] (L) -- ++(0,1.00);
\draw[line] (L) -- ++(-0.85,-0.72);
\draw[line] (R) -- ++(0.85,1.00);
\draw[line] (R) -- ++(0,-1.00);

\draw[brane] (SL) -- ++(0,0.84);
\draw[brane] (SR) -- ++(0,-0.84);
\node[branepoint] at (SL) {};
\node[branepoint] at (SR) {};

\draw[xmark] ($(X)+(-0.10,-0.10)$) -- ($(X)+(0.10,0.10)$);
\draw[xmark] ($(X)+(-0.10,0.10)$) -- ($(X)+(0.10,-0.10)$);

\node at (1.30,-1.35) {$\mathrm{Sym}^w_0$};
\end{scope}

\begin{scope}[shift={(8.30,0)}]
\coordinate (L) at (0,0);
\coordinate (R) at (2.60,0);
\coordinate (X) at ($(L)!0.5!(R)$);

\draw[line] (L) -- (R);
\draw[line] (L) -- ++(0,1.00);
\draw[line] (L) -- ++(-0.85,-0.72);
\draw[line] (R) -- ++(0.85,1.00);
\draw[line] (R) -- ++(0,-1.00);

\draw[brane] ($(X)+(0,-0.62)$) -- ($(X)+(0,0.62)$);

\draw[xmark] ($(X)+(-0.10,-0.10)$) -- ($(X)+(0.10,0.10)$);
\draw[xmark] ($(X)+(-0.10,0.10)$) -- ($(X)+(0.10,-0.10)$);

\node at (1.30,-1.35) {$S_\pm$};
\end{scope}
\end{tikzpicture}%
}
\caption{The dual brane/antibrane configurations for the antisymmetric, symmetric traceless, and spinor Wilson lines of \(Spin(2N)_k\) in the \(\Omega\sigma^{SO}_-\) orientifold of the resolved conifold.}
\label{fig:so2n-brane-configurations}
\end{figure}
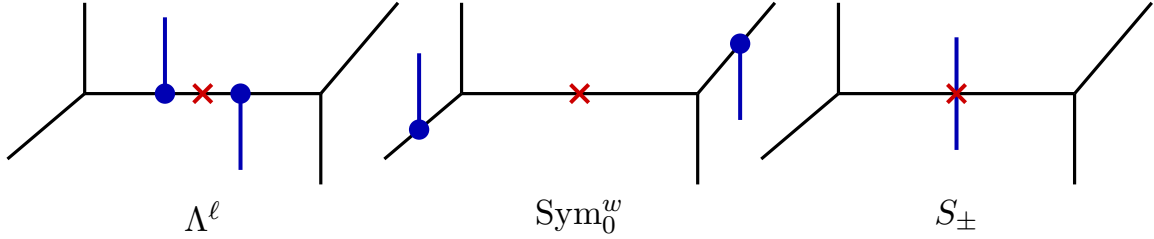

\paragraph{Brane/Antibrane for a general representation.}

For \(Spin(2N)_k\) the highest weight for a general representation in orthogonal coordinates can be written as, \footnote{It turns out that the topological string amplitudes for both $+\lambda_N$ and $-\lambda_N$ are the same, so without loss of generality we will focus on $+\lambda_N$.}
\[
\lambda=(\lambda_1,\ldots,\lambda_N)=\sum_{i=1}^N \lambda_i e_i .
\]
with the dominance condition,
\[
\lambda_1\ge \lambda_2\ge\cdots\ge \lambda_{N-1}\ge |\lambda_N|.
\]
The weight-lattice requires,
\[
\lambda_i\in \mathbb Z\ \quad \forall i
\qquad\text{or}\qquad
\lambda_i\in \mathbb Z+\frac12\quad \forall i .
\]
One cannot mix integer and half-integer coordinates. Consider an integer representation  whose highest weight in the orthogonal coordinates is given by,
\[
\lambda_R=(R_1,R_2,\ldots,R_P,0,\ldots,0),
\]
where \(R_i \in \mathbb{Z}\) and $P < N$. This can be interpreted as a Young tableau with $P$ rows. Similar to the $SU(N)_k$, this can given a description in terms of \(P\) antibranes in the external-edge along with its image pairs with moduli \(y_i\) as, 

\[
\widehat t = g_s(2N-1-2P)
\qquad
y_i=g_s\left(R_i-i+P+\tfrac12\right)
\]

\begin{equation}
\begin{split}
\mathcal{F}_{\Omega\sigma^{SO}_-}^{\rm inst}(\widehat t\, )
 \;+
&\sum_{i=1}^{P}
\sum_{m=1}^{\infty}
\frac{1}{m}\,
\frac{e^{-m y_i}-e^{-m(y_i+\widehat t)}}{[m]}-
\sum_{1\leq i<j\leq P}
\sum_{m=1}^{\infty}
\frac{e^{-m(y_i-y_j)}}{m}\\ 
&-
\frac12
\sum_{i,j=1}^{P}
\sum_{m=1}^{\infty}
\frac{e^{-m(y_i+y_j+\widehat t)}}{m}-
\sum_{i=1}^{P}
\sum_{\substack{m=1\\ m\ {\rm odd}}}^{\infty}
\frac{e^{-m(y_i+\widehat t/2)}}{m}
\end{split}
\end{equation}

Equivalently, in the compact notation this is,\footnote{Here and below,
\(\displaystyle\mathcal M_{\rm odd}(u):=
\sum_{\substack{m\ge1\\ m\ {\rm odd}}}\frac{e^{-mu}}{m}\).}
\begin{equation}
\mathcal{F}_{\Omega\sigma^{SO}_-}^{\rm inst}(\widehat t\, )
\;+\;\SymP(y;\that)
\;-\;\sum_{i=1}^{P}
\left(\frac12\D(\that+2y_i)+\mathcal M_{\rm odd}(y_i+\tfrac{\that}{2})\right).
\end{equation}
Notice that we exclude \(P=N\), since in that case the flux of the internal-edge would become negative. Equivalently, using the transpose of the Young tableau,
\[
R^T=(R^T_1,R^T_2,\ldots,R^T_M),
\]
where \(R^T_i \leq N\) are the column lengths, we can give the internal-edge brane description of the wilson line in terms of \(M\) branes along with its orientifold imagewith moduli \(x_i\) as,

\[
\widehat t = g_s(2N-1+2M)
\qquad
x_i=g_s\left(R^T_i-i+M+\tfrac12\right)
\]

\begin{equation}
\begin{aligned}
\mathcal{F}_{\Omega\sigma^{SO}_-}^{\rm inst}(\widehat t\, )
 \;+
&\sum_{i=1}^{M}
\sum_{m=1}^{\infty}
\frac{1}{m}\,
\frac{e^{-m x_i}+e^{-m(\widehat t-x_i)}}{[m]}
-
\sum_{1\leq i<j\leq M}  
\sum_{m=1}^{\infty}
\frac{e^{-m(x_i-x_j)}}{m}
\\
&-
\frac12
\sum_{i,j=1}^{M}
\sum_{m=1}^{\infty}
\frac{e^{-m(\widehat t-x_i-x_j)}}{m}
+
\sum_{i=1}^{M}
\sum_{\substack{m=1\\ m\ {\rm odd}}}^{\infty}
\frac{e^{-m(\widehat t/2-x_i)}}{m}
\end{aligned}
\end{equation}
Equivalently, in compact notation this is
\begin{equation}
\mathcal{F}_{\Omega\sigma^{SO}_-}^{\rm inst}(\widehat t\, )
\;+\;\AntiP(x;\that)
\;-\;\sum_{i=1}^{M}
\left(\frac12\D(\that-2x_i)-\mathcal M_{\rm odd}(\tfrac{\that}{2}-x_i)\right).
\end{equation}

In the brane description, when \(P=N\), we notice that two branes sit on the stuck locus. Usually, branes are not allowed to be on top of each other due to interaction between them through the annulus term. This can be made precise by noticing that the annulus contribution makes the amplitude vanish when two branes are brought together. For two branes at positions \(x_1\) and \(x_2\),
\begin{equation}
\exp\left(
- \sum_{m\geq 1}
{\frac{e^{-m(x_1 - x_2)}} {m}}
\right)
=
1-e^{-(x_1 - x_2)}
\end{equation}
which is goes to zero as $x_1 \rightarrow x_2$. However, the situation is different in the orientifold case. The two objects that are brought on top of each other are not two independent branes, but a brane and its orientifold image. Therefore, the annulus contribution is half the usual value. In addition, the orientifold theory also contains a M\"obius sector. This M\"obius contribution combines with the half-annulus term to remove the would-be zero. Therefore, after including the M\"obius contribution, with $z = e^{-(t/2 - x)}$, we have
\footnote{Notice the sign in front of the M\"obius contribution. In the \(Sp\) case the sign is opposite, and the total amplitude vanishes.}

\begin{equation}
\exp\left(
-\frac 1 2
\sum_{m\geq 1}
{\frac{z^{2m}} {m}}
\right)\exp\left(+
\sum_{ \substack{m\geq 1\\ m\ {\rm odd}}}
{\frac{z^m} {m}}
\right)
=
(1-z^2)^{1/2}
\left({\frac{1+z} {1-z}}\right)^{1/2}
=
1+z
\end{equation}
which does not vanish as $z \rightarrow 1$. This allows for at most two branes to sit on top of each other  at the stuck locus. If one tries to put more branes on the stuck locus, then the additional branes are no longer just image pairs. The ordinary annulus interaction remains, and the usual Vandermonde zero makes the amplitude vanish.

Consider now a half-integer highest weight of \(Spin(2N)_k\),
\[
\lambda \in \left(\mathbb Z+\frac12\right)^N .
\]
The half-integer weight can be written as the sum,
\[
\lambda
=
\left(
R_1+\frac12,
R_2+\frac12,
\ldots,
R_P+\frac12,
\frac12,\ldots,\frac12
\right) =
\lambda_R+\lambda_{S_+},
\qquad
\]
where $\lambda_R$ is an ordinary integral highest weight, equivalently a Young tableau with \(P\) nonzero rows. The interpretation is that the integral part \(\lambda_R\) is described by the usual brane or antibrane system, while the universal $\lambda_{S_+}$ is described by a stuck brane sitting at the orientifold stuck locus. We have the  external edge antibrane description with,
\[
\widehat t
=
g_s(2N-2P) \qquad y_i
=
g_s\left(R_i-i+P+\frac12\right)
\qquad
i = 1, 2 \hdots P
\]
\begin{equation}
\begin{aligned}
\mathcal{F}_{\Omega\sigma^{SO}_-}^{\rm inst}(\widehat t\, )
+&
\sum_{m=1}^{\infty}
\frac{1}{m}\,
\frac{e^{-m\widehat t/2}}{[m]} + 
\sum_{i=1}^{P}
\sum_{m=1}^{\infty}
\frac{1}{m}\,
\frac{e^{-m y_i}-e^{-m(y_i+\widehat t)}}{[m]}
\\ 
&-
\sum_{1\leq i<j\leq P}
\sum_{m=1}^{\infty}
\frac{e^{-m(y_i-y_j)}}{m}
-
\frac12
\sum_{i,j=1}^{P}
\sum_{m=1}^{\infty}
\frac{e^{-m(y_i+y_j+\widehat t)}}{m}
\\ 
&-
\sum_{i=1}^{P}
\sum_{\substack{m=1\\ m\ {\rm odd}}}^{\infty}
\frac{e^{-m(y_i+\widehat t/2)}}{m}
+
\sum_{i=1}^{P}
\sum_{m=1}^{\infty}
\frac{e^{-m(y_i+\widehat t/2)}}{m}.
\end{aligned}
\end{equation}
Equivalently, in compact notation this is\footnote{Here
\(\mathcal B_{\rm st}\left( \widehat t \,  \right):=
\sum_{m\ge1}\frac{e^{-m\that/2}}{m[m]}\) denotes the one boundary open string contribution of the stuck brane.}
\begin{equation}
\mathcal{F}_{\Omega\sigma^{SO}_-}^{\rm inst}(\widehat t\, )
\;+\;\mathcal B_{\rm st}(\that)
\;+\;\SymP(y;\that)
\;-\;\sum_{i=1}^{P}
\left(\frac12\D(\that+2y_i)+\mathcal M_{\rm odd}(y_i+\tfrac{\that}{2})-\D(y_i+\tfrac{\that}{2})\right).
\end{equation}
Compared to the integer representation, we have an additional one-boundary contribution for the stuck brane and its annuli with the external edge antibranes. With a sign determined by the overlap given by, 
\[
-\langle \bar{\mathcal{B}}|{\mathcal{B}} \rangle= -(-1)(+1) = +1
\]
Equivalently, one may use the internal-edge brane description with \(R^T_i\), the column lengths.
\[
\widehat t
=
g_s(2N+2M) \qquad
x_i
=
g_s\left(R^T_i-i+M+\frac12\right),
\qquad
i=1,\ldots,M.
\]

\begin{equation}
\begin{aligned}
\mathcal{F}_{\Omega\sigma^{SO}_-}^{\rm inst}(\widehat t\, )
+&
\sum_{m=1}^{\infty}
\frac{1}{m}\,
\frac{e^{-m\widehat t/2}}{[m]}
+
\sum_{i=1}^{M}
\sum_{m=1}^{\infty}
\frac{1}{m}\,
\frac{e^{-m x_i}+e^{-m(\widehat t-x_i)}}{[m]}
\\
&-
\sum_{1\leq i<j\leq M}
\sum_{m=1}^{\infty}
\frac{e^{-m(x_i-x_j)}}{m}
-
\frac12
\sum_{i,j=1}^{M}
\sum_{m=1}^{\infty}
\frac{e^{-m(\widehat t-x_i-x_j)}}{m}
\\
&+
\sum_{i=1}^{M}
\sum_{\substack{m=1\\ m\ {\rm odd}}}^{\infty}
\frac{e^{-m(\widehat t/2-x_i)}}{m}
-
\sum_{i=1}^{M}
\sum_{m=1}^{\infty}
\frac{e^{-m(\widehat t/2-x_i)}}{m}.
\end{aligned}
\end{equation}
Equivalently, in compact notation this is
\begin{equation}
\mathcal{F}_{\Omega\sigma^{SO}_-}^{\rm inst}(\widehat t\, )
\;+\;\mathcal B_{\rm st}(\that)
\;+\;\AntiP(x;\that)
\;-\;\sum_{i=1}^{M}
\left(\frac12\D(\that-2x_i)-\mathcal M_{\rm odd}(\tfrac{\that}{2}-x_i)+\D(\tfrac{\that}{2}-x_i)\right).
\end{equation}
The last term is the ordinary annulus between the internal brane at \(x_i\) and the stuck brane, with the opposite sign as earlier. Notice that the a column of length $N$  no longer places additional brane at the stuck locus, due to the change in flux by the stuck brane. This can also be seen as a consequence of the new annulus-stuck brane interaction. Setting $w_i = e^{\frac{ \widehat t}{2} - x}$, including the self-image annulus, the Möbius, and the ordinary annulus to the stuck brane contributions, we get
\begin{equation}
\exp\left(
-\frac12\sum_{m\geq1}\frac{w_i^{2m}}{m}
\right)
\exp\left(
\sum_{\substack{m\geq1\\ m\ {\rm odd}}}
\frac{w_i^m}{m}
\right)
\exp\left(
-\sum_{m\geq1}\frac{w_i^m}{m}
\right)
=
1-w_i^2
\end{equation}
We see that as the brane approaches the  stuck locus 
\(
w_i\to 1
\)
The amplitude vanishes. \footnote{The remaining terms from the Weyl product also take the expected form of $\mathcal{F}^0(N - P) + \log\frac{\zeta(q)}{\zeta(q^2)}$ for the case of P antibranes and a stuck brane. Replace $P$ with $M$ in the case of $M$ branes and a stuck brane}

\subsubsection{\boldmath Branes/Antibranes for \(Spin(2N+1)_k\)}

The \(Spin(2N+1)_k\) case is closely related to the \(Spin(2N)_k\) case. 
The main change is the shift of the background flux,
\[
2N \longrightarrow 2N+1,
\]
together with a different nonperturbative completion given by
\[
\mathcal{F}^{0}(N)
=
\frac12
\log
\frac{M(q)}
{\zeta(q^2)\,\zeta(q)^{2N-3}\,\zeta(q^{1/2})^2}.
\]

The highest weights are also analogous to the \(Spin(2N)_k\) case, but with one important difference. For \(Spin(2N+1)\), the Weyl group can flip the sign of the
orthogonal coordinate from
\(
(\lambda_1,\lambda_2,\ldots,-|\lambda_N|)
\)
to
\(
(\lambda_1,\lambda_2,\ldots,+|\lambda_N|).
\)
Thus, the dominant chamber may be chosen as
\[
\lambda_1\geq \lambda_2\geq \cdots \geq \lambda_N\geq 0,
\]
Therefore, \(Spin(2N+1)\) has a unique spinor representation. The insertion of the single stuck brane associated with this unique spinor shifts the
prefactor by
\[
\mathcal F^0(N)
+
\log\frac{\zeta(q^{1/2})\zeta(q^2)}{\zeta(q)^2}.
\]
Therefore, the Wilson lines of \(Spin(2N+1)_k\) are described topological string amplitudes that are the same brane/antibrane amplitudes of \(Spin(2N)_k\) but with the correct non-perturbative terms and the corresponding stuck-brane shift taken into account.

\subsubsection{\boldmath Branes/Antibranes
 for $Sp(N)_k$}
\label{sec:sp-Chern--Simons-theory}
The dual branes/antibranes corresponding to the representation of \(Sp(N)_k\) are analogous to the tensor-representation of \(Spin(2N)_k\), with the replacement
\(
2N-1\longrightarrow 2N+1 
\) 
and the appropriate change of signs for the unoriented contributions arising from the change of crosscap state. We will only focus on the instanton part of the free energy, with the predictable prefactor of $\mathcal{F}^0(N - M)$ or $\mathcal{F}^0(N-P)$ for $M$ brane-image pairs or $P$ antibrane-image pairs, where
\begin{equation}
\mathcal{F}^0(N) = \frac12
\log
\frac{M(q)}
{\zeta(q)^{2N-1}\zeta(q^2)}
\end{equation}
A highest weight of \(Sp(N)_k\) in orthogonal coordinates is written as
\[
\lambda=(\lambda_1,\ldots,\lambda_N)=\sum_{i=1}^N \lambda_i e_i .
\]
with the dominance condition
\[
\lambda_1\ge \lambda_2\ge\cdots\ge \lambda_N\ge 0 .
\]
The weight lattice requires
\[
\lambda_i\in \mathbb Z \qquad \forall i .
\]
Thus the allowed highest weights of \(Sp(N)_k\) are ordinary partitions with at most \(N\) parts. In particular, there is no half-integer weights. For an integer highest weight represented by a Young tableau with $P$ rows, 
\[
\lambda_R=(R_1,\ldots,R_P,0,\ldots,0),
\qquad
P \leq N,
\]
We can give a description in terms of $P$ antibrane in the external edge, 
\[
\widehat t=g_s(2N+1-2P),
\qquad
y_i=g_s\left(R_i-i+P+\frac12\right).
\]
\begin{equation}
\begin{split}
\mathcal F_{\Omega\sigma^{Sp}_{+}}^{\rm inst}(\widehat t\, )
&+
\sum_{i=1}^{P}
\sum_{m\ge1}
\frac{1}{m}
\frac{e^{-m y_i}-e^{-m(y_i+\widehat t)}}{[m]}
-
\sum_{1\le i<j\le P}
\sum_{m\ge1}
\frac{e^{-m(y_i-y_j)}}{m}
\\-& 
\frac12
\sum_{i,j=1}^{P}
\sum_{m\ge1}
\frac{e^{-m(y_i+y_j+\widehat t)}}{m}
+
\sum_{i=1}^{P} 
\sum_{\substack{m\ge1\\ m\ {\rm odd}}}
\frac{e^{-m(y_i+\widehat t/2)}}{m}.
\end{split}
\end{equation}
Equivalently, in compact notation this is
\begin{equation}
\mathcal F_{\Omega\sigma^{Sp}_{+}}^{\rm inst}(\widehat t\, )
\;+\;\SymP(y;\that)
\;-\;\sum_{i=1}^{P}
\left(\frac12\D(\that+2y_i)-\mathcal M_{\rm odd}(y_i+ \tfrac{\that}{2})\right).
\end{equation}
Alternatively, using the transpose Young tableau with $M$ columns, 
\[
R^T=(R^T_1,\ldots,R^T_M),
\]
We can give a description in terms of \(M\) branes internal-edge as, 
\[
\widehat t=g_s(2N+1+2M),
\qquad
x_i=g_s\left(R^T_i-i+M+\frac12\right).
\]

\begin{equation}
\begin{split}
\mathcal F_{\Omega\sigma^{Sp}_{+}}^{\rm inst}(e^{-\widehat t})
+&
\sum_{i=1}^{M}
\sum_{m\ge1}
\frac{1}{m}
\frac{e^{-m x_i}+e^{-m(\widehat t-x_i)}}{[m]}
-
\sum_{1\le i<j\le M}
\sum_{m\ge1}
\frac{e^{-m(x_i-x_j)}}{m}
\\& 
-
\frac12
\sum_{i,j=1}^{M}
\sum_{m\ge1}
\frac{e^{-m(\widehat t-x_i-x_j)}}{m}
-
\sum_{i=1}^{M}
\sum_{\substack{m\ge1\\ m\ {\rm odd}}}
\frac{e^{-m(\widehat t/2-x_i)}}{m}.
\end{split}
\end{equation}
Equivalently, in compact notation this is
\beq
\mathcal F_{\Omega\sigma^{Sp}_{+}}^{\rm inst}(\widehat t\, )
\;+\;\AntiP(x;\that)
\;-\;\sum_{i=1}^{M}
\left(\frac12\D(\that-2x_i)+\mathcal M_{\rm odd}(\tfrac{\that}{2}-x_i)\right).
\eeq
The characteristic feature of the \(Sp(N)_k\) case is that there is no stuck-brane configurations. All representations are described by ordinary branes together with their orientifold images. A single brane on the stuck locus would require a one-dimensional symplectic Chan--Paton space, which does not exist. The smallest possible symplectic Chan--Paton space is two-dimensional, corresponding to an \(Sp(1)\simeq SU(2)\) gauge theory, but even this pair of branes doesnot exist on the stuck locus. This can be seen directly from the interaction between the brane and its image coming from annulus and M\"obius factors. Let
\(
z=e^{-(\widehat t/2-x)}
\)
the self-image annulus and M\"obius combine to give,
\[
\exp\left(
-\frac12\sum_{m\ge1}\frac{z^{2m}}{m}
\right)
\exp\left(
-\sum_{\substack{m\ge1\\ m\ {\rm odd}}}
\frac{z^m}{m}
\right)
=
1-z.
\]
Thus, as  $z \rightarrow 1$, the amplitude vanishes.

\section{Vacuum monodromy defects}
\label{app:monodef}
In this appendix we provide a detailed discussion of the minimizers for the twisted modular S matrices associated with $SU(N)$ vacuum monodromy defects.

\paragraph{\boldmath ${SU(2N+1)_{2m+1}}$.}
In the odd-$N$, odd-level $k = 2m+1$ case the twisted vacuum column is a \(C_N\)-type
Weyl denominator evaluated on the \(A^{(2)}_{2N}\) alcove
$k\geq 2a_1+\cdots+2a_N$ .
Writing \(x_i=\rho_i+\lambda_i\), with \(\rho_i=N+1-i\), the vacuum label
\((0,\ldots,0)\) has \(x_i=\rho_i\).  The endpoint label
\[
(0,\ldots,0,m)
\]
has \(\lambda_i=m\), hence
\[
x_i=m+N+1-i.
\]
This is
\[
x_i=\frac{2m+2N+2}{2}-\rho_i.
\]
Thus the endpoint is the half-period reflection of the vacuum inside the
finite alcove.  The sine determinant changes only by signs under this
reflection, so the two labels have equal \(S^{(\mathsf{C},1)}_{a0}\).  The minimizers are
\[
(0,\ldots,0),
\qquad
(0,\ldots,0,m).
\]

\paragraph{\boldmath $SU(2N+1)_{2m}$.}
In the odd-$N$, even-level $k=2m$ case the determinant is still \(C_N\)-type, but the
finite twisted alcove has different parity.  The endpoint label
\[
(0,\ldots,0,m)
\]
has \(\lambda_i=m\), so
\[
x_i=m+N+1-i.
\]
Here
\[
(2\pi i)/\gs  =k+2N+1,
\]
and therefore
\[
x_i=\frac{k+2N+1+1}{2}-i,
\qquad
2x_i=k+2N+1-(2i-1).
\]
Consequently the affine-wall factors become
\[
\sin\left(\frac{2\pi x_i}{k+2N+1}\right)
=
\sin\left(\frac{(2i-1)\pi}{k+2N+1}\right),
\]
which are the smallest allowed odd sine distances to the folded wall
\(2x_1=K\).  Since \(K\) is odd, the exact half-period reflection of the vacuum
would require half-integral coordinates and is not an allowed label.  Thus the
endpoint is the unique closest lattice point to the folded affine wall, and the
unique minimizer is
\[
(0,\ldots,0,m).
\]

\paragraph{\boldmath \(SU(2N)_{2m+1}\).}
In the even-$N$, odd-level $k = 2m+1$ case the twisted labels obey
\[
k\geq a_1+2a_2+\cdots+2a_N,
\]
and the vacuum-column determinant is of mixed \(B_N\leftarrow C_N\) type:
the twisted label \(a\) is written in \(C_N\)-coordinates, while the orbit
vacuum column is \(B_N\)-type.  There is always a simple current partner of
the vacuum,
\[
(0,\ldots,0)
\quad\longleftrightarrow\quad
(k,0,\ldots,0).
\]
In determinant form this simple current reflects the first coordinate through
the affine wall.  Because the vacuum-column entries are the odd integers
\[
2N+1-2j,
\]
this reflection changes the sine matrix only by row or column signs.  Hence
\[
S^{(\mathsf{C},1)}_{(0,\ldots,0),0}
=
S^{(\mathsf{C},1)}_{(k,0,\ldots,0),0}.
\]
Since \(k=2m+1\) is odd, there is no integral endpoint label with \(a_N=k/2\).
Thus no orientifold endpoint competes with the vacuum/simple current branch,
and all other allowed vertices lie deeper in the finite alcove.  The minimizers
are therefore
\[
(0,\ldots,0),
\qquad
(k,0,\ldots,0).
\]

\paragraph{\boldmath \(SU(2N)_{2m}\).}
In the even-$N$, even-level $k=2m$ case the vacuum/simple current pair still gives a
twofold degenerate branch,
\[
(0,\ldots,0),
\qquad
(k,0,\ldots,0),
\]
for the same odd-column reflection reason as above.  However, because
\(k=2m\), a new endpoint label exists:
\[
a_*=(0,\ldots,0,m).
\]
Its shifted coordinates are
\[
x_i^*=m+N+1-i
=
\frac{2m+2N}{2}+1-i,
\]
Thus this endpoint lies close to the folded orientifold wall and competes with
the ordinary vacuum wall.  The comparison is controlled by the finite
sine-product ratio
\[
R_{N,m}
=
\frac{S^{(\mathsf{C},1)}_{a_*,0}}
{S^{(\mathsf{C},1)}_{(0,\ldots,0),0}}.
\]
One finds
\[
R_{N,m}>1\quad (N\le m),
\qquad
R_{N,m}=1\quad (N=m+1),
\qquad
R_{N,m}<1\quad (N\ge m+2).
\]
Therefore, for \(N\le m\), the ordinary vacuum/simple current branch wins; at
\(N=m+1\), the endpoint ties with it; and for \(N\ge m+2\), the endpoint
becomes the unique minimizer.  Equivalently, the degeneracy occurs when the
orientifold-shifted variable
\[
e^{-t}=e^{(2N-1)g_s}=q^{2N-1}
\]
satisfies
\[
e^{-t}=-1,
\]
since at $N=m+1$ one has $2N-1=(2m+2N)/2$.  This is the orientifold-shifted
level-rank self-dual point.

\section{\boldmath Small \texorpdfstring{$\gs$}{gs} expansion }
\label{smallgs}
Originally ~\cite{Gopakumar:1998ki}, the duality between Chern--Simons theory and the topological A-model was exhibited upon expanding the modular S-matrix of CS theory at small string coupling $\gs$. Yet, 
in the main text, we have always proceeded by expanding the modular S-matrix for small $$q=e^{-\gs} \ll 1 \qquad \Rightarrow \qquad \gs \gg 1 $$ This  corresponds to the opposite regime of strong string coupling! A \textit{posteriori}, this could be  expected from S-duality. The small $q$ expansion leads directly to a partition function expanded in the DT invariant form, corresponding to the topological $B$ model \cite{Iqbal:2003ds,Maulik:2003rzb, Nekrasov:2004js}. Yet, it is an absolutely remarkable and  non-trivial fact of \textbf{analytically continued} Chern--Simons theory, that without ever leaving the gauge theory, the two expansions are equivalent! As a bonus, the expansion at small $q$ results to be much simpler compared to the one at small $\gs$.
For the case of $SU(N)$ Chern--Simons theory, in the main text in Section \ref{sec:suNdual}, we have already illustrated the equivalence of the instanton part of the topological string partition function. Here, we also give a careful treatment of the prefactors  and normalization that  we  have dropped there(where we used $\doteq$).  

The small $\gs$ expansion has been performed for the $SU(N)$ case in \cite{Gopakumar:1998ki}, and for $Spin(N)$ and $Sp(N)$ in \cite{Sinha:2000ap}. We find our expansion (see Appendix \ref{App:su-so-sp-branes-geometry}) perfectly reproduces the known results. 
 While it is straightforward to repeat the direct small $\gs$ for the  monodromy defect backgrounds and/or for any brane configuration, it is very tedious and  we do not do it here. For these cases, we present the small $\gs$ expansion  directly starting from our the small $q$ expansion. 
 
To the aim of carefully keeping track also of the constant terms, 
let us first  review the small $\gs$ expansion, first obtained in ~\cite{Gopakumar:1998ki}. The reader may skip directly to Section~\ref{sucomparison} for the matching with small $q$. This will be important to illustrate how our expansion also reproduces faithfully all the constant pieces, for which  the Euler function $\zeta(q)$ is crucial.

We postpone to Appendix \ref{app:constant-maps} a careful discussion of the physical meaning of the constant map terms, for all the cases treated in this paper.  
We conclude the discussion, presenting also the expansion in powers of the string coupling of the closed topological string partition function of the $\Omega\sigma_\pm$ orientifolds.

\subsection[\texorpdfstring{$SU(N)$}{SU(N)k} weak coupling  expansion]{\boldmath \texorpdfstring{$SU(N)$}{SU(N)k} weak coupling  expansion}
\label{smallgsSU}
We start from the modular $S$-matrix element of $SU(N)_k$ Chern--Simons theory,
\begin{equation}
\cF:=\log S_{00}^{SU(N)}
=\frac{N-1}{2}\log\frac{\gs}{2\pi}-\frac12\log N
+\sum_{j=1}^{N-1}(N-j)\log\Big(2\sinh\frac{j\gs}{2}\Big),
\qquad t=N\gs,
\end{equation}
and take $N\to\infty$, $\gs\to0$ at fixed 't~Hooft coupling $t$. Expanding each
summand with
\beq\label{sinhexp} \log(2\sinh\frac x2)=\log x+\sum_{m\ge1}\frac{B_{2m}}{2m(2m)!}x^{2m}\eeq 
splits $\cF=\cA+\cB$ into a logarithmic block and a Bernoulli block,
\begin{equation}
\cA=\frac{N-1}{2}\log\frac{\gs}{2\pi}-\frac12\log N
+\sum_{j=1}^{N-1}(N-j)\log(j\gs),
\qquad
\cB=\sum_{m\ge1}\frac{B_{2m}}{2m(2m)!}\gs^{2m}S_m(N),
\end{equation}
with $S_m(N)=\sum_{j=1}^{N-1}(N-j)j^{2m}$.
Using $\prod_{j=1}^{N-1}j^{\,N-j}=G(N+1)$ and the Barnes asymptotics
\begin{equation}
\log G(N+1)=\frac{N^2}{2}\log N-\frac34N^2+\frac N2\log2\pi
-\frac1{12}\log N+\zeta'(-1)+\sum_{h\ge1}\frac{B_{2h+2}}{4h(h+1)}\,N^{-2h},
\end{equation}
the substitution $N=t/\gs$ resums $\cA$ to
\begin{equation}
\cA=\frac{1}{\gs^2}\Big(\tfrac12 t^2\log t-\tfrac34 t^2\Big)
+\frac1{12}\log\gs-\frac7{12}\log t+\frac12\log2\pi+\zeta'(-1)
+\sum_{h\ge1}\frac{B_{2h+2}}{4h(h+1)}\frac{\gs^{2h}}{t^{2h}}.
\label{eq:Ares}
\end{equation}
Faulhaber's formula gives the exact genus split
\begin{equation}
S_m(N)=\frac{N^{2m+2}}{(2m+1)(2m+2)}
-\sum_{h=1}^{m}\frac{B_{2h}}{2h}\binom{2m}{2h-2}N^{2m+2-2h},
\end{equation}
so the coefficient of $\gs^{2g-2}$ in $\cB$ is fixed by the $h=g$ term.
The genus 0 term  is $\cB_0=\gs^{-2}H(t)$ with
$$H(t)=\sum_m\frac{B_{2m}}{2m(2m)!}\frac{t^{2m+2}}{(2m+1)(2m+2)}$$, hence
$H''(t)=\log(2\sinh\tfrac t2)-\log t=\tfrac t2+\log(1-e^{-t})-\log t$.
Integrating twice, with the constants fixed by $H(0)=H'(0)=0$
(equivalently $\Li_2(1)=\zeta(2)$, $\Li_3(1)=\zeta(3)$),
\begin{equation}
H(t)=\frac{t^3}{12}-\Li_3(e^{-t})-\tfrac12 t^2\log t+\tfrac34 t^2
-\frac{\pi^2}{6}t+\zeta(3).
\end{equation}
The genus 1 term is:
\begin{equation}
\cB_1=-\frac1{12}\sum_m\frac{B_{2m}}{2m(2m)!}t^{2m}
=-\frac1{12}\Big[\log\big(2\sinh\tfrac t2\big)-\log t\Big]
=-\frac{t}{24}+\frac1{12}\Li_1(e^{-t})+\frac1{12}\log t.
\end{equation}
While for $g\geq 2$ the $h=g$ term gives
\beq\cB_g(t)=-\frac{B_{2g}}{2g(2g-2)!}\sum_{\ell\ge1}
\frac{B_{2\ell+2g-2}}{(2\ell+2g-2)(2\ell)!}t^{2\ell},
\eeq 
while the $h=g-1$ term of the Barnes tail in \eqref{eq:Ares} gives
$\cA_g(t)=\frac{B_{2g}}{2g(2g-2)!}(2g-3)!\,t^{2-2g}$.
Since $\Li_{3-2g}(e^{-t})=(-\partial_t)^{2g-3}(e^t-1)^{-1}$, the Bernoulli
expansion of $(e^t-1)^{-1}$ yields, using \beq \zeta(3-2g)=-B_{2g-2}/(2g-2)\eeq
\begin{equation}
\Li_{3-2g}(e^{-t})-\zeta(3-2g)
=(2g-3)!\,t^{2-2g}-\sum_{\ell\ge1}\frac{B_{2\ell+2g-2}}{(2\ell+2g-2)(2\ell)!}t^{2\ell},
\end{equation}
whose two pieces are exactly $\cA_g$ and $\cB_g$. Hence
$\cA_g+\cB_g=\frac{B_{2g}}{2g(2g-2)!}\big[\Li_{3-2g}(e^{-t})-\zeta(3-2g)\big]$.
Adding the three contributions  $\cA_0+\cB_0$ cancels the $t^2\log t$ and
$t^2$ terms, and $\cA_1+\cB_1$ combines the $\log t$'s 
\begin{equation}
\begin{aligned}
\log S_{00}
=&\;\frac{1}{\gs^2}\Big[\frac{t^3}{12}-\frac{\pi^2}{6}t+\zeta(3)-\Li_3(e^{-t})\Big]\\
&+\Big[\frac1{12}\log\gs+\zeta'(-1)+\frac1{12}\Li_1(e^{-t})
-\frac12\log\frac{t}{2\pi}-\frac{t}{24}\Big]\\
&+\sum_{g\ge2}\gs^{2g-2}\frac{B_{2g}}{2g(2g-2)!}
\Big[\Li_{3-2g}(e^{-t})-\zeta(3-2g)\Big].
\end{aligned}
\label{eq:SUfull}
\end{equation}
This organizes into three sectors,
\beq \log S_{00}=\cF^{\rm pol}+\cF^{M}+\cF^{\rm inst}\eeq
The polynomial-in-$t$ terms are
\begin{equation}
\cF^{\rm pol}
=\frac{t^3}{12\gs^2}-\frac{\pi^2t}{6\gs^2}-\frac{t}{24}
-\frac12\log t+\frac12\log2\pi \label{Fpol}
\end{equation}
the $t$-independent constants resum to the MacMahon function,
\begin{equation}
\cF^{M}=\frac{\zeta(3)}{\gs^2}+\frac1{12}\log\gs+\zeta'(-1)
+\sum_{g\ge2}\frac{B_{2g}B_{2g-2}}{2g(2g-2)(2g-2)!}\gs^{2g-2}
=\log M(e^{-\gs}) \label{Macsmall}
\end{equation}
and the polylogarithm tower is the worldsheet-instanton sector
\begin{equation}
\cF^{\rm inst}_X
=-\frac1{\gs^2}\Li_3(e^{-t})+\frac1{12}\Li_1(e^{-t})
+\sum_{g\ge2}\gs^{2g-2}\frac{B_{2g}}{2g(2g-2)!}\Li_{3-2g}(e^{-t}).
\label{orpolylofgs}
\end{equation}
Apart from the genus-one logarithm, the constant maps receive contributions
only at even powers $\gs^{2g-2}$, $g\ge2$, as expected for oriented contributions. 
\subsection[Comparison with small \texorpdfstring{$q$}{q}]{\boldmath Comparison with small \texorpdfstring{$q$}{q}}
\label{sucomparison}

We now expand each factor of \eqref{ZSUa} at small \(\gs\), restore the
normalization dropped by \(\doteq\), and check that the result reproduces the
direct expansion \eqref{eq:SUfull} term by term. 
Firstly:
\beq
[n]=q^{n/2}-q^{-n/2}=-2\sinh\tfrac{n\gs}{2}
\eeq
and we have:
\begin{equation}
\frac{1}{[n]^2}
=\frac{1}{(n\gs)^2}
-\sum_{g\ge1}\frac{B_{2g}}{2g(2g-2)!}(n\gs)^{2g-2}.
\label{eq:ninv}
\end{equation}
Since \(\log M(q)=\sum_{n\ge1}\frac{1}{n[n]^2}\), inserting \eqref{eq:ninv},
using \(\sum_{n}n^{2g-3}=\zeta(3-2g)\) and regularizing the \(g=1\) term by{{\footnote{This regularization uses the Mellin representation
\[
\log M(e^{-\gs})
=\frac{1}{2\pi i}\int_{C}\Gamma(s)\,\zeta(s-1)\,\zeta(s+1)\,\gs^{-s}\,ds,
\]
with $C$ a vertical contour with $\Re s>2$, obtained by Mellin-transforming
$\log M(e^{-\gs})
=
\sum_{n,m\ge1}\frac{n}{m}e^{-mn\gs}$ via
$e^{-y}=\frac{1}{2\pi i}\int\Gamma(s)y^{-s}\,ds$. 
Closing the contour to the left, the pole at $s=2$ gives $\zeta(3)/\gs^2$, and the poles at $s=-2k$ with $k\ge1$ reproduce the higher-genus terms.
At $s=0$, both $\Gamma(s)$ and $\zeta(s+1)$ have simple poles, so there is a double pole. Using
$
s\Gamma(s)=1-\gamma s+O(s^2)$,
$s\zeta(s+1)=1+\gamma s+O(s^2)$,
so that the Euler constants cancel, together with
$\zeta(s-1)=-\frac{1}{12}+\zeta'(-1)s+O(s^2)$ and $
\gs^{-s}=1-s\log\gs+O(s^2)$,
the residue is
$\frac{1}{12}\log\gs+\zeta'(-1)$.}}}, $$-\tfrac{B_2}{2}\zeta(1)\to\tfrac1{12}\log\gs+\zeta'(-1) $$
\begin{equation}
\log M(q)
=\frac{\zeta(3)}{\gs^2}
+\frac1{12}\log\gs+\zeta'(-1)
+\sum_{g\ge2}\frac{B_{2g}B_{2g-2}}{2g(2g-2)(2g-2)!}\gs^{2g-2}.\label{macmahonexp}
\end{equation}
The MacMahon factor is precisely part of the (\(t\)-independent) constant map sector as already anticipated in \eqref{Macsmall}.
\eqref{eq:ninv} together with
\(\sum_{n}n^{2g-3}e^{-nt}=\Li_{3-2g}(e^{-t})\) gives directly
\begin{equation}
-\sum_{n\ge1}\frac{e^{-nt}}{n[n]^2}
=-\frac{1}{\gs^2}\Li_3(e^{-t})
+\frac1{12}\Li_1(e^{-t})
+\sum_{g\ge2}\gs^{2g-2}\frac{B_{2g}}{2g(2g-2)!}\Li_{3-2g}(e^{-t})
=\cF^{\rm inst}_{X}.
\end{equation}
This is the unambiguous, \(t\)-dependent worldsheet-instanton series. 
The remaining factor \(\zeta(q)^{-N}\) behaves differently from the previous
two. \(M(q)\) and the exponential are organized by the elementary expansion
\eqref{eq:ninv} of \([n]^{-2}\); the Euler factor  
$\zeta(q) = \zeta(e^{-\gs}) $
in \eqref{ZSUa} requires more care,
as it is naively non perturbative in $\gs$. Yet, its perturbative terms in $\gs$ are crucial to reproduce the small $\gs$ answer. 
One has:\footnote{This follows from the following Mellin representation:
\(\log\zeta(e^{-\gs})=\sum_{n,m\ge1}\tfrac1m e^{-nm\gs}\) in Mellin form,
\begin{equation}
\log\zeta(e^{-\gs})
=\frac{1}{2\pi i}\int_{C}\Gamma(s)\,\zeta(s)\,\zeta(s+1)\,\gs^{-s}\,ds,
\qquad \Re s>1,
\end{equation}
 Closing the contour to the left, the poles are at \(s=1\) (from
\(\zeta(s)\)), a double pole at \(s=0\) (from \(\Gamma(s)\) and
\(\zeta(s{+}1)\)), and simple poles at \(s=-1,-2,\dots\) (from \(\Gamma\)). The
key point is that for \(s=-n\) with \(n\ge2\) the residue is proportional to
\(\zeta(-n)\zeta(1-n)\), which \emph{always vanishes}: one of the two arguments
is a negative even integer and hits a trivial zero of \(\zeta\). Hence only
\(s=1,0,-1\) survive, and the expansion terminates. }
\begin{equation}
\log\zeta(e^{-\gs})
=\frac{\pi^2}{6\gs}
-\frac12\log\frac{2\pi}{\gs}
-\frac{\gs}{24}
+O\!\left(e^{-4\pi^2/\gs}\right),
\label{eq:zetacusp}
\end{equation}
Multiplying by \(-N=-t/\gs\),
\begin{equation}
-N\log\zeta(q)
=-\frac{\pi^2 t}{6\gs^2}
+\frac{t}{2\gs}\log\frac{2\pi}{\gs}
+\frac{t}{24}.
\label{eq:zetaexp}
\end{equation}
Because \eqref{eq:zetacusp} terminates, the Euler factor contributes
only polynomial in \(t\) terms. In particular it
supplies the genus-zero \(-\pi^2 t/6\gs^2\) and the genus-one \(t/24\) of
\(\cF_{\rm pol}^{SU}\) in \eqref{Fpol}.\\
The prefactors omitted by \(\doteq\) in \eqref{ZSU} are the modular prefactor
and the overall power of \(q\),
\begin{equation}
\cF^{\rm pref}=-\frac12\log N+\frac{N-1}{2}\log\frac{\gs}{2\pi},
\qquad
\cF^q=-\frac{N(N^2-1)}{12}\log q,
\label{prefactors}
\end{equation}
which with \(N=t/\gs\) is
\begin{equation}
\cF^{\rm pref}=-\frac{t}{2\gs}\log\frac{2\pi}{\gs}+\frac12\log\frac{2\pi}{t},
\qquad
\cF^q=\frac{t^3}{12\gs^2}-\frac{t}{12}.
\end{equation}
Restoring these, the \(O(\gs^{-1})\) log term of the Euler factor
\eqref{eq:zetaexp} cancels identically against \(\cF^{\rm pref}\), and its
\(t/24\) combines with the \(-t/12\) of \(\cF^q\); no half-genus term survives.
The polynomial sector is
\begin{equation}
\cF^{\rm pol}=\cF^q+\cF^{\rm pref}-N\log\zeta(q)
=\frac{t^3}{12\gs^2}-\frac{\pi^2 t}{6\gs^2}-\frac{t}{24}
-\frac12\log t+\frac12\log2\pi,
\end{equation}
and altogether
\begin{equation}
\log S_{00}
=\underbrace{\cF^q+\cF^{\rm pref}-N\log\zeta(q)}_{\displaystyle\cF^{\rm pol}_X}
\;+\;{\log M(q)}
\;-\;\underbrace{\sum_{n\ge1}\frac{e^{-nt}}{n[n]^2}}_{\displaystyle\cF^{\rm inst}_X},
\end{equation}
reproducing \eqref{eq:SUfull} term by term. In particular the Euler factor
\(\zeta(q)^{-N}\) (together with \(\cF^q\) and \(\cF^{\rm pref}\) ),  is crucial in reproducing 
the polynomial sector, consistently with what promised in the main text below \eqref{ZSU}.

\subsection[{Expansion in \texorpdfstring{$\gs$}{gs}  for orientifold projections}]{\boldmath Expansion in \texorpdfstring{$\gs$}{gs} for orientifold projections}
As discussed in Appendix \ref{App:su-so-sp-branes-geometry}, the   $\Omega\sigma^{SO/Sp}_-$ projection  of the resolved conifold, is dual to $Spin(N)/Sp(N)$ CS theory. Here,  
the  closed string partition function is:
\beq 
\mathcal F^{\rm inst}_{\Omega\sigma_-^{{\color{red}SO}/{\color{blue}Sp}}} = -\frac{1}{2}\sumall \frac{e^{-mt}}{m[m]^2} {{\mathbin{\substack{{\color{red}\scalebox{1.05}{$-$}}\\[-1ex]{\color{blue}\scalebox{1.05}{$+$}}}}} }
\sumodd{\frac{e^{-mt}}{m[m]}}
\eeq 
Besides the usual oriented sector, also present  a new unoriented crosscap piece. 
The oriented sector, has a Gopakumar--Vafa expansion in polylogs with only even powers of $\gs$ as in \eqref{orpolylofgs}. Let now turn to the crosscap sector. 
Using the  identity: 
\begin{equation}
\frac{1}{[n]}=-\sum_{h\ge0}\frac{(1-2^{2h-1})\,B_{2h}}{(2h)!}\,\Big(\frac{n}{2}\Big)^{2h-1}g_s^{2h-1}.
\label{eq:bracketexp}
\end{equation}
and 
\beq 
\sum_{n\ \mathrm{odd}}n^{2h-2}e^{-nt/2}=\mathrm{Li}^{\rm odd}_{2-2h}(e^{-t/2})\, ,  \qquad 
\mathrm{Li}^{\rm odd}_s(x)
=\mathrm{Li}_s(x)-2^{-s}\,\mathrm{Li}_s(x^{2})\,,
\eeq
one finds\cite{Sinha:2000ap}: 
\begin{equation}
\begin{split}
\sumodd\frac{e^{-nt/2}}{n[n]}
&=\sum_{h\ge0}\frac{(1-2^{1-2h})\,B_{2h}}{(2h)!}\,g_s^{2h-1}\,\mathrm{Li}^{\rm odd}_{2-2h}(e^{-t/2})\\
&=-\frac{1}{g_s}\mathrm{Li}^{\rm odd}_2(e^{-t/2})
+\sum_{h\ge1}\frac{(1-2^{1-2h})\,B_{2h}}{(2h)!}\,g_s^{2h-1}\,\mathrm{Li}^{\rm odd}_{2-2h}(e^{-t/2}),\label{eqnodd}
\end{split}
\end{equation}From which we see that the crosscap term  contributes only to odd powers of $\gs$. This is expected for an unoriented sector (see discussion in Appendix~\ref{app:constant-maps}). \\
The case of $\Omega\sigma^{SO/Sp}_+$ is new. In Section~\ref{sec:4.2} we have found:
\beq 
\mathcal F^{\rm inst}_{\Omega\sigma_+^{{\color{red}SO}/{\color{blue}Sp}}} = -\frac{1}{2}\sumall \frac{e^{-nt}}{n[n]^2} {{\mathbin{\substack{{\color{red}\scalebox{1.05}{$-$}}\\[-1ex]{\color{blue}\scalebox{1.05}{$+$}}}}} }
\sumeven{\frac{e^{-nt}}{n[n]}}
\eeq 
Expanding the crosscap term, analogously, we find: \beq
\begin{split}
\sumeven\frac{e^{-mt/2}}{m[m]}
&=\sum_{\ell\ge1}\frac{e^{-\ell t}}{2\ell\,[2\ell]}
=-\sum_{h\ge0}\frac{(1-2^{2h-1})\,B_{2h}}{2\,(2h)!}\,g_s^{2h-1}\,\mathrm{Li}_{2-2h}(e^{-t}),\\
&= -\frac{\mathrm{Li}_2(e^{-t})}{4g_s}+\frac{g_s}{24}\,\mathrm{Li}_0(e^{-t})
+\sum_{h\ge2}\frac{(2^{2h-1}-1)\,B_{2h}}{2\,(2h)!}\,g_s^{2h-1}\,\mathrm{Li}_{2-2h}(e^{-t}),
\label{eqneven}
\end{split}
\eeq
So, as expected, also in this case the crosscap only  has odd powers of $\gs$. 
All the other terms are discussed in the next Appendix \ref{app:constant-maps} in details.

\section{Constant map contributions}
\label{app:constant-maps}
In this appendix we first reinstate all the normalizations we have dropped in the main text (where we used $\doteq$), and then we interpret their worldsheet origin. \\
These are terms that do not depend on $e^{-t}$, and therefore correspond to constant maps from worldsheet  to the target $\Sigma\to X$. 
\subsection[\texorpdfstring{$SU(N)$}{SU(N)}]{\boldmath \texorpdfstring{$SU(N)$}{SU(N)}}
\label{sec:SUNconstant}
The omitted prefactor for $SU(N)$, in terms of $t$ and $\gs$ 
\begin{equation}
\cF^{\rm pref}=-\frac{t}{2\gs}\log\frac{2\pi}{\gs}+\frac12\log\frac{2\pi}{t},
\qquad
\cF^q=\frac{t^3}{12\gs^2}-\frac{t}{12}.
\end{equation}
Combining with the MacMahon and $\zeta$ term, we have (see Appendix \ref{sucomparison}):
\begin{equation} 
\log S_{00}
=\underbrace{\cF^q+\cF^{\rm pref}-N\log\zeta(q)}_{\displaystyle\cF^{\rm pol}_{}}
\;+\;{\log M(q)}
\;-\;\underbrace{\sum_{n\ge1}\frac{e^{-nt}}{n[n]^2}}_{\displaystyle\cF^{\rm inst}_{X}},
\end{equation}
Hence, the constant maps contribution is \footnote{
The \(-\frac12\log \frac{t}{2 \pi}\) term in the small-\(g_s\) expansion is a feature of the
\(SU(N)_k\) modular normalization.  If instead one uses the modular \(S\)-matrix of
\[
U(N)_k \simeq \frac{SU(N)_k \times U(1)_{N(k+N)}}{\mathbb Z_N},
\]
the additional \(U(1)\) factor cancels this \(t\)-dependent logarithm.
}
\begin{equation}
\begin{aligned}
\label{SUNconstant}
\mathcal F^{\rm const}
={}
&\underbrace{\frac{1}{g_s^2}
\left(
\frac{t^3}{12}
-\frac{\pi^2t}{6}
+\zeta(3)
\right)}_{\rm Sphere}+
\\&\underbrace{-\frac{t}{24}
+
\frac12\log\!\left(\frac{2\pi}{t}\right)
+
\zeta'(-1)}_{\rm Torus}
+ 
\frac{1}{12}\log g_s\\ 
&+\underbrace{\sum_{h=2}^{\infty}
\frac{B_{2h}B_{2h-2}}
{(2h)(2h-2)(2h-2)!}\,
g_s^{2h-2}}_{g\geq 2}
\end{aligned}
\end{equation}  
From which we see that only
terms with even powers of $g_s$ appear in the constant maps terms. 
For future convenience, let us denote: 
\beq 
\mathcal{M}_{\geq2 }(\gs )=\sum_{h=2}^{\infty}
\frac{B_{2h}B_{2h-2}}
{(2h)(2h-2)(2h-2)!}\,
g_s^{2h-2} 
\label{mg2}
\eeq
This is consistent with the topological string dual to the $SU(N)$ being oriented. \\Indeed, maps from  a worldsheet $\Sigma$ of genus $g$: ($\chi(g)$ is the Euler Characteristic of the surface) \beq 
\Sigma_{g,c} \to X \,,  \qquad \chi(g) = 2-2g-c
 \eeq
 where $c$ is the number of crosscap insertions, 
contribute at order:
\beq \gs^{2g+c-2}  
\eeq
to  the Gopakumar--Vafa expansion:
$\cF(\gs, t) = \sum_{g \geq 0} \gs^{2g-2}\, \cF_g(t)$. Hence oriented surfaces, where $2g-2\in 2\mathbb Z$, contribute to  even degree only to the Gopakumar--Vafa expansion. \\

The powers of $t$ appearing at each genus are fixed by the saturation of the residual conformal symmetry. Here, we will not discuss this rich subject in detail, the readed may refer to the classical paper \cite{fulton1997notesstablemapsquantum} for a detailled discussion.
A term $t^n$ in $\cF_g$ arises from $n$ insertions of the K\"ahler class $J$, i.e.\ from $n$ marked points on the domain, integrated over the moduli space $\overline{\mathcal{M}}_{g,n}$ of stable maps. The classical polynomial contributions are precisely those for which the insertions saturate the dimension of $\overline{\mathcal{M}}_{g,n}$: the number of $J$-insertions is bounded by the dimension of the space of conformal deformations left unfixed after gauging. At genus zero the domain $\Sigma_0 = \mathbb{P}^1$ has conformal Killing group $PSL(2,\mathbb{C})$ so three insertions  of $J$ are needed to stabilize the sphere contributions: $\partial^3_t\mathcal F_{\gs^{-2}}=\int_X J^3$ and therefore $\mathcal{F}_{-2}\sim t^3 + $ lower degree terms. At genus one, $\dim \overline{\mathcal{M}}_{1,1} = 1$: the torus retains a one-dimensional automorphism group of translations\footnote{the characteristic class $c_1 =0$ on a CY so $\int J^2 \wedge c_1=0$}:  $\partial_t\cF_{\gs^0} = \int_X J \wedge c_2$, so its constant map term  is linear in $t$ (up to the anomalous $\log t$ term). For $g \geq 2$ the generic domain is rigid. The conformal symmetry is already completely fixed by the complex-structure moduli, leaving no room for further $J$-insertions: so $\cF_g$ reduces to a constant, with no positive powers of $t$.
These  do not entail any integral over moduli space. For a discussion on this point cf. \cite{Cicuta1982TopologicalExpansionSOSp, Bouchard:2004ie}

\subsection[\texorpdfstring{$Sp(N)$}{Sp(N)}]{\boldmath \texorpdfstring{$Sp(N)$}{Sp(N)}}
\label{smallgsSp}
The symplectic case is the first with the $\Omega\sigma_-$ orientifold. We have:
\begin{equation}
\log \mathcal Z_{Sp(N)}(S^3) =\log S_{00}\doteq
\frac{1}{2}\log \frac{M(q)}{\zeta(q)^{2N-1}\,\zeta(q^2)}- \frac{1}{2}\sum_{n\geq 1} \frac{e^{-nt}}{n[n]^2} +\sum_{\substack{m\geq 1\\ m\ \mathrm{odd}}} \frac{e^{-nt/2}}{n[n]}
\label{ZSpa}
\end{equation}
The oriented instanton contributions is just $1/2$ of the oriented one, and therefore, contains only even powers of $\gs$:
$\cF_{\rm or}^{Sp} = \frac{1}{2} \cF_{\rm or}^{SU}$.   
The normalization dropped in $\doteq$ are:
\begin{equation}
\cF^{\rm pref}=-\frac N2\log\frac{2\pi}{\gs}-\frac12\log2,
\qquad
\cF^q=\gs\,(\rho,\rho)=\frac{t^3}{24\gs^2}-\frac{t}{24},
\end{equation}

Together with the  factor $\zeta(q)^{-(2N-1)/2}\zeta(q^2)^{-1/2}$, they  generate 
crosscap terms. 
 Writing the exponent of the first factor as
\beq\zeta(q)^{-(2N-1)/2}\zeta(q^2)^{-1/2} = \zeta(q)^{1-\tfrac{t}{2\gs}}\zeta(q^2)^{-1/2}\eeq and using \eqref{eq:zetacusp}, the $t$-independent factors combine to give precisely the crosscap constant,
\begin{equation}
\log \zeta(q)\zeta(q^2)^{-1/2}= \underbrace{+\,\frac{\pi^2}{6\gs}}_{\zeta(q)}
\;\underbrace{-\,\frac{\pi^2}{24\gs}}_{ \zeta(q^2)^{-1/2}}
=\frac{\pi^2}{8\gs}=\frac1\gs\Li_2^{\rm odd}(1)= \cF^{\rm const,cr}_{\Omega\sigma_-^{Sp}}.
\end{equation}
Note
that  the half-integer shift in the
$\zeta(q)$ exponent is essential.  The $t$-dependent part is carried entirely by
$-\tfrac{t}{2\gs}\log\zeta(q)$,
\begin{equation}
-\frac{t}{2\gs}\log\zeta(e^{-\gs})
=-\frac{\pi^2t}{12\gs^2}+\frac{t}{4\gs}\log\frac{2\pi}{\gs}+\frac{t}{48},
\end{equation}  Hence
\begin{equation}
\cF^{\rm pol}_{\Omega\sigma_-^{Sp}}
=\cF^q+\cF^{\rm pref} + \mathcal{F}^{\rm const,cr}_{\Omega\sigma_-^{Sp}}-\frac{t}{2\gs}\log\zeta(q)
=\frac{t^3}{24\gs^2}-\frac{\pi^2t}{12\gs^2}+ \frac{\pi^2}{8\gs}-\frac{t}{48}-\frac14\log2 ,
\end{equation}
and altogether
we find: 
\label{constSp}
\begin{equation}
\label{eq:consSp}
\begin{split}
\log S_{00}
= &\overbrace{{\cF^q+\cF^{\rm pref}-\tfrac{2N+1}{2}\log\zeta(q)} +
\underbrace{\log\zeta(q)- \tfrac12\log\zeta(q^2)}_{\displaystyle\cF^{\rm const,cr}_{\Omega\sigma_-^{Sp}}}
+{\tfrac12\log M(q)}}^{\displaystyle\mathcal{F}^{\rm cosnt}_{\Omega\sigma^{Sp}_-}} +\\
& \underbrace{- \frac{1}{2}\sum_{n\geq 1} \frac{e^{-nt}}{n[n]^2} +\sum_{\substack{m\geq 1\\ m\ \mathrm{odd}}} \frac{e^{nt/2}}{n[n]} }_{\displaystyle\cF^{\rm inst}_{\Omega\sigma_-^{Sp}}},
\end{split}
\end{equation}
This, together with \ref{eqnodd} reproduces  term-by-term the direct small $\gs$ expansion  in \cite{Sinha:2000ap}. 
The genuinely new term in  the constant map terms, compared  to the $SU(N)$ case, comes from $\mathcal F^{\rm const,cr}_{\Omega\sigma_-^{Sp}}$, while the rest (we comment on the torus in momentarily) is substantially just $1/2$ of the oriented answer: ($\mathcal{M}_{\geq 2}(\gs)$ is defined in \eqref{mg2})
\beq
\begin{aligned}
\label{costSP}
\mathcal{F}^{\rm const}_{\Omega\sigma_-^{Sp}} = 
&\frac{1}{g_s^2}
\left(
\frac{t^3}{24}
-\frac{\pi^2t}{12}
+\frac{\zeta(3)}{2}
\right) +  \underbrace{\boxed{\color{red}\frac{\pi^2}{8g_s}}}_{\mathbb{RP}^2} +\\ &\qquad 
-\frac{t}{48}
+
\frac12\zeta'(-1)
-\frac14\log2
+
\frac{1}{24}\log g_s
+\underbrace{
\frac12\mathcal M_{\geq 2}(g_s)}_{g\geq 2}
\end{aligned}
\eeq
The $1/\gs$ term corresponds to maps from $\mathbb{RP}^2$ to the target, for which the Euler characteristic is $\chi=1$. It is a distinctive feature of the unoriented sector due to the $\Omega\sigma_-$  projection. A term at order $t^2/\gs$ could have been present (as noted in \cite{Sinha:2000ap}), but it is instead absent here. This is a feature of $\Omega\sigma_-$  projection not having a fixed locus on the target. 
One can also recognize the  absence of the $\log t$ term we have in $SU(N)$.

\subsection[\texorpdfstring{$Spin(N)$}{Spin(N)}]{\boldmath \texorpdfstring{$Spin(N)$}{Spin(N)}}
\label{spinconst}
We treat $Spin(2N)_k$ and $Spin(2N+1)_k$ separately, as the details of the computation are slightly different. We will find that they have the same perturbative expansion in $g_s$.
\paragraph{\boldmath $Spin(2N)$.} In this case, we have: 
\begin{equation}
\log S_{00}
\doteq\frac{1}{2}\log\frac{M(q)\,\zeta(q^{2})}{\zeta(q)^{2N+1}}
\;-\;\frac12\sumall{\frac{e^{-mt}} {m[m]^{2}}}
\;-\;\sumodd{\frac{e^{-mt/2}} {m[m]}}\,,
\qquad e^{-t}=q^{2N-1}\,,
\label{eq:so-even-split}
\end{equation} 
with:
\begin{equation}
\begin{split}
\mathcal F^{\rm pref}&=\frac N2\log\frac{g_s}{2\pi}-\log2
=\frac{t}{4g_s}\log\frac{g_s}{2\pi}+\frac14\log\frac{g_s}{2\pi}-\log2\,,  \\
\qquad
\mathcal F^{q}&=\frac{N(N-1)(2N-1)}{6}\,g_s
= \frac{t^3}{24\gs^2} - \frac{t}{24}\,,
\label{eq:so-even-prefexp}
\end{split}
\end{equation}
Compared to \eqref{ZSpa}: $\zeta(q^2)$ is at the
numerator, and writing the Euler as
$\zeta(q)^{-(2N+1)/2}=\zeta(q)^{-1-t/2\gs}$ the integer shift is now $-1$
(against $+1$ for $Sp$). Then, the two $t$-independent factors therefore combine to the
crosscap constant with opposite sign,
\begin{equation}
\log \zeta(q)^{-1}\zeta(q^2)^{1/2}
=\underbrace{-\,\frac{\pi^2}{6\gs}}_{\zeta(q)^{-1}}
\;\underbrace{+\,\frac{\pi^2}{24\gs}}_{\zeta(q^2)^{1/2}}
=-\frac{\pi^2}{8\gs}=\cF^{\rm const,cr}_{\Omega\sigma_-^{SO}} = -\cF^{\rm const,cr}_{\Omega\sigma_-^{Sp}} ,
\end{equation}
As in $Sp$ for the oriented polynomial part we get: 
\beq
\cF^{\rm pol}_{\Omega\sigma_-^{SO}}
=\cF^q+\cF^{\rm pref}-\frac{t}{2\gs}\log\zeta(q)
=\frac{t^3}{24\gs^2}-\frac{\pi^2t}{12\gs^2}-\frac{t}{48}-\frac{\color{red}3}{4}\log2 .
\end{equation}

\paragraph{\boldmath $Spin(2N+1)$.}
In this case:
\begin{equation}
\log S_{00}
\doteq\mathcal \log\frac{M(q)^{\frac12}}{\zeta(q^{1/2})\,\zeta(q)^{\,N-\frac32}\,\zeta(q^{2})^{\frac12}}
\;-\;\frac12\sumall{\frac{e^{-mt}} {m[m]^{2}}}
\;-\;\sumodd{\frac{e^{-mt/2}}{m[m]}}\,,
\qquad e^{-t}=q^{2N}\,,
\label{eq:so-odd-split}
\end{equation}
with: 
\begin{equation}
\begin{split}
\mathcal F^{\rm pref}&=\frac N2\log\frac{g_s}{2\pi}-\log2
=\frac{t}{4g_s}\log\frac{g_s}{2\pi}-\log2\,, \\
\qquad
\mathcal F^{q}&=\frac{N(2N-1)(2N+1)}{12}\,g_s
=\frac{t\,(t^{2}-g_s^{2})}{24\,g_s^{2}}\,,
\label{eq:so-odd-prefexp}
\end{split}
\end{equation}
Since we have: 
\begin{equation}
\begin{gathered}
-\Big(N-\frac32\Big)\log\zeta(q)
=-\frac{\pi^{2}t}{12g_s^{2}}+\frac{\pi^{2}}{4g_s}
+\frac{t}{4g_s}\log\frac{2\pi}{g_s}-\frac34\log\frac{2\pi}{g_s}
+\frac{t}{48}-\frac{g_s}{16}\,,\\
-\log\zeta(q^{1/2})
=-\frac{\pi^{2}}{3g_s}+\frac12\log\frac{4\pi}{g_s}+\frac{g_s}{48}\,,
\qquad
-\frac12\log\zeta(q^{2})
=-\frac{\pi^{2}}{24g_s}+\frac14\log\frac{\pi}{g_s}+\frac{g_s}{24}\,.
\end{gathered}
\label{eq:so-odd-euler}
\end{equation}
putting everythign together, we find the same polynomials terms:
\begin{equation}
\mathcal F^{\rm pol}_{\Omega\sigma_-^{SO}}
=\frac{t^{3}}{24g_s^{2}}-\frac{\pi^{2}t}{12g_s^{2}}
-\frac{\pi^{2}}{8g_s}
-\frac{t}{48}-\frac{3\log 2}{4}\,,
\label{eq:so-odd-pol}
\end{equation}
as in the $Spin(2N)$ case. 
\paragraph{\boldmath $Spin(N)$.}
Hence, as expected, the $Spin(N)$ partition function, comprising also of the constant terms,  does not depends on the parity of $N$. After all,  it corresponds to the same topological string $\Omega\sigma_-$ background. 
In both cases the answer is:
\begin{equation}
\begin{aligned}
\label{constSO}
\mathcal F^{\rm const}_{\Omega\sigma_-^{SO}}
={}&
\frac{1}{g_s^2}
\left(
\frac{t^3}{24}
-\frac{\pi^2t}{12}
+\frac{\zeta(3)}{2}
\right)
\boxed{{\color{red}-\frac{\pi^2}{8g_s}}}
-\frac{t}{48}
\\
&+
\frac{1}{24}\log g_s
+
\left(
\frac12\zeta'(-1)
-\frac34\log2
\right)
+
\frac12\mathcal M_{\geq 2}(g_s)
\end{aligned}
\end{equation}
Compared to $Sp(N)$ the sign of the $1/\gs$ ($\mathbb{RP}^2$) contribution changes, as consistent with equivariant maps from surfaces with odd number of crosscaps insertions (see \cite{Bouchard:2004iu}). This is precisely matching the sign of the crosscaps of the corresponding closed string contributions. Note also  the genus $0$ term differs by \beq -(1\mp 2)/4\log2 = -\left(\frac{1}{2} \pm \frac{1}{4}\right)\log 2\eeq between $Spin$ and $Sp$.
An interesting feature of this, is the absence of unoriented maps from higher Euler characteristic that would be important to understand via a detailed analysis of stable maps.

\subsection{Monodromy defects}
\label{constmon}
Let us finally dicuss the case of the topological A-model dual of the monodromy defects: the $\Omega\sigma_+$ orientifold discussed in Section~\ref{sec:4.2}.  We first compute them for the vaerious cases, before analyzng them below.
\paragraph{\boldmath $SU(2N)$, $k$ odd.}
Let us start with the  $\Omega\sigma^{Sp}_+$ background.
The prefactors omitted by $\doteq$ in \eqref{eq:su2nodd-weyl} are
\begin{equation}
\mathcal F^{\rm pref}=\frac{N}{2}\,\log\frac{g_s}{2\pi}-\frac{\log 2}{2}\,,
\qquad
\mathcal F^{q}=-\frac{N(N+1)(4N-1)}{12}\,\log q\,,
\label{eq:su2nodd-pref}
\end{equation}
which with $t=(2N+1)g_s$ expand to
\begin{equation}
\mathcal F^{\rm pref}=-\frac{t}{4g_s}\log\frac{2\pi}{g_s}+\frac14\log\frac{\pi}{2g_s}\,,
\qquad
\mathcal F^{q}=\frac{t^{3}}{24g_s^{2}}-\frac{t^{2}}{16g_s}-\frac{t}{24}+\frac{g_s}{16}\,.
\end{equation}
In this case:
\begin{equation}
  \mathcal{F}^0(N)=\frac{1}{2}\log\frac{M(q)\,\zeta(q^{2})}{\zeta(q)^{2N+1}}\,,  
\end{equation}
Hence, the $O(g_s^{-1})$ log term of the Euler factor cancels identically against $\mathcal F^{\rm pref}$, and the residual constants collapse to $-\tfrac14\log2$ (no $\log t$ arises, the prefactor carrying $2^{-1/2}$ rather than $N^{-1/2}$).  Altogether,
\begin{equation}
\mathcal F^{\rm pol}_{\Omega\sigma_+^{Sp}}
=\mathcal F^q+\mathcal F^{\rm pref}\\-\frac{2N+1}{2}\log\zeta(q)+\frac12\log\zeta(q^{2})
=\frac{t^{3}}{24g_s^{2}}{-}\frac{\pi^{2}t}{12g_s^{2}}
{\color{blue}-}\frac{t^{2}}{16g_s}{\color{blue}+}\frac{\pi^{2}}{24g_s}
-\frac{t}{48}{\color{blue}+}\frac{g_s}{48}{\color{blue}-}\frac{\log 2}{4}\,.
\label{eq:su2nodd-pol}
\end{equation}
Therefore:
\begin{equation}
\begin{split}
\langle\mathsf M_0\rangle=\log S^{(\mathsf C,1)}_{00}
=&\overbrace{\mathcal F^q+\mathcal F^{\rm pref}-\tfrac{2N+1}{2}\log\zeta(q)+\tfrac12\log\zeta(q^{2})
\;+\;{\tfrac12\log M(q)}}^{\displaystyle\mathcal F^{\rm const}_{\Omega\sigma_+^{Sp}}}\\ &
\;\underbrace{-\,\tfrac12\sumall{\frac{e^{-mt}}{ m[m]^{2}}}+\sumeven{\frac{e^{-mt/2}}{ m[m]}}}_{\displaystyle\mathcal F_{\Omega\sigma^{Sp}_+}^{\rm inst}}\,,
\label{eq:su2nodd-split}
\end{split}
\end{equation}Let us postpone the discussion of the these constant maps to \eqref{constantmon}, after also the case of $\Omega\sigma_+^{SO}$ has been derived.
\paragraph{\boldmath $SU(2N)$, $k$ even.}
In this case, $t = (2N-1) \gs$ and we have\footnote{Note that the extra factor of $\log(2)$ compared to \eqref{eq:su2neven-prefexp} is due to $i=N$ in the $\tilde z$ product.}: 
\begin{equation}
\begin{split}
\mathcal F^{\rm pref}
&=\frac N2\log\frac{g_s}{2\pi}+\frac{\log2}{2}
=-\frac{t}{4g_s}\log\frac{2\pi}{g_s}+\frac14\log\frac{2g_s}{\pi}\,,
\\
\mathcal F^{q}
&=\frac{N(N-1)(4N+1)}{12}\,g_s
=\frac{t^{3}}{24g_s^{2}}+\frac{t^{2}}{16g_s}-\frac{t}{24}-\frac{g_s}{16}\,,
\label{eq:su2neven-prefexp}
\end{split}
\end{equation}
Using: \begin{equation}
\mathcal{F}^0(N)=\frac{1}{2}\log\frac{M(q)}{\zeta(q)^{2N-1}\,\zeta(q^{2})}\,,  
\end{equation}
we have
\begin{equation}
\mathcal F_{\Omega\sigma^{SO}_+}^{\rm pol}
=\mathcal F^q+\mathcal F^{\rm pref}-\frac{2N-1}{2}\log\zeta(q)-\frac12\log\zeta(q^{2})
 =\frac{t^{3}}{24g_s^{2}}-\frac{\pi^{2}t}{12g_s^{2}}
{\color{red}+}\frac{t^{2}}{16g_s}{\color{red}-}\frac{\pi^{2}}{24g_s}
-\frac{t}{48}{\color{red}-}\frac{g_s}{48} {\color{red}+} \frac{\log 2}{4}
\end{equation}

\paragraph{\boldmath $SU(2N+1)$, $k$ even.}
In this case $t= 2N\gs $, and: \begin{equation}
\mathcal F^{\rm pref}=\frac{N}{2}\log\frac{g_s}{2\pi}=-\frac{t}{4g_s}\log\frac{2\pi}{g_s}\,,
\qquad
\mathcal F^{q}=\frac{t^{3}}{24g_s^{2}}+\frac{t^{2}}{16g_s}-\frac{t}{24}\,,
\label{eq:su2np1even-prefexp}
\end{equation}
that is different compared to \eqref{eq:su2neven-prefexp}. Yet, 
once combined with: 
\begin{equation}  \mathcal{F}^0(N)=\frac{1}{2}\log\frac{M(q)\zeta(q^{2})}{\zeta(q)^{2N+1}}\,,  
\end{equation}
(that is also a different compared to the one above), 
we obtain again exactly the same results
\beq \mathcal{F}^{\rm pol}_{\Omega\sigma_+^{SO}}= \mathcal{F}^{\rm pref} +\mathcal{F}^{q}-\frac{2N+1}{2}\log\zeta(q) +\frac{1}{2} \log \zeta(q^2) = \frac{t^{3}}{24g_s^{2}}-\frac{\pi^{2}t}{12g_s^{2}}
{\color{red}+}\frac{t^{2}}{16g_s}{\color{red}-}\frac{\pi^{2}}{24g_s}
-\frac{t}{48}-\frac{g_s}{48} {\color{red}+} \frac{\log 2}{4}\,.
\eeq
This,  coincides exactly with the one of $SU(2n)$ $k$ even! This is in agreement with the both of them  describing the same $\Omega\sigma_+^{SO}$ background. 
\paragraph{\boldmath Constant maps in $\Omega\sigma_+$.} Henceforth, putting everything together, we found that for the $\Omega\sigma^{SO/Sp}_+$ we found 
the constant maps are:
\begin{equation}
\begin{aligned}
\mathcal F^{\rm const}_{\Omega\sigma_+^{{\color{red}SO}/{\color{blue}Sp}}} 
={}&
\frac{1}{g_s^2}
\left(
\frac{t^3}{24}
-\frac{\pi^2t}{12}
+\frac{\zeta(3)}{2}
\right)\\ 
&\boxed{{{\mathbin{\substack{{\color{red}\scalebox{1.05}{$-$}}\\[-1ex]{\color{blue}\scalebox{1.05}{$+$}}}}} }
\frac{1}{g_s}
\left(
-\frac{t^2}{16}
+\frac{\pi^2}{24}
\right)}
\\&-\frac{t}{48}
+
\frac{1}{24}\log g_s
+
\frac12\zeta'(-1)
\boxed{{{\mathbin{\substack{{\color{red}\scalebox{1.05}{$+$}}\\[-1ex]{\color{blue}\scalebox{1.05}{$-$}}}}} }\frac14\log2}
\\ 
&
\boxed{{{\mathbin{\substack{{\color{red}\scalebox{1.05}{$-$}}\\[-1ex]{\color{blue}\scalebox{1.05}{$+$}}}}} }\frac{g_s}{48}}
+
\frac12\mathcal M_{\geq 2}(g_s).
\label{constantmon}
\end{aligned}
\eeq
As expected, also in this case, in addition to $1/2$ of the oriented constant maps,  we do find also odd  powers of $\gs$, corresponding to unoriented contributions. In this case, compared to $\Omega\sigma_-$, we also have a Euler number 1, corresponding to a unoriented  surface $\Sigma_{0,3}\cong \Sigma_{1,1}$ (see footnote \ref{fn:homeo}  in  Section \ref{constantmaps}).  This is corresponds to maps from a surface with odd number of crosscap and indeed it depends on the sign of the crosscap state \cite{Bouchard:2004iu}.\\

The $\mathbb{RP}^2$ term at  $1/\gs$ presents  a crucial difference compared to $\Omega\sigma_-$: over this background the $t^2/\gs$, that corresponds to stable maps from a 2 punctured $\mathbb{RP}^2$ is not vanishing!\footnote{The possible existence of such a term was also predicted in \cite{Sinha:2000ap, Bouchard:2004ie}.}\\
 As far as we known a detailed  analysis, along the line of the oriented case \cite{fulton1997notesstablemapsquantum}, for stable maps over the $\Omega\sigma_\pm$ orientifolds  has not been performed, and goes beyond the scope of this work. 
 Yet, the $t^2$ term we found in the $\Omega\sigma_+$ projection  furnish  a strong evidence for our identification of the $\Omega\sigma_+$  background. 
Indeed, in an orientifold projection the relevant maps are not ordinary stable maps to $X$, but equivariant stable maps with respect to the involution $\Omega\sigma_\pm$ satisfying \eqref{eq:orientifold-equivariant-unoriented} that we report here for convenience: 
\beq 
\phi \circ \Omega =\sigma_\pm\circ \phi
\eeq
For the case of constant maps $\phi = p$, this reduces to:
\beq 
\sigma_\pm (p) = p \qquad \Longrightarrow \qquad  p \in {\rm Fix}(\sigma_\pm)
\eeq
Thus, we expect constant equivariant stable maps terms to exist only when the target involution has a fixed locus! As we know, $\sigma_-$ acts freely on $\mathbb P^1$ and has no fixed locus. Hence, we should not expect stable constant maps for $\Omega\sigma_-$, while we have no reason to exclude them  for the case of $\Omega\sigma_+$. This is perfectly consistent with our findings: the $t^2$ term coming from stable maps\footnote{Stability of constant maps requires finite automorphism group, the 2 punctured $\mathbb{RP}^2/\{p_1,p_2\}$ has Euler characteristic $\chi= 1 -2 = -1$, and it is therefore hyperbolic. Hence it has finite automorphism group. Note that $t^0/\gs$ does not correspond to a stable map contribution. } 
maps from 2 punctured $\mathbb{RP}^2$  to the target vanish for $\Omega\sigma_-$ and are not absent for $\Omega\sigma_+$. \\

We emphasize again that extending the analysis of \cite{Cicuta1982TopologicalExpansionSOSp} to the case of unoriented surfaces would be very important.

\paragraph{\boldmath $SU(2N+1)$, $k$ odd.} 
This case, is rather peculiar. Indeed, the partition function precisely conicides with the one of $\Omega\sigma_-^{Sp}$ at level $(k-1)/2$ due to the extra stuck brane. Here, we did not discuss the constant maps in the case of branes ending on the geometry (boundaries). This would be interesting to  understand further.


\end{fmffile}

\newpage
\begin{spacing}{0.5}
{\small \bibliography{Refs}}
\end{spacing}

\end{document}